\documentclass[iop,apj,twocolumn,revtex4]{emulateapj}
\usepackage{courier}
\usepackage{afterpage}
\usepackage{url}
\usepackage{natbib}

\slugcomment{submitted to The Astrophysical Journal Supplements}

\newcommand{\tbmax}{\ensuremath{t_{B {\rm max}}}}

\newcommand{\dm}{\ensuremath{\Delta m_{15}(B)}}

\newcommand{\threesig}{$3$-$\sigma$}

\newcommand{\jhk}{\ensuremath{JHK_{s}}}

\newcommand{\snIa}{\mbox{SN~Ia}}
\newcommand{\snIax}{\mbox{SN~Iax}}
\newcommand{\snIbc}{\mbox{SN~Ib/c}}
\newcommand{\snII}{\mbox{SN~II}}

\newcommand{\scale}{1}

\newcommand{\ncfafive}{15}

\newcommand{\ncspir}{73}

\newcommand{\ncspobsy}{829}
\newcommand{\ncspobsj}{776} 
\newcommand{\ncspobsh}{705} 
\newcommand{\ncspobsk}{124} 
\newcommand{\ncspobsir}{2434}

\newcommand{\nptelcsp}{18} 
\newcommand{\nwvcsp}{9}
\newcommand{\nfcspnew}{9}
\newcommand{\nptelcspnormal}{14}
 
\newcommand{\nptelcspsuperch}{2}

\newcommand{\nIa}{118}

\newcommand{\nsnIacfair}{94} 
\newcommand{\nsnIaxcfair}{4} 
\newcommand{\nIanogogo}{23} 
\newcommand{\nIanogo}{23} 

\newcommand{\nsnIacfairopt}{92} 
\newcommand{\nsnIaredo}{21} 
\newcommand{\nsnIawv}{21}

\newcommand{\nsnIanorm}{88}
\newcommand{\nsnIanormtempl}{86}

\newcommand{\npretbmaxpercent}{55\%}  
\newcommand{\npretbmaxfivepercent}{34\%} 

\newcommand{\firstepoch}{13}   
\newcommand{\lastepoch}{127}

\newcommand{\jmedmagerrlcpeak}{0.032} 
\newcommand{\hmedmagerrlcpeak}{0.053}
\newcommand{\kmedmagerrlcpeak}{0.115}

\newcommand{\jmedmagerr}{0.086}  
\newcommand{\hmedmagerr}{0.122}
\newcommand{\kmedmagerr}{0.175}

\newcommand{\jmeanepochs}{18}
\newcommand{\hmeanepochs}{17}
\newcommand{\kmeanepochs}{13}

\newcommand{\jmaxepochs}{46}

\newcommand{\jmedsntemp}{4}
\newcommand{\hmedsntemp}{4}
\newcommand{\kmedsntemp}{3}

\newcommand{\jmaxsntemp}{11}

\newcommand{\obsmulting}{doubling}

\newcommand{\nptelobs}{4637}  
\newcommand{\nptelobsj}{1733}  
\newcommand{\nptelobsh}{1636}  
\newcommand{\nptelobsk}{1268}  

\newcommand{\nirIa}{200}

\newcommand{\nwvobs}{1087}

\newcommand{\nII}{20}

\newcommand{\inprep}{$in\ prep.$}

\newcommand{\etal}{et al.}

\newcommand{\nminstar}{5}

\newcommand{\nwelliso}{20} %

\newcommand{\galsubwmeanj}{$-0.0009$}
\newcommand{\galsubwmeanh}{$0.0006$}
\newcommand{\galsubwmeank}{$0.0007$}

\newcommand{\galsubewmeanj}{$0.0016$}
\newcommand{\galsubewmeanh}{$0.0019$}
\newcommand{\galsubewmeank}{$0.0026$}

\newcommand{\meanrestpj}{0.0014}
\newcommand{\meanrestph}{0.0014}
\newcommand{\meanrestpk}{-0.0055}
\newcommand{\meanrestpjerr}{0.0006}
\newcommand{\meanrestpherr}{0.0007}
\newcommand{\meanrestpkerr}{0.0007}

\newcommand{\La}{\Lambda}

\newcommand{\sn}{SN~} 

\newcommand{\WV}{WV08} 
\newcommand{\F}{CfAIR2}

\newcommand{\PTL}{PAIRITEL}

\newcommand{\swarp}{\texttt{SWarp}}
\newcommand{\dophot}{\texttt{DoPHOT}}
\newcommand{\snoopy}{\texttt{SNooPy}}
\newcommand{\bayesn}{\texttt{BayeSN}}

\newcommand{\M}{M09}

\newcommand{\MO}{M11}

\newcommand{\CFAFIVE}{CfA5 \inprep}
\newcommand{\FRIED}{Friedman \etal{} 2015a}
\newcommand{\WVF}{Wood-Vasey \etal{} 2008}

\newcommand{\HOWIEII}{Marion \etal{} 2015b \inprep} 

\newcommand{\jhcsp}{ (J-H)_{CSP} }

\newcommand{\fig}{Fig.}
\newcommand{\figs}{Figs.}
\newcommand{\eq}{Eq.}
\newcommand{\eqs}{Eqs.}

\newcommand{\nt}{N_{\rm T}}

\newcommand{\nnt}{\rm NNT}

\newcommand{\snt}{\rm SNTEMP}

\newcommand{\ptel}{PTL}

\newcommand{\twomass}{{\rm 2M}}

\newcommand{\sgphot}{\sigma_{\rm do}}

\newcommand{\sgphotcorr}{\tilde{\sgphot}}

\newcommand{\sgnnt}{\sigma_{\rm NNT}}
\newcommand{\sgnntcorr}{\tilde{\sgnnt}}

\newcommand{\colordir}{./}

\begin{document}

\title{CfAIR2: Near Infrared Light Curves of \nsnIacfair{} Type Ia Supernovae}
\shorttitle{CfAIR2: NIR LCs of \nsnIacfair{} SN Ia}
\shortauthors{Friedman et al.}


\author{
{Andrew~S.~Friedman}\altaffilmark{1,2}, 
{W.~M.~Wood-Vasey}\altaffilmark{3},
{G.~H.~Marion}\altaffilmark{1,4}, 
{Peter~Challis}\altaffilmark{1},  
{Kaisey~S.~Mandel}\altaffilmark{1},
{Joshua~S.~Bloom}\altaffilmark{5},
{Maryam~Modjaz}\altaffilmark{6},  
{Gautham~Narayan}\altaffilmark{1,7,8}, 
{Malcolm~Hicken}\altaffilmark{1}, 
{Ryan~J.~Foley}\altaffilmark{9,10}, 
{Christopher~R.~Klein}\altaffilmark{5}, 
{Dan~L.~Starr}\altaffilmark{5}, 
{Adam~Morgan}\altaffilmark{5}, 
{Armin~Rest}\altaffilmark{11},
{Cullen~H.~Blake}\altaffilmark{12}, 
{Adam~A.~Miller}\altaffilmark{13},
{Emilio~E.~Falco}\altaffilmark{1}, 
{William~F.~Wyatt}\altaffilmark{1},
{Jessica~Mink}\altaffilmark{1},
{Michael~F.~Skrutskie}\altaffilmark{14},
and
{Robert~P.~Kirshner}\altaffilmark{1} 
}

\email{asf@mit.edu,
wmwv@pitt.edu, 
hman@astro.as.utexas.edu, 
pchallis@cfa.harvard.edu, 
kmandel@cfa.harvard.edu,
joshbloom@berkeley.edu,
mmodjaz@nyu.edu,  
gnarayan@noao.edu, 
malcolmhicken@hotmail.com, 
rfoley@illinois.edu, 
cklein@berkeley.edu, 
dstarr1@gmail.com, 
amorgan@astro.berkeley.edu, 
arest@stsci.edu,  
chblake@sas.upenn.edu,
amiller@astro.caltech.edu, 
efalco@cfa.harvard.edu, 
wfw781kra@gmail.com,
jmink@cfa.harvard.edu, 
skrutskie@virginia.edu,
rkirshner@cfa.harvard.edu}


\altaffiltext{1}{Harvard-Smithsonian Center for Astrophysics, 60 Garden Street, Cambridge, MA 02138, USA}
\altaffiltext{2}{Center for Theoretical Physics and Department of Physics, Massachusetts Institute of Technology, Cambridge, MA 02139, USA}
\altaffiltext{3}{Department of Physics and Astronomy, University of Pittsburgh, 100 Allen Hall, 3941 O'Hara St, Pittsburgh, PA 15260, USA}
\altaffiltext{4}{Astronomy Department, University of Texas at Austin, Austin, TX 78712, USA}
\altaffiltext{5}{Department of Astronomy, University of California Berkeley, Berkeley, CA 94720, USA}
\altaffiltext{6}{Center for Cosmology and Particle Physics, New York University, Meyer Hall of Physics, 4 Washington Pl., Room 529, New York, NY 10003, USA}
\altaffiltext{7}{Physics Department, Harvard University, 17 Oxford Street, Cambridge, MA 02138, USA}
\altaffiltext{8}{National Optical Astronomy Observatory, 950 N. Cherry Ave., Tucson, AZ 85719, USA}
\altaffiltext{9}{Department of Astronomy, University of Illinois at Urbana-Champaign, 1002 W.\ Green St, Urbana, IL 61801, USA}
\altaffiltext{10}{Department of Physics, University of Illinois at Urbana-Champaign, 1110 W.\ Green Street, Urbana, IL 61801, USA}
\altaffiltext{11}{Space Telescope Science Institute, STScI, 3700 San Martin Drive, Baltimore, MD 21218, USA}
\altaffiltext{12}{University of Pennsylvania, Department of Physics and Astronomy, 209 South 33rd St., Philadelphia, PA 19104, USA}
\altaffiltext{13}{Jet Propulsion Laboratory, California Institute of Technology, Pasadena, CA 91125, USA, Hubble Fellow}
\altaffiltext{14}{Department of Astronomy, P.O. Box 400325, 530 McCormick Road Charlottesville, VA 22904, USA}

\date{\today}


\begin{abstract}

\F{} is a large homogeneously reduced set of near-infrared (NIR) light curves for Type Ia supernovae (\snIa) obtained with the 1.3m Peters Automated InfraRed Imaging TELescope (PAIRITEL).  This data set includes \nptelobs{} measurements of \nsnIacfair{} \snIa{} and \nsnIaxcfair{} additional \snIax{} 
observed from 2005-2011 at the Fred Lawrence Whipple Observatory on Mount Hopkins, Arizona.  \F{} includes \jhk{} photometric measurements for 88 normal and 6 spectroscopically peculiar \snIa{}
in the nearby universe, with a median redshift of $z\sim0.021$ for the normal \snIa.  
\F{} data span the range from -\firstepoch{} days to +\lastepoch{} days from $B$-band maximum.  More than half of the light curves begin before the time of maximum and the coverage typically contains $\sim 13$--$18$ epochs of observation, 
depending on the filter. We present extensive tests that verify the fidelity of the \F{} data pipeline, including comparison to the excellent data of the Carnegie Supernova Project.  \F{} contributes to a firm local anchor for supernova cosmology studies in the NIR.  Because \snIa{} are more nearly standard candles in the NIR and are less vulnerable to the vexing problems of extinction by dust, \F{} will help the supernova cosmology community develop more precise and accurate extragalactic distance probes to improve our knowledge of cosmological parameters, including dark energy and its potential time variation.
\end{abstract}


\keywords{distance scale -- supernovae: general, infrared observations, photometry}
\section{Introduction}
\label{sec:intro}


Optical observations of Type Ia Supernovae (\snIa{}) were crucial to the surprising 1998 discovery of the acceleration of cosmic expansion  (\citealt{riess98,schmidt98,perlmutter99}). Since then, several independent cosmological techniques have confirmed the \snIa{} results (see \citealt{frieman08,weinberg13} for reviews), while \snIa{} provide increasingly accurate and precise measurements of extragalactic distances and dark energy (see \citealt{kirshner10,goobar11,kirshner13} for reviews). Increasing evidence suggests that \snIa{} observations at rest-frame NIR wavelengths yield more accurate and more precise distance estimates to \snIa{} host galaxies than optical data alone (\citealt{krisciunas04b,krisciunas07,woodvasey08, mandel09, mandel11,contreras10,folatelli10,burns11,stritzinger11,phillips12,kattner12,barone12,weyant14,mandel14a,burns14}).  

This work presents \F{}, a densely sampled, low-redshift photometric data set including \nsnIacfair{} \snIa{} NIR \jhk{}-band light curves (LCs) 
observed from 2005--2011 with the f/13.5 
\PTL{} 1.3-m telescope at the Fred Lawrence Whipple Observatory (FLWO) on Mount Hopkins, Arizona. Combining low-redshift NIR \snIa{} data sets like \F{} with higher redshift samples will play a crucial role in ongoing and future supernova cosmology experiments, from the ground and from space, which hope to reveal whether dark energy behaves like Einstein's cosmological constant $\La$ or some other phenomenon that may vary over cosmic history.  

While \snIa{} observed at optical wavelengths have been shown to be excellent {\it standardizeable} candles using a variety of sophisticated methods correlating luminosity with LC shape and color, \snIa{} are very nearly {\it standard} candles at NIR wavelengths, even before correction for LC shape or reddening (e.g., {\WVF{}; \citealt{kattner12}; hereafter \WV{} and K12). Compared to the optical, \snIa{} in the NIR are both better standard candles and relatively immune to the effects of extinction and reddening by dust.   Systematic distance errors from photometric calibration uncertainties, uncertain dust estimates, and intrinsic variability of un-reddened \snIa{} colors are outstanding problems with using \snIa{} for precise cosmological measurements of dark energy with optical data alone (\citealt{wang06,jha07,conley07,guy07,woodvasey07,hicken09a,kessler09,guy10,conley11,campbell13,narayan13,rest14,betoule14,scolnic14a,scolnic14b}). 
By contrast, many of the systematic uncertainties and discrepancies between the most prominent optical LC fitting and distance estimation methods are avoided with the incorporation of NIR data (\citealt{mandel11}; hereafter \MO{}; \citealt{folatelli10,burns11}; K12; \citealt{mandel14a}).
The most promising route toward understanding the dust in other galaxies and mitigating systematic distance errors in supernova cosmology comes from NIR observations.  

\F{} \jhk{} observations with \PTL{} are part of a systematic multi-wavelength program of CfA supernova observations at FLWO. We follow up nearby supernovae as they are discovered to obtain densely sampled, high signal-to-noise ratio (S/N) optical and NIR LCs of hundreds of nearby low-redshift SN in $UBVRIr'i'\jhk{}$.  Whenever possible, \PTL{} NIR data were observed for targets with additional optical photometry at the FLWO 1.2-m, optical spectroscopy at the 1.5-m Tillinghast telescope with the FAST spectrograph, and/or late time spectroscopy at the MMT (\citealt{matheson08,hicken09,hicken09b,blondin12,hicken12}).  By obtaining concurrent optical photometry and spectroscopy for many objects observed with \PTL{}, we considerably increase the value of the \F{} data set.  
Of the 98 \F{} objects, \nsnIacfairopt{} have complementary optical observations from the CfA or other groups, including unpublished data.\footnote{All 10 spectroscopically peculiar \snIa{} and \snIax{} have optical data from the CfA or other groups, including unpublished CfA5 optical data. Of the 88 spectroscopically normal \F{} \snIa{} in Table~\ref{tab:general}, 64 have published optical data from the CfA or other groups, and 12 have unpublished CfA5 optical data. An additional 4 have CfA optical observations but no successfully reduced LCs yet: \sn{}2010jv, \sn{}2010ex, \sn{}2010ew, \sn{}2009fw.  In addition, 2 objects have unpublished optical data from other groups PTF10icb (PTF: \citealt{parrent11}: only spectra included), PTF10bjs (PTF, CfA4: only natural system r$^{\prime}$i$^{\prime}$). 
Six objects currently have no optical photometry, according to our search of the literature: \sn{}2010dl, \sn{}2009im, \sn{}2008hy, \sn{}2008fx, \sn{}2005ch, \sn{}2005ao.} Table~\ref{tab:general} lists general properties of the 94 \snIa{}.

\renewcommand{\arraystretch}{0.8}
\begin{table*}
\begin{center}
\caption[General Properties of \nsnIacfair{} \PTL{} \snIa{}]{General Properties of \nsnIacfair{} \PTL{} \snIa{} \\} 
\tiny
\begin{tabular}{@{}l@{}r@{}r@{}l@{}l@{}l@{}l@{}c@{}c@{}c@{}c@{}l@{}}
\hline
SN & \multicolumn{1}{c}{RA\tablenotemark{a}} & \multicolumn{1}{c}{DEC\tablenotemark{a}} & Host\tablenotemark{b} & Morphology\tablenotemark{c} & \multicolumn{1}{c}{$z_{\rm helio}$\tablenotemark{d}} & \multicolumn{1}{c}{$\sigma_{z_{\rm helio}}$\tablenotemark{d}} & $z$ \tablenotemark{d} & Discovery\tablenotemark{b} & Discoverer(s)\tablenotemark{e} & Type\tablenotemark{f} & Type\tablenotemark{g} \\
Name & \multicolumn{1}{c}{$\alpha(2000)$} & \multicolumn{1}{c}{$\delta(2000)$} & Galaxy &  & &  & Ref. & Reference &  & Reference &  \\
\hline
\sn{}2005ao           \ & \ 266.20653    \ & \ 61.90786     \ & \ NGC 6462                          \ & \ SABbc             \ & \ 0.038407    \ & \ 0.000417    \ & \ 1    \ & \ CBET 115     \ & \ POSS                  \ & \ IAUC 8492    \ & \ Ia     \\
\sn{}2005bl           \ & \ 181.05098    \ & \ 20.40683     \ & \ NGC 4070                          \ & \  \nodata          \ & \ 0.02406     \ & \ 0.00008     \ & \ 1    \ & \ IAUC 8515    \ & \ LOSS, POSS            \ & \ IAUC 8514    \ & \ Iap    \\
\sn{}2005bo           \ & \ 192.42099    \ & \ -11.09663    \ & \ NGC 4708                          \ & \ SA(r)ab pec?      \ & \ 0.013896    \ & \ 0.000027    \ & \ 1    \ & \ CBET 141     \ & \ POSS                  \ & \ CBET 142     \ & \ Ia     \\
\sn{}2005cf           \ & \ 230.38906    \ & \ -7.44874     \ & \ MCG -01-39-3                      \ & \ S0 pec            \ & \ 0.006461    \ & \ 0.000037    \ & \ 1    \ & \ CBET 158     \ & \ LOSS                  \ & \ IAUC 8534    \ & \ Ia     \\
\sn{}2005ch           \ & \ 215.52815    \ & \ 1.99316      \ & \ 1                                            \ & \  \nodata          \ & \ 0.027       \ & \ 0.005       \ & \ 3    \ & \ CBET 166     \ & \ ROTSE-III             \ & \ CBET 167     \ & \ Ia     \\
\sn{}2005el           \ & \ 77.95316     \ & \ 5.19417      \ & \ NGC 1819                          \ & \ SB0               \ & \ 0.01491     \ & \ 0.000017    \ & \ 1    \ & \ CBET 233     \ & \ LOSS                  \ & \ CBET 235     \ & \ Ia     \\
\sn{}2005eq           \ & \ 47.20575     \ & \ -7.03332     \ & \ MCG -01-9-6                       \ & \ SB(rs)cd?         \ & \ 0.028977    \ & \ 0.000073    \ & \ 1    \ & \ IAUC 8608    \ & \ LOSS                  \ & \ IAUC 8610    \ & \ Ia     \\
\sn{}2005eu           \ & \ 36.93011     \ & \ 28.17698     \ & \ 2                                               \ & \  \nodata          \ & \ 0.03412     \ & \ 0.000046    \ & \ 1    \ & \ CBET 242     \ & \ LOSS                  \ & \ CBET 244     \ & \ Ia     \\
\sn{}2005iq           \ & \ 359.63517    \ & \ -18.70914    \ & \ MCG -03-1-8                       \ & \ Sa                \ & \ 0.034044    \ & \ 0.000123    \ & \ 1    \ & \ IAUC 8628    \ & \ LOSS                  \ & \ CBET 278     \ & \ Ia     \\
\sn{}2005ke           \ & \ 53.76810     \ & \ -24.94412    \ & \ NGC 1371                          \ & \ (R')SAB(r'l)a     \ & \ 0.00488     \ & \ 0.000007    \ & \ 1    \ & \ IAUC 8630    \ & \ LOSS                  \ & \ IAUC 8631    \ & \ Iap    \\
\sn{}2005ls           \ & \ 43.56630     \ & \ 42.72480     \ & \ MCG +07-7-1                       \ & \ Spiral            \ & \ 0.021118    \ & \ 0.000117    \ & \ 1    \ & \ IAUC 8643    \ & \ Armstrong             \ & \ CBET 324     \ & \ Ia     \\
\sn{}2005na           \ & \ 105.40287    \ & \ 14.13304     \ & \ UGC 3634                          \ & \ SB(r)a            \ & \ 0.026322    \ & \ 0.000083    \ & \ 1    \ & \ CBET 350     \ & \ POSS                  \ & \ CBET 351     \ & \ Ia     \\
\sn{}2006D            \ & \ 193.14111    \ & \ -9.77519     \ & \ MCG -01-33-34                     \ & \ SAB(s)ab pec?     \ & \ 0.008526    \ & \ 0.000017    \ & \ 1    \ & \ CBET 362     \ & \ BRASS                 \ & \ CBET 366     \ & \ Ia     \\
\sn{}2006E            \ & \ 208.36880    \ & \ 5.20619      \ & \ NGC 5338                          \ & \ SB0               \ & \ 0.002686    \ & \ 0.000005    \ & \ 2    \ & \ CBET 363     \ & \ POSS, LOSS, CROSS     \ & \ ATEL 690     \ & \ Ia     \\
\sn{}2006N            \ & \ 92.13021     \ & \ 64.72362     \ & \ MCG +11-8-12                      \ & \  \nodata          \ & \ 0.014277    \ & \ 0.000083    \ & \ 1    \ & \ CBET 375     \ & \ Armstrong             \ & \ IAUC 8661    \ & \ Ia     \\
\sn{}2006X            \ & \ 185.72471    \ & \ 15.80888     \ & \ NGC 4321                          \ & \ SAB(s)bc          \ & \ 0.00524     \ & \ 0.000003    \ & \ 1    \ & \ IAUC 8667    \ & \ Suzuki, CROSS         \ & \ CBET 393     \ & \ Ia     \\
\sn{}2006ac           \ & \ 190.43708    \ & \ 35.06872     \ & \ NGC 4619                          \ & \ SB(r)b pec?       \ & \ 0.023106    \ & \ 0.000037    \ & \ 1    \ & \ IAUC 8669    \ & \ LOSS                  \ & \ CBET 398     \ & \ Ia     \\
\sn{}2006ax           \ & \ 171.01434    \ & \ -12.29156    \ & \ NGC 3663                          \ & \ SA(rs)bc pec      \ & \ 0.016725    \ & \ 0.000019    \ & \ 2    \ & \ CBET 435     \ & \ LOSS                  \ & \ CBET 437     \ & \ Ia     \\
\sn{}2006cp           \ & \ 184.81198    \ & \ 22.42723     \ & \ UGC 7357                          \ & \ SAB(s)c           \ & \ 0.022289    \ & \ 0.000002    \ & \ 1    \ & \ CBET 524     \ & \ LOSS                  \ & \ CBET 528     \ & \ Ia     \\
\sn{}2006cz           \ & \ 222.15254    \ & \ -4.74193     \ & \ MCG -01-38-2                      \ & \ SA(s)cd?          \ & \ 0.0418      \ & \ 0.000213    \ & \ 1    \ & \ IAUC 8721    \ & \ LOSS                  \ & \ CBET 550     \ & \ Ia     \\
\sn{}2006gr           \ & \ 338.09445    \ & \ 30.82871     \ & \ UGC 12071                         \ & \ SBb               \ & \ 0.034597    \ & \ 0.00003     \ & \ 1    \ & \ CBET 638     \ & \ LOSS                  \ & \ CBET 642     \ & \ Ia     \\
\sn{}2006le           \ & \ 75.17457     \ & \ 62.25525     \ & \ UGC 3218                          \ & \ SAb               \ & \ 0.017432    \ & \ 0.000023    \ & \ 1    \ & \ CBET 700     \ & \ LOSS                  \ & \ CBET 702     \ & \ Ia     \\
\sn{}2006lf           \ & \ 69.62286     \ & \ 44.03379     \ & \ UGC 3108                          \ & \ S?                \ & \ 0.013189    \ & \ 0.000017    \ & \ 2    \ & \ CBET 704     \ & \ LOSS                  \ & \ CBET 705     \ & \ Ia     \\
\sn{}2006mq           \ & \ 121.55157    \ & \ -27.56262    \ & \ ESO 494-G26                       \ & \ SAB(s)b pec       \ & \ 0.003229    \ & \ 0.000003    \ & \ 1    \ & \ CBET 721     \ & \ LOSS                  \ & \ CBET 724     \ & \ Ia     \\
\sn{}2007S            \ & \ 150.13010    \ & \ 4.40702      \ & \ UGC 5378                          \ & \ Sb                \ & \ 0.01388     \ & \ 0.000033    \ & \ 1    \ & \ CBET 825     \ & \ POSS                  \ & \ CBET 829     \ & \ Ia     \\
\sn{}2007ca           \ & \ 202.77451    \ & \ -15.10175    \ & \ MCG -02-34-61                     \ & \ Sc pec sp         \ & \ 0.014066    \ & \ 0.00001     \ & \ 1    \ & \ CBET 945     \ & \ LOSS                  \ & \ CBET 947     \ & \ Ia     \\
\sn{}2007co           \ & \ 275.76493    \ & \ 29.89715     \ & \ MCG +05-43-16                     \ & \  \nodata          \ & \ 0.026962    \ & \ 0.00011     \ & \ 1    \ & \ CBET 977     \ & \ Nicolas               \ & \ CBET 978     \ & \ Ia     \\
\sn{}2007cq           \ & \ 333.66839    \ & \ 5.08017      \ & \ 3                                             \ & \  \nodata          \ & \ 0.026218    \ & \ 0.000167    \ & \ 3    \ & \ CBET 983     \ & \ POSS                  \ & \ CBET 984     \ & \ Ia     \\
\sn{}2007fb           \ & \ 359.21827    \ & \ 5.50886      \ & \ UGC 12859                         \ & \ Sbc               \ & \ 0.018026    \ & \ 0.000007    \ & \ 2    \ & \ CBET 992     \ & \ LOSS                  \ & \ CBET 993     \ & \ Ia     \\
\sn{}2007if           \ & \ 17.71421     \ & \ 15.46103     \ & \ 4                                              \ & \  \nodata          \ & \ 0.0745      \ & \ 0.00015     \ & \ 5    \ & \ CBET 1059    \ & \ ROTSE-III             \ & \ CBET 1059    \ & \ Iap    \\
\sn{}2007le           \ & \ 354.70186    \ & \ -6.52269     \ & \ NGC 7721                          \ & \ SA(s)c            \ & \ 0.006728    \ & \ 0.000002    \ & \ 1    \ & \ CBET 1100    \ & \ Monard                \ & \ CBET 1101    \ & \ Ia     \\
\sn{}2007nq           \ & \ 14.38999     \ & \ -1.38874     \ & \ UGC 595                           \ & \ E                 \ & \ 0.045031    \ & \ 0.000053    \ & \ 1    \ & \ CBET 1106    \ & \ ROTSE-III             \ & \ CBET 1106    \ & \ Ia     \\
\sn{}2007qe           \ & \ 358.55408    \ & \ 27.40916     \ & \ 5                                         \ & \  \nodata          \ & \ 0.024       \ & \ 0.001       \ & \ 6    \ & \ CBET 1138    \ & \ ROTSE-III             \ & \ CBET 1138    \ & \ Ia     \\
\sn{}2007rx           \ & \ 355.04908    \ & \ 27.42097     \ & \ 6                                           \ & \  \nodata          \ & \ 0.0301      \ & \ 0.001       \ & \ 7    \ & \ CBET 1157    \ & \ ROTSE-III             \ & \ CBET 1157    \ & \ Ia     \\
\sn{}2007sr           \ & \ 180.46995    \ & \ -18.97269    \ & \ NGC 4038                          \ & \ SB(s)m pec        \ & \ 0.005417    \ & \ 0.000017    \ & \ 2    \ & \ CBET 1172    \ & \ CSS                   \ & \ CBET 1173    \ & \ Ia     \\
\sn{}2008C            \ & \ 104.29794    \ & \ 20.43723     \ & \ UGC 3611                          \ & \ S0/a              \ & \ 0.016621    \ & \ 0.000013    \ & \ 1    \ & \ CBET 1195    \ & \ POSS                  \ & \ CBET 1197    \ & \ Ia     \\
\sn{}2008Z            \ & \ 145.81364    \ & \ 36.28439     \ & \ 7                                              \ & \  \nodata          \ & \ 0.02099     \ & \ 0.000226    \ & \ 1    \ & \ CBET 1243    \ & \ POSS                  \ & \ CBET 1246    \ & \ Ia     \\
\sn{}2008af           \ & \ 224.86846    \ & \ 16.65325     \ & \ UGC 9640                          \ & \ E                 \ & \ 0.033507    \ & \ 0.000153    \ & \ 1    \ & \ CBET 1248    \ & \ Boles                 \ & \ CBET 1253    \ & \ Ia     \\
SNF20080514-002    \ & \ 202.30350    \ & \ 11.27236     \ & \ UGC 8472                          \ & \ S0                \ & \ 0.022095    \ & \ 0.00009     \ & \ 1    \ & \ ATEL 1532    \ & \ SNF                   \ & \ ATEL 1532    \ & \ Ia     \\
SNF20080522-000    \ & \ 204.19796    \ & \ 5.14200      \ & \ SDSS?                             \ & \  \nodata          \ & \ 0.04526     \ & \ 0.0002      \ & \ 9    \ & \ SNF          \ & \ SNF                   \ & \ B09          \ & \ Ia     \\
SNF20080522-011    \ & \ 229.99519    \ & \ 4.90454      \ & \ SDSS?                             \ & \  \nodata          \ & \ 0.03777     \ & \ 0.00006     \ & \ 9    \ & \ SNF          \ & \ SNF                   \ & \ B09          \ & \ Ia     \\
\sn{}2008fr           \ & \ 17.95488     \ & \ 14.64068     \ & \ 8                                                \ & \  \nodata          \ & \ 0.039       \ & \ 0.002       \ & \ 8    \ & \ CBET 1513    \ & \ ROTSE-III             \ & \ CBET 1513    \ & \ Ia     \\
\sn{}2008fv           \ & \ 154.23873    \ & \ 73.40986     \ & \ NGC 3147                          \ & \ SA(rs)bc          \ & \ 0.009346    \ & \ 0.000003    \ & \ 1    \ & \ CBET 1520    \ & \ Itagaki               \ & \ CBET 1522    \ & \ Ia     \\
\sn{}2008fx           \ & \ 32.89166     \ & \ 23.87998     \ & \ 9                                                  \ & \  \nodata          \ & \ 0.059       \ & \ 0.003       \ & \ 3    \ & \ CBET 1523    \ & \ CSS                   \ & \ CBET 1525    \ & \ Ia     \\
\sn{}2008gb           \ & \ 44.48821     \ & \ 46.86566     \ & \ UGC 2427                          \ & \ Sbc               \ & \ 0.037626    \ & \ 0.000041    \ & \ 3    \ & \ CBET 1527    \ & \ POSS                  \ & \ CBET 1530    \ & \ Ia     \\
\sn{}2008gl           \ & \ 20.22829     \ & \ 4.80531      \ & \ UGC 881                           \ & \ E                 \ & \ 0.034017    \ & \ 0.000117    \ & \ 1    \ & \ CBET 1545    \ & \ CHASE                 \ & \ CBET 1547    \ & \ Ia     \\
\sn{}2008hm           \ & \ 51.79540     \ & \ 46.94421     \ & \ 2MFGC 02845                       \ & \ Spiral            \ & \ 0.019664    \ & \ 0.000077    \ & \ 1    \ & \ CBET 1586    \ & \ LOSS                  \ & \ CBET 1594    \ & \ Ia     \\
\sn{}2008hs           \ & \ 36.37335     \ & \ 41.84311     \ & \ NGC 910                           \ & \ E+                \ & \ 0.017349    \ & \ 0.000073    \ & \ 2    \ & \ CBET 1598    \ & \ LOSS                  \ & \ CBET 1599    \ & \ Ia     \\
\sn{}2008hv           \ & \ 136.89178    \ & \ 3.39240      \ & \ NGC 2765                          \ & \ S0                \ & \ 0.012549    \ & \ 0.000067    \ & \ 1    \ & \ CBET 1601    \ & \ CHASE                 \ & \ CBET 1603    \ & \ Ia     \\
\sn{}2008hy           \ & \ 56.28442     \ & \ 76.66533     \ & \ IC 334                            \ & \ S?                \ & \ 0.008459    \ & \ 0.000023    \ & \ 1    \ & \ CBET 1608    \ & \ POSS                  \ & \ CBET 1610    \ & \ Ia     \\
\hline
\end{tabular}
\vspace{0.3cm}
\tablecomments{ \scriptsize
\\
{\bf (a)} SN RA, DEC positions [in decimal degrees] are best fit SN centroids appropriate for forced \dophot{} photometry at fixed coordinates.
\\
{\bf (b)} Host Galaxy Names, discovery references, and discovery group/individual credits from NASA/IPAC Extragalactic Database (NED; \url{http://ned.ipac.caltech.edu/}) and NASA/ADS 
(\url{http://adswww.harvard.edu/abstract\_service.html}). Also see IAUC List of Supernovae: \url{http://www.cbat.eps.harvard.edu/lists/Supernovae.html}.  
For \snIa{} with non-standard IAUC names, we found the associated host galaxy from IAUC/CBET/ATel notices or the literature and searched for the recession velocity with NED.  When the \snIa{} is associated with a faint host not named in any major catalogs (NGC, UGC, \ldots) but named in a large galaxy survey (e.g., SDSS, 2MASS), we include the host name from the large survey rather than ``Anonymous".  However, to fit the table on a single page, long galaxy names are numbered.  \\
1: APMUKS(BJ) B141934.25+021314.0 (\sn{}2005ch), 2: NSF J022743.32+281037.6 (\sn{}2005eu), 3: 2MASX J22144070+0504435 (\sn{}2007cq), 4: J011051.37+152739 (\sn{}2007if), 5: NSF J235412.09+272432.3 (\sn{}2007qe), 6: BATC J234012.05+272512.23 (\sn{}2007rx), 7: SDSS J094315.36+361709.2 (\sn{}2008Z), 8: SDSS J011149.19+143826.5 (\sn{}2008fr), 9: 2MASX J02113233+2353074 (\sn{}2008fx).
The machine readable version of this table has full galaxy names.
\\
{\bf (c)} Host galaxy morphologies taken from NED where available. 
Hosts with unknown morphologies denoted by \nodata
\\
{\bf (d)} Heliocentric redshift $z_{\rm helio}$, $\sigma_{z_{\rm helio}}$ references are from 1: NED Host galaxy name, 2: NED 21-cm or optical with smallest uncertainty, 3: CfA FAST spectrum on Tillinghast 1.5-m telescope, 4:  \citealt{rest14}: PanSTARRS1, 5:  \citealt{childress11}, 6: CBET 1176, 7:  \citealt{hicken09a}, 8: CBET 1513, 9:  \citealt{childress13a}. For \sn{}2008fr, the NED redshift incorrectly lists the redshift of \sn{}2008fs (see CBET 1513). Heliocentric redshifts have not been corrected for any local flow models.
\\
{\bf (e)} Discovery References/URLs: LOSS: Lick Observatory Supernova Search (see \citealt{li00,filippenko05}, and references therein); Tenagra II (\url{http://www.tenagraobservatories.com/Discoveries.htm}); ROTSE-III \citep{quimby06}; POSS: Puckett Observatory Supernova Search (\url{http://www.cometwatch.com/search.html}); BRASS:  (\url{http://brass.astrodatabase.net}); SDSS-II: Sloan Digital Sky Survey II \citep{frieman08a}; CSS: Catalina Sky Survey (\url{http://www.lpl.arizona.edu/css/}); SNF: Nearby Supernova Factory (\url{http://snfactory.lbl.gov/}); CHASE: CHilean Automatic Supernova sEarch (\url{http://www.das.uchile.cl/proyectoCHASE/}); CRTS: Catalina Real-Time Transient Survey (\url{http://crts.caltech.edu/}); Itagaki (\url{http://www.k-itagaki.jp/}); Boles: Coddenham Astronomical Observatory, U.K. (\url{http://www.coddenhamobservatories.org/}); CROSS (\url{http://wwww.cortinasetelle.it/snindex.htm}); LSSS: La Sagra Sky Survey (\url{http://www.minorplanets.org/OLS/LSSS.html}); PASS: Perth Automated Supernova Search (\url{http://www.perthobservatory.wa.gov.au/research/spps.html}); \citealt{williams97}); PIKA: Comet and Asteroid Search Program (\url{http://www.observatorij.org/Pika.html}); PanSTARRS1: (\url{http://pan-starrs.ifa.hawaii.edu/public/}); THCA Supernova Survey (\url{http://www.thca.tsinghua.edu.cn/en/index.php/TUNAS})
\\
{\bf (f)} Spectroscopic type reference. B09=\citealt{bailey09}.
\\
{\bf (g)} Spectroscopic type of SN Ia = spectroscopically normal \snIa{}. Spectroscopically peculiar \snIa{}: including 91bg-like and 06gz-like objects.  
}
\label{tab:general}
\end{center}
\end{table*}
\renewcommand{\arraystretch}{1}

It has only recently become understood that \sn{}2002cx-like objects, 
which we categorize as \snIax{} (e.g., \citealt{foley13}), are significantly distinct both from normal \snIa{} and spectroscopically peculiar \snIa{} \citep{li03,branch04,chornock06,jha06b,phillips07,sahu08,maund10,mcclelland10,narayan11,kromer13,foley09b,foley10a,foley10b,foley13,foley14a,foley14b,foley15,mccully14a,mccully14b,stritzinger15}.   Throughout, we treat the \nsnIaxcfair{} \snIax{} included in \F{} (\sn{}2005hk, \sn{}2008A, \sn{}2008ae, \sn{}2008ha) as a separate class of objects from \snIa{} (see Table~\ref{tab:general3}).

This work is a report on photometric data from \PTL{} which improves upon and supersedes a previously published subset including 20 \snIa{} \jhk{} LCs from (\WV{}; implicitly ``CfAIR1"), 1 \snIax{} LC from \WV{} (\sn{}2005hk), and 1 \snIax{} LC from \citealt{foley09b} (\sn{}2008ha), along with work presented in \citealt{friedman12} (hereafter F12).\footnote{\tiny F12 PDF available at \url{http://search.proquest.com/docview/1027769281}} Data points for these 20 objects have been reprocessed using our newest mosaic and photometry pipelines and are presented as part of this CfAIR2 data release. The CfAIR1 (\WV{}) and CfAIR2 NIR data sets complement previous CfA optical studies of \snIa{} (CfA1: \citealt{riess99}; CfA2: \citealt{jha06}; CfA3: \citealt{hicken09b}; and CfA4: \citealt{hicken12}) and CfA5 (to be presented elsewhere).  CfA5 will include optical data for at least \ncfafive{} \F{} objects and additional optical LCs for non-\F{} objects.

The \nptelobs{} individual \F{} \jhk{} data points represent the largest homogeneously observed and reduced set of NIR \snIa{} and \snIax{} observations to date. Simultaneous \jhk{} observing provided nightly cadence for the most densely sampled LCs and extensive time coverage, ranging from \firstepoch{} days before to \lastepoch{} days after the time of $B$-band maximum brightness ($\tbmax$). \F{} data have means of $\jmeanepochs$, $\hmeanepochs$, and $\kmeanepochs$ observed epochs for each LC in \jhk{}, respectively, as well as $\jmaxepochs$ epochs for the most extensively sampled LC. 
\F{} LCs have significant early-time coverage. 
Out of 98 \F{} objects, \npretbmaxpercent{} have NIR observations before $\tbmax$, while \npretbmaxfivepercent{} have observations at least $5$ days before $\tbmax$.  The highest S/N LC points for each \F{} object have median uncertainties of $\sim \jmedmagerrlcpeak$, $\hmedmagerrlcpeak$, and $\kmedmagerrlcpeak$ mag in \jhk{}, respectively.  The median uncertainties of all \F{} LC points are $\jmedmagerr$, $\hmedmagerr$, and $\kmedmagerr$ mag in \jhk{}, respectively.

Of the 98 \F{} objects, 88 are spectroscopically normal \snIa{} and 86 will be useful for supernova cosmology (\sn{}2006E and \sn{}2006mq were discovered late and lack precise \tbmax{} estimates). The 6 spectroscopically peculiar \snIa{} and \nsnIaxcfair{} \snIax{} are not standardizable candles using existing LC fitting techniques, and currently must be excluded from Hubble diagrams.

\subsection{Previous Results with NIR \snIa{}}
\label{sec:prev}

For optical \snIa{} LCs, many sophisticated methods are used to reduced the scatter in distance estimates. These include $\dm{}$  (\citealt{phillips93,hamuy96,phillips99,prieto06}), multicolor light-curve shape (MLCS; \citealt{riess96,riess98,jha06,jha07}), ``stretch''  \citep{perlmutter97,goldhaber01}, Bayesian Adapted Template Match (BATM; \citealt{tonry03}), color-magnitude intercept calibration (CMAGIC; \citealt{wang03}), spectral adaptive template (SALT; \citealt{guy05,astier06,guy07}), empirical methods (e.g., SiFTO; \citealt{conley08}), and \bayesn{}, a novel hierarchical Bayesian method developed at the CfA (\M, \MO). 

Unlike optical \snIa{}, which are {\it standardizable} candles after a great deal of effort, spectroscopically normal NIR \snIa{} appear to be nearly {\it standard} candles at the $\sim0.15$--$0.2$ mag level or better, depending on the filter 
(\citealt{meikle00,krisciunas04a,krisciunas05a,krisciunas07,folatelli10,burns11,phillips12}; \WV; \M; \MO; K12). Overall, \snIa{} are superior standard candles and distance indicators in the NIR compared to optical wavelengths, with a narrow distribution of peak \jhk{} magnitudes and $\sim$5--11 times less sensitivity to reddening than optical $B$-band data alone. 

Following \citet{meikle00}, pioneering work by \citet{krisciunas04a} (hereafter K04a) demonstrated that \snIa{} have a narrow luminosity range in \jhk{} at \tbmax{} with smaller scatter than in $B$ and $V$.  Using 16 NIR \snIa{}, K04a found no correlation between optical LC shape and intrinsic NIR luminosity. K04a measured \jhk{} absolute magnitude distributions with 1-$\sigma$ uncertainties of only $\sigma_J=0.14$, $\sigma_H=0.18$, and $\sigma_{K_s}=0.12$~mag. While K04a used a small, inhomogeneous, sample of 16 LCs, in \WV{}, we presented \nwvobs{} \jhk{} photometric observations of \nsnIaredo{} objects (including 20 \snIa{} and 1 \snIax), the largest homogeneously observed low-$z$ sample at the time. NIR data from \WV{} and the literature strengthened the evidence that normal \snIa{} are excellent NIR standard candles, especially in the $H$-band, where absolute magnitudes have an intrinsic root-mean-square (RMS) of $0.15$--$0.16$ mag, {\it without applying any reddening or LC shape corrections}, comparable to the scatter in optical data corrected for both. 

\WV{} suggested that LC shape variation, especially in the $J$-band, might provide additional information for correcting NIR LCs and improving distance determinations. In \M{}, we applied a novel hierarchical Bayesian framework and a model accounting for variations in the $J$-band LC shape to NIR \snIa{} data, constraining the marginal scatter of the NIR peak absolute magnitudes to $0.17$, $0.11$, and $0.19$ mag, in \jhk{}, 
respectively (see Fig. 9 of \M{}). \citealt{folatelli10} obtained similar dispersions of $0.12$--$0.16$ mag in $Y\jhk$, after correcting for NIR LC shape. Using 13 well-sampled, low extinction, normal NIR \snIa{} LCs from the CSP, K12 find scatters in absolute magnitude of $0.12$, $0.12$, and $0.09$ mag in $YJH$, respectively. K12 also confirm that NIR LC shape correlates with intrinsic NIR luminosity, finding evidence for a non-zero correlation between the peak absolute $JH$-maxima and the decline rate parameter $\Delta m_{15}$, 
with only marginal dependence in $Y$. For a set of 12 \snIa{} with $JH$ LCs, \citealt{barone12} find a very small $JH$-band scatter of only $0.116$ and $0.085$ mag respectively, although their data set only includes $3$-$5$ LC points for each of the 12 objects.  Similarly, \citealt{weyant14} use only 1-3 data points for each of 13 low-$z$ NIR \snIa{} to infer an $H$-band dispersion of 0.164 mag.  Both \citealt{barone12} and \citealt{weyant14} use auxiliary optical data to estimate \tbmax{}. All of these results suggest that NIR data will be crucial for maximizing the utility of \snIa{} as cosmological distance indicators. 

\subsection{Organization of Paper}
\label{sec:org}

This paper is organized as follows.   In \S\ref{sec:observations}, we discuss the current sample of nearby NIR \snIa{} data including \F{}, describe the technical specifications of \PTL{}, and outline our follow-up campaign.  In \S\ref{sec:reduction} we describe the data reduction process, including mosaicked image creation, sky subtraction, host galaxy subtraction, and our photometry pipeline.  In \S\ref{sec:phot}, we present tests of \PTL{} photometry, emphasizing internal calibration with 2MASS field star observations, tests for potential systematic errors, and external consistency checks for objects observed both by \PTL{} and the CSP.  Throughout \S\ref{sec:observations}-\ref{sec:phot}, we frequently reference F12, where many additional technical details can be found.  In \S\ref{sec:results}, we present the principal data products of this paper, which include \jhk{} LCs of \nsnIacfair{} \snIa{} and \nsnIaxcfair{} \snIax. Further analysis of this data will be presented elsewhere. \PTL{} and CSP comparison is discussed further in \S\ref{sec:disc}. Conclusions and directions for future work are summarized in \S\ref{sec:conc}. Additional details are included in a mathematical appendix (also see \S7 of F12).  

 
\addtocounter{table}{-1} 
\renewcommand{\arraystretch}{0.8}
\begin{table*}
\begin{center}
\caption[General Properties of \nsnIacfair{} \PTL{} \snIa{} (continued)]{General Properties of \nsnIacfair{} \PTL{} \snIa{} (continued) \\} 
\tiny
\begin{tabular}{@{}l@{}r@{}r@{}l@{}l@{}l@{}l@{}c@{}c@{}c@{}c@{}l@{}}
\hline
SN & \multicolumn{1}{c}{RA\tablenotemark{a}} & \multicolumn{1}{c}{DEC\tablenotemark{a}} & Host\tablenotemark{b} & Morphology\tablenotemark{c} & \multicolumn{1}{c}{$z_{\rm helio}$\tablenotemark{d}} & \multicolumn{1}{c}{$\sigma_{z_{\rm helio}}$\tablenotemark{d}} & $z$ \tablenotemark{d} & Discovery\tablenotemark{b} & Discoverer(s)\tablenotemark{e} & Type\tablenotemark{f} & Type\tablenotemark{g} \\
Name & \multicolumn{1}{c}{$\alpha(2000)$} & \multicolumn{1}{c}{$\delta(2000)$} & Galaxy &  & &  & Ref. & Reference &  & Reference &  \\
\hline
\sn{}2009D            \ & \ 58.59495     \ & \ -19.18194    \ & \ MCG -03-10-52                     \ & \ Sb                \ & \ 0.025007    \ & \ 0.000033    \ & \ 1    \ & \ CBET 1647    \ & \ LOSS                  \ & \ CBET 1647    \ & \ Ia     \\
\sn{}2009Y            \ & \ 220.59865    \ & \ -17.24675    \ & \ NGC 5728                          \ & \ (R$ 1$)SAB(r)a    \ & \ 0.009316    \ & \ 0.000026    \ & \ 2    \ & \ CBET 1684    \ & \ PASS, LOSS            \ & \ CBET 1688    \ & \ Ia     \\
\sn{}2009ad           \ & \ 75.88914     \ & \ 6.66000      \ & \ UGC 3236                          \ & \ Sbc               \ & \ 0.0284      \ & \ 0.000005    \ & \ 1    \ & \ CBET 1694    \ & \ POSS                  \ & \ CBET 1695    \ & \ Ia     \\
\sn{}2009al           \ & \ 162.84201    \ & \ 8.57833      \ & \ NGC 3425                          \ & \ S0                \ & \ 0.022105    \ & \ 0.00008     \ & \ 1    \ & \ CBET 1705    \ & \ CSS                   \ & \ CBET 1708    \ & \ Ia     \\
\sn{}2009an           \ & \ 185.69715    \ & \ 65.85145     \ & \ NGC 4332                          \ & \ SB(s)a            \ & \ 0.009228    \ & \ 0.000004    \ & \ 2    \ & \ CBET 1707    \ & \ Cortini+, Paivinen    \ & \ CBET 1709    \ & \ Ia     \\
\sn{}2009bv           \ & \ 196.83538    \ & \ 35.78433     \ & \ MCG +06-29-39                     \ & \  \nodata          \ & \ 0.036675    \ & \ 0.000063    \ & \ 1    \ & \ CBET 1741    \ & \ PIKA                  \ & \ CBET 1742    \ & \ Ia     \\
\sn{}2009dc           \ & \ 237.80042    \ & \ 25.70790     \ & \ UGC 10064                         \ & \ S0                \ & \ 0.021391    \ & \ 0.00007     \ & \ 1    \ & \ CBET 1762    \ & \ POSS                  \ & \ CBET 1768    \ & \ Iap    \\
\sn{}2009do           \ & \ 188.74310    \ & \ 50.85108     \ & \ NGC 4537                          \ & \ S                 \ & \ 0.039734    \ & \ 0.00008     \ & \ 1    \ & \ CBET 1778    \ & \ LOSS, POSS            \ & \ CBET 1778    \ & \ Ia     \\
\sn{}2009ds           \ & \ 177.26706    \ & \ -9.72892     \ & \ NGC 3905                          \ & \ SB(rs)c           \ & \ 0.019227    \ & \ 0.000021    \ & \ 2    \ & \ CBET 1784    \ & \ Itagaki               \ & \ CBET 1788    \ & \ Ia     \\
\sn{}2009fw           \ & \ 308.07711    \ & \ -19.73336    \ & \ ESO 597-6                         \ & \ SA(rs)0-?         \ & \ 0.028226    \ & \ 0.00011     \ & \ 1    \ & \ CBET 1836    \ & \ CHASE                 \ & \ CBET 1849    \ & \ Ia     \\
\sn{}2009fv           \ & \ 247.43430    \ & \ 40.81153     \ & \ NGC 6173                          \ & \ E                 \ & \ 0.0293      \ & \ 0.00005     \ & \ 1    \ & \ CBET 1834    \ & \ POSS                  \ & \ CBET 1846    \ & \ Ia     \\
\sn{}2009ig           \ & \ 39.54843     \ & \ -1.31257     \ & \ NGC 1015                          \ & \ SB(r)a            \ & \ 0.00877     \ & \ 0.000021    \ & \ 1    \ & \ CBET 1918    \ & \ LOSS                  \ & \ CBET 1918    \ & \ Ia     \\
\sn{}2009im           \ & \ 53.34204     \ & \ -4.99903     \ & \ NGC 1355                          \ & \ S0 sp             \ & \ 0.0131      \ & \ 0.0001      \ & \ 1    \ & \ CBET 1925    \ & \ Itagaki               \ & \ CBET 1934    \ & \ Ia     \\
\sn{}2009jr           \ & \ 306.60846    \ & \ 2.90889      \ & \ IC 1320                           \ & \ SB(s)b?           \ & \ 0.016548    \ & \ 0.00006     \ & \ 1    \ & \ CBET 1964    \ & \ Arbour                \ & \ CBET 1968    \ & \ Ia     \\
\sn{}2009kk           \ & \ 57.43441     \ & \ -3.26447     \ & \ 2MFGC 03182                       \ & \  \nodata          \ & \ 0.012859    \ & \ 0.00015     \ & \ 1    \ & \ CBET 1991    \ & \ CSS                   \ & \ CBET 1991    \ & \ Ia     \\
\sn{}2009kq           \ & \ 129.06316    \ & \ 28.06711     \ & \ MCG +05-21-1                      \ & \ Spiral            \ & \ 0.011698    \ & \ 0.00002     \ & \ 1    \ & \ CBET 2005    \ & \ POSS                  \ & \ ATEL 2291    \ & \ Ia     \\
\sn{}2009le           \ & \ 32.32152     \ & \ -23.41242    \ & \ ESO 478-6                         \ & \ Sbc               \ & \ 0.017792    \ & \ 0.000009    \ & \ 2    \ & \ CBET 2022    \ & \ CHASE                 \ & \ CBET 2025    \ & \ Ia     \\
\sn{}2009lf           \ & \ 30.41513     \ & \ 15.33290     \ & \ 10                                              \ & \  \nodata          \ & \ 0.045       \ & \ 0.002       \ & \ 3    \ & \ CBET 2023    \ & \ CSS                   \ & \ CBET 2025    \ & \ Ia     \\
\sn{}2009na           \ & \ 161.75577    \ & \ 26.54364     \ & \ UGC 5884                          \ & \ SA(s)b            \ & \ 0.020979    \ & \ 0.000006    \ & \ 2    \ & \ CBET 2098    \ & \ POSS                  \ & \ CBET 2103    \ & \ Ia     \\
\sn{}2010Y            \ & \ 162.76658    \ & \ 65.77966     \ & \ NGC 3392                          \ & \ E?                \ & \ 0.01086     \ & \ 0.000103    \ & \ 1    \ & \ CBET 2168    \ & \ Cortini               \ & \ CBET 2168    \ & \ Ia     \\
PS1-10w            \ & \ 160.67450    \ & \ 58.84392     \ & \ Anonymous                         \ & \  \nodata          \ & \ 0.031255    \ & \ 0.0001      \ & \ 4    \ & \ R14    \ & \ PanSTARRS1             \ & \ R14    \ & \ Ia     \\
PTF10bjs         \ & \ 195.29655    \ & \ 53.81604     \ & \ MCG +09-21-83                     \ & \  \nodata          \ & \ 0.030027    \ & \ 0.000073    \ & \ 1    \ & \ ATEL 2453    \ & \ PTF                   \ & \ ATEL 2453    \ & \ Ia     \\
\sn{}2010ag           \ & \ 255.97330    \ & \ 31.50152     \ & \ UGC 10679                         \ & \ Sb(f)             \ & \ 0.033791    \ & \ 0.000175    \ & \ 2    \ & \ CBET 2195    \ & \ POSS                  \ & \ CBET 2196    \ & \ Ia     \\
\sn{}2010ai           \ & \ 194.84999    \ & \ 27.99646     \ & \ 11                                            \ & \ E                 \ & \ 0.018369    \ & \ 0.000123    \ & \ 1    \ & \ CBET 2200    \ & \ ROTSE-III, Itagaki    \ & \ CBET 2200    \ & \ Ia     \\
\sn{}2010cr           \ & \ 202.35442    \ & \ 11.79637     \ & \ NGC 5177                          \ & \ S0                \ & \ 0.02157     \ & \ 0.000097    \ & \ 1    \ & \ CBET 2281    \ & \ Itagaki, PTF          \ & \ ATEL 2580    \ & \ Ia     \\
\sn{}2010dl           \ & \ 323.75440    \ & \ -0.51345     \ & \ IC 1391                           \ & \  \nodata          \ & \ 0.030034    \ & \ 0.00015     \ & \ 1    \ & \ CBET 2296    \ & \ CSS                   \ & \ CBET 2298    \ & \ Ia     \\
PTF10icb         \ & \ 193.70484    \ & \ 58.88198     \ & \ MCG +10-19-1                      \ & \  \nodata          \ & \ 0.008544    \ & \ 0.000008    \ & \ 2    \ & \ ATEL 2657    \ & \ PTF                   \ & \ ATEL 2657    \ & \ Ia     \\
\sn{}2010dw           \ & \ 230.66775    \ & \ -5.92125     \ & \ 12                                        \ & \  \nodata          \ & \ 0.03812     \ & \ 0.00015     \ & \ 1    \ & \ CBET 2310    \ & \ PIKA                  \ & \ CBET 2311    \ & \ Ia     \\
\sn{}2010ew           \ & \ 279.29933    \ & \ 30.63026     \ & \ CGCG 173-018                      \ & \ S                 \ & \ 0.025501    \ & \ 0.000127    \ & \ 1    \ & \ CBET 2345    \ & \ POSS                  \ & \ CBET 2356    \ & \ Ia     \\
\sn{}2010ex           \ & \ 345.04505    \ & \ 26.09894     \ & \ CGCG 475-019                      \ & \ Compact           \ & \ 0.022812    \ & \ 0.000005    \ & \ 1    \ & \ CBET 2348    \ & \ Ciabattari+           \ & \ CBET 2353    \ & \ Ia     \\
\sn{}2010gn           \ & \ 259.45832    \ & \ 40.88128     \ & \ 13                                           \ & \ Disk Gal          \ & \ 0.0365      \ & \ 0.0058      \ & \ 1    \ & \ ATEL 2718    \ & \ PTF                   \ & \ CBET 2386    \ & \ Ia     \\
\sn{}2010iw           \ & \ 131.31205    \ & \ 27.82325     \ & \ UGC 4570                          \ & \ SABdm             \ & \ 0.021498    \ & \ 0.000017    \ & \ 1    \ & \ CBET 2505    \ & \ CSS                   \ & \ CBET 2511    \ & \ Ia?     \\
\sn{}2010ju           \ & \ 85.48321     \ & \ 18.49746     \ & \ UGC 3341                          \ & \ SBab              \ & \ 0.015244    \ & \ 0.000013    \ & \ 1    \ & \ CBET 2549    \ & \ LOSS                  \ & \ CBET 2550    \ & \ Ia     \\
\sn{}2010jv           \ & \ 111.86051    \ & \ 33.81143     \ & \ NGC 2379                          \ & \ SA0               \ & \ 0.013469    \ & \ 0.000083    \ & \ 1    \ & \ CBET 2549    \ & \ LOSS                  \ & \ CBET 2550    \ & \ Ia     \\
\sn{}2010kg           \ & \ 70.03505     \ & \ 7.34995      \ & \ NGC 1633                          \ & \ SAB(s)ab          \ & \ 0.016632    \ & \ 0.000007    \ & \ 2    \ & \ CBET 2561    \ & \ LOSS                  \ & \ CBET 2561    \ & \ Ia     \\
\sn{}2011B            \ & \ 133.95016    \ & \ 78.21693     \ & \ NGC 2655                          \ & \ SAB(s)0/a         \ & \ 0.00467     \ & \ 0.000003    \ & \ 1    \ & \ CBET 2625    \ & \ Itagaki               \ & \ CBET 262     \ & \ Ia     \\
\sn{}2011K            \ & \ 71.37662     \ & \ -7.34808     \ & \ 14                                            \ & \  \nodata          \ & \ 0.0145      \ & \ 0.001       \ & \ 3    \ & \ CBET 2636    \ & \ CSS                   \ & \ CBET 2636    \ & \ Ia     \\
\sn{}2011aa           \ & \ 114.17727    \ & \ 74.44319     \ & \ UGC 3906                          \ & \ S                 \ & \ 0.012512    \ & \ 0.000033    \ & \ 2    \ & \ CBET 2653    \ & \ POSS                  \ & \ CBET 2653    \ & \ Iap?    \\
\sn{}2011ae           \ & \ 178.70514    \ & \ -16.86280    \ & \ MCG -03-30-19                     \ & \  \nodata          \ & \ 0.006046    \ & \ 0.000019    \ & \ 1    \ & \ CBET 2658    \ & \ CSS                   \ & \ CBET 2658    \ & \ Ia     \\
\sn{}2011ao           \ & \ 178.46267    \ & \ 33.36277     \ & \ IC 2973                           \ & \ SB(s)d            \ & \ 0.010694    \ & \ 0.000002    \ & \ 2    \ & \ CBET 2669    \ & \ POSS                  \ & \ CBET 2669    \ & \ Ia     \\
\sn{}2011at           \ & \ 142.23977    \ & \ -14.80573    \ & \ MCG -02-24-27                     \ & \ SB(s)d            \ & \ 0.006758    \ & \ 0.00002     \ & \ 1    \ & \ CBET 2676    \ & \ POSS                  \ & \ CBET 2676    \ & \ Ia     \\
\sn{}2011by           \ & \ 178.93951    \ & \ 55.32592     \ & \ NGC 3972                          \ & \ SA(s)bc           \ & \ 0.002843    \ & \ 0.000005    \ & \ 1    \ & \ CBET 2708    \ & \ Jin+                  \ & \ CBET 2708    \ & \ Ia     \\
\sn{}2011de           \ & \ 235.97179    \ & \ 67.76196     \ & \ UGC 10018                         \ & \ (R')SB(s)bc       \ & \ 0.029187    \ & \ 0.000017    \ & \ 2    \ & \ CBET 2728    \ & \ POSS                  \ & \ CBET 2728    \ & \ Iap?    \\
\sn{}2011df           \ & \ 291.89008    \ & \ 54.38632     \ & \ NGC 6801                          \ & \ SAcd              \ & \ 0.014547    \ & \ 0.000019    \ & \ 2    \ & \ CBET 2729    \ & \ POSS                  \ & \ CBET 2729    \ & \ Ia     \\
\hline
\end{tabular}
\tablecomments{ \scriptsize
\\
{\bf (a)} See caption in first part of Table~\ref{tab:general}.
\\
{\bf (b)} Host Galaxy Names, discovery references, and discovery group/individual credits from NASA/IPAC Extragalactic Database (NED; \url{http://ned.ipac.caltech.edu/}) and NASA/ADS 
(\url{http://adswww.harvard.edu/abstract\_service.html}). Also see IAUC List of Supernovae: \url{http://www.cbat.eps.harvard.edu/lists/Supernovae.html}.  
For \snIa{} with non-standard IAUC names, we found the associated host galaxy from IAUC/CBET/ATel notices or the literature and searched for the recession velocity with NED.  When the \snIa{} is associated with a faint host not named in any major catalogs (NGC, UGC, \ldots) but named in a large galaxy survey (e.g., SDSS, 2MASS), we include the host name from the large survey rather than ``Anonymous".  However, to fit the table on a single page, long galaxy names are numbered.  \\
10: 2MASX J02014081+151952 (\sn{}2009lf), 11: SDSS J125925.04+275948.2 (\sn{}2010ai), 12: 2MASX J15224062-0555214 (\sn{}2010dw), 13: SDSS J171750.05+405252.5 (\sn{}2010gn), 14: CSS J044530.38-072054.7 (\sn{}2011K). The machine readable version of this table has full galaxy names.
\\
{\bf (c)}--{\bf (e)} See caption in first part of Table~\ref{tab:general}.
\\
{\bf (f)} Spectroscopic type reference. R14=\citealt{rest14}.
\\
{\bf (g)} Spectroscopic type of SN Ia = spectroscopically normal \snIa{}. Spectroscopically peculiar \snIa{}: including 91bg-like and 06gz-like objects.  
Uncertain spectroscopic types are denoted with a question mark ($?$): 
\sn{}2011de: classified as normal Ia in CBET 2728.  But NIR LC morphology is consistent with a slow declining object (e.g., \sn{}2009dc-like). We classify it as Ia-pec.; 
\sn{}2011aa: classified as \sn{}1998aq-like normal Ia in CBET 2653.  But \citealt{brown14} identified it as a Super Chandrasekhar mass candidate, and NIR LC morphology is consistent with a slow declining object (e.g., \sn{}2009dc-like). We classify it as Ia-pec.
\sn{}2010iw: classified as \sn{}2000cx-like, peculiar Ia in CBET 2511.  But the NIR LC has the double peaked morphology of normal Ia. We classify it as a normal Ia.
}
\label{tab:general}
\end{center}
\end{table*}
\renewcommand{\arraystretch}{1}

\renewcommand{\arraystretch}{0.8}
\begin{table*}
\begin{center}
\caption[General Properties of \nsnIaxcfair{} \PTL{} \snIax{}]{General Properties of \nsnIaxcfair{} \PTL{} \snIax{}\\} 
\tiny
\begin{tabular}{@{}l@{}r@{}r@{}l@{}l@{}l@{}l@{}c@{}c@{}c@{}c@{}l@{}}
\hline
SN & \multicolumn{1}{c}{RA\tablenotemark{a}} & \multicolumn{1}{c}{DEC\tablenotemark{a}} & Host\tablenotemark{b} & Morphology\tablenotemark{c} & \multicolumn{1}{c}{$z_{\rm helio}$\tablenotemark{d}} & \multicolumn{1}{c}{$\sigma_{z_{\rm helio}}$\tablenotemark{d}} & $z$ \tablenotemark{d} & Discovery\tablenotemark{b} & Discoverer(s)\tablenotemark{e} & Type\tablenotemark{f} & Type\tablenotemark{g} \\
Name & \multicolumn{1}{c}{$\alpha(2000)$} & \multicolumn{1}{c}{$\delta(2000)$} & Galaxy &  & &  & Ref. & Reference &  & Reference &  \\
\hline
\sn{}2005hk           \ & \ 6.96187      \ & \ -1.19819     \ & \ UGC 272                           \ & \ SAB(s)d       \ & \ 0.012993    \ & \ 0.000041    \ & \ 1    \ & \ IAUC 8625    \ & \ SDSS-II, LOSS         \ & \ CBET 269; Ph07     \ & \ Iax    \\
\sn{}2008A            \ & \ 24.57248     \ & \ 35.37029     \ & \ NGC 634                           \ & \ Sa                \ & \ 0.016455    \ & \ 0.000007    \ & \ 2    \ & \ CBET 1193    \ & \ Ichimura              \ & \ CBET 1198; F13; Mc14b    \ & \ Iax    \\
\sn{}2008ae           \ & \ 149.01322    \ & \ 10.49965     \ & \ IC 577                            \ & \ S?                  \ & \ 0.03006     \ & \ 0.000037    \ & \ 2    \ & \ CBET 1247    \ & \ POSS                  \ & \ CBET 1250; F13    \ & \ Iax    \\
\sn{}2008ha           \ & \ 353.71951    \ & \ 18.22659     \ & \ UGC 12682                     \ & \ Im                \ & \ 0.004623    \ & \ 0.000002    \ & \ 2    \ & \ CBET 1567    \ & \ POSS                  \ & \ CBET 1576; F09    \ & \ Iax    \\
\hline
\end{tabular}
\tablecomments{ \scriptsize
\\
{\bf (a)}--{\bf (e)} See Table~\ref{tab:general} caption.
\\
{\bf (f)} Spectroscopic type reference, Ph07: \citealt{phillips07}; F09: \citealt{foley09b}; F13: \citealt{foley13}; Mc14b: \citealt{mccully14a}.
\\
{\bf (g)} Spectroscopic type.  Iax (\citealt{foley13}).
}
\label{tab:general3}
\end{center}
\end{table*}
\renewcommand{\arraystretch}{1}

\section{Observations}
\label{sec:observations}

In~\S\ref{sec:NIR_Ia}, we provide recent historical context for \F{} by describing the growing low-$z$ sample of NIR \snIa{} LCs. In \S\ref{sec:pobs}-\ref{sec:strategy}, we overview CfA NIR SN observations, describe \PTL{}'s observing capabilities, and detail our follow up strategy to observe \snIa{} in \jhk{}.

\subsection{Low-$z$ NIR Light Curves of \snIa{}}
\label{sec:NIR_Ia}

Technological advances in infrared detector technology have recently made it possible to obtain high quality NIR photometry for large numbers of \snIa{}.  \citet{phillips12} provides an excellent recent review of the cosmological and astrophysical results derived from NIR \snIa{} observations made over the past three decades.  Early NIR observations of \snIa{} were made by \citet{kirshner73,elias81,elias85,frogel87}, and were particularly challenging as a result of the limited technology of the time. In addition, the flux contrast between the host galaxy and the \snIa{} is typically smaller in the NIR than at optical wavelengths,
making high S/N observations possible only for the brightest NIR objects with the detectors available in the 1970s and 1980s.  While this situation has improved somewhat in the subsequent decades, NIR photometry is still significantly more challenging than at optical wavelengths. \citet{elias85} was the first to present a NIR Hubble diagram for 6 \snIa{}. Although these 6 \snIa{} LCs were not classified spectroscopically, \citet{elias85} was also the first to use what became the modern spectroscopic nomenclature of Type Ia instead of Type I to distinguish between Type Ia and Type Ib SN; SN Ib are now thought to be core collapse supernovae of stars that have lost their outer Hydrogen envelopes (see \citealt{modjaz14} and references therein).

In the late 1990s and early 2000s, panoramic NIR arrays made it possible to obtain NIR photometry comparable in quantity and quality to optical photometry for nearby \snIa{}.   The first early-time NIR photometry with modern NIR detectors observed before \tbmax{} was presented for \sn{}1998bu \citep{jha99,hernandez00}.  Since the first peak in the \jhk-band occurs $\sim 3$--$5$ days {\it before} \tbmax{}, depending on the filter, \snIa{} must generally be discovered by optical searches at least $\sim 5$--$8$ days before \tbmax{} in order to be observed before the NIR maximum (F12; see~\S\ref{sec:strategy}).  

Pioneering early work was performed in the early 2000s in Chile at the Las Campanas Observatory (LCO) and the Cerro Tololo Inter-American Observatory (CTIO), spearheaded by the work of \citet{krisciunas00,krisciunas01,krisciunas03,krisciunas04b,krisciunas04c}.  
K04a presented the largest Hubble diagram of its kind to date with 16 \snIa{}.  
Before \WV{} published \nsnIawv{} \PTL{} NIR LCs observed by the CfA at FLWO, a handful of other NIR observations, usually for individual or small numbers of \snIa{} or \snIax{} of particular interest were presented in \citep{cuadra02,dipaola02,valentini03,candia03,benetti04,garnavich04,sollerman04,krisciunas05a,krisciunas06,krisciunas07,phillips06,phillips07,pastorello07a,pastorello07b,stritzinger07,stanishev07,eliasrosa06,eliasrosa08,pignata08,wangx08,taubenberger08}.  
The largest NIR \snIa{} sample prior to \F{} was obtained by the Carnegie Supernova Project (CSP: \citealt{freedman05,hamuy06}) at LCO, including observations of 59 normal and 14 peculiar NIR \snIa{} LCs \citep{schweizer08,contreras10,stritzinger10,stritzinger11,taubenberger11}.\footnote{The CSP work did not yet distinguish \snIax{} as a separate subclass from \snIa{}.}
Other \snIa{} or \snIax{} papers with published NIR data since \WV{} include 
\citep{krisciunas09,leloudas09,yamanaka09,krisciunas11,barone12,biscardi12,matheson12,taddia12,silverman13,stritzinger14,weyant14,cartier14,foley14b,amanullah14,goobar14,marion15a,stritzinger15}.  See Table~\ref{tab:NIRcensus2} for a fairly comprehensive listing of \snIa{} and \snIax{} with NIR observations in the literature or presented in this paper.

Overall, while $\sim1000$ nearby \snIa{} have been observed at optical wavelengths, prior to \F{}, only 147 total unique nearby objects have at least 1 NIR band of published $Y$\jhk{} data obtained with modern NIR detectors (from \sn{}1998bu onwards). These include 121 normal \snIa{}, 22 peculiar \snIa{}, and 4 \snIax{}. 
\F{} adds 66 new unique objects, including 62 normal \snIa{}.  By this measure, \F{} increases the world published NIR sample of total unique objects by $66/147 \approx 45$\% and normal \snIa{} by $62/121 \approx 51$\%. 12 additional \F{} objects have new data which supersedes previously published \PTL{} LCs and 
no data published by other groups.  If we include these, \F{} adds 78 total objects and 73 normal \snIa{} to the literature.  By this measure, \F{} increases the world published sample of NIR objects by $78/135 \approx 58$\% and the sample of normal \snIa{} by $72/110 \approx 65$\%.  See Table~\ref{tab:NIRcensus2}.

\renewcommand{\arraystretch}{0.01}
\begin{table*}
\begin{center}
\caption[\snIa{} and \snIax{} with NIR Photometry]{\snIa{} and \snIax{} with Published NIR Photometry\\}
\tiny
\begin{tabular}{lll | lll | lll}
\hline
SN Name           & Type$^a$ &  NIR Photometry  & SN Name              & Type$^a$ & NIR Photometry & SN Name              & Type$^a$ & NIR Photometry\\
                            &                    &   References$^b$ &                       &                  & References$^b$ &                          &                  & References$^b$\\
\hline
\sn{}2012Z            & Iax       & S15                      &         \sn{}2007nq           & Ia         & CfAIR2; S11    & \sn{}2007as    & Ia        & S11 \\
\sn{}2014J            & Ia         & A14; Go14; F14b    & \sn{}2007le    & Ia        & CfAIR2; S11     & \sn{}2007ax      & Ia-pec     & S11                \\
\sn{}2013bh           & Ia-pec     & Si13                & \sn{}2007if    & Ia-pec    & CfAIR2; S11                & \sn{}2007ba      & Ia-pec     & S11                \\
\sn{}2011fe           & Ia         & M12                 & \sn{}2007fb    & Ia        & CfAIR2                     & \sn{}2007bc      & Ia         & S11                \\
\sn{}2010ae           & Iax        & S14                 & \sn{}2007cq    & Ia        & CfAIR2; WV08               & \sn{}2007bd      & Ia         & S11                \\
\sn{}2008J            & Ia         & Ta12                & \sn{}2007co    & Ia        & CfAIR2                     & \sn{}2007bm      & Ia         & S11                \\
\sn{}2011df           & Ia         & CfAIR2              & \sn{}2007ca    & Ia        & CfAIR2; S11                & \sn{}2007hx      & Ia         & S11                \\
\sn{}2011de           & Ia-pec?    & CfAIR2              & \sn{}2007S     & Ia        & CfAIR2; S11                & \sn{}2007jg      & Ia         & S11                \\
\sn{}2011by           & Ia         & CfAIR2              & \sn{}2006mq    & Ia        & CfAIR2                     & \sn{}2007on      & Ia         & S11                \\
\sn{}2011at           & Ia         & CfAIR2              & \sn{}2006lf    & Ia        & CfAIR2; WV08               & \sn{}2008R       & Ia         & S11                \\
\sn{}2011ao           & Ia         & CfAIR2              & \sn{}2006le    & Ia        & CfAIR2; WV08               & \sn{}2008bc      & Ia         & S11                \\
\sn{}2011ae           & Ia         & CfAIR2              & \sn{}2006gr    & Ia        & CfAIR2; WV08               & \sn{}2008bq      & Ia         & S11                \\
\sn{}2011aa           & Ia-pec?    & CfAIR2              & \sn{}2006cz    & Ia        & CfAIR2                     & \sn{}2008fp      & Ia         & S11                \\
\sn{}2011K            & Ia         & CfAIR2              & \sn{}2006cp    & Ia        & CfAIR2; WV08               & \sn{}2008gp      & Ia         & S11                \\
\sn{}2011B            & Ia         & CfAIR2              & \sn{}2006ax    & Ia        & CfAIR2; WV08; C10          & \sn{}2008ia      & Ia         & S11                \\
\sn{}2010kg           & Ia         & CfAIR2              & \sn{}2006ac    & Ia        & CfAIR2; WV08               & \sn{}2009F       & Ia-pec     & S11                \\
\sn{}2010jv           & Ia         & CfAIR2              & \sn{}2006X     & Ia        & CfAIR2; WV08; C10; WX08    & \sn{}2004eo      & Ia         & C10; Pa07b         \\
\sn{}2010ju           & Ia         & CfAIR2              & \sn{}2006N     & Ia        & CfAIR2; WV08               & \sn{}2004S       & Ia         & K07                \\
\sn{}2010iw           & Ia?        & CfAIR2              & \sn{}2006E     & Ia        & CfAIR2                     & \sn{}2003hv      & Ia         & L09                \\
\sn{}2010gn           & Ia         & CfAIR2              & \sn{}2006D     & Ia        & CfAIR2; WV08; C10          & \sn{}2003gs      & Ia-pec     & K09                \\
\sn{}2010ex           & Ia         & CfAIR2              & \sn{}2005na    & Ia        & CfAIR2; WV08; C10          & \sn{}2003du      & Ia         & St07               \\
\sn{}2010ew           & Ia         & CfAIR2              & \sn{}2005ls    & Ia        & CfAIR2                     & \sn{}2003cg      & Ia         & ER06               \\
\sn{}2010dw           & Ia         & CfAIR2              & \sn{}2005ke    & Ia-pec    & CfAIR2; WV08; C10          & \sn{}2002fk      & Ia         & Ca14               \\
PTF10icb         & Ia         & CfAIR2              & \sn{}2005iq    & Ia        & CfAIR2; WV08; C10          & \sn{}2002dj      & Ia         & P08                \\
\sn{}2010dl           & Ia         & CfAIR2              & \sn{}2005hk    & Iax       & CfAIR2; WV08; Ph07         & \sn{}2002cv      & Ia         & ER08; DP02         \\
\sn{}2010cr           & Ia         & CfAIR2              & \sn{}2005eu    & Ia        & CfAIR2; WV08               & \sn{}2002bo      & Ia         & K04c; B04          \\
\sn{}2010ai           & Ia         & CfAIR2              & \sn{}2005eq    & Ia        & CfAIR2; WV08; C10          & \sn{}2001el      & Ia         & K03; S07           \\
\sn{}2010ag           & Ia         & CfAIR2              & \sn{}2005el    & Ia        & CfAIR2; WV08; C10          & \sn{}2001cz      & Ia         & K04c               \\
PTF10bjs         & Ia         & CfAIR2              & \sn{}2005ch    & Ia        & CfAIR2; WV08               & \sn{}2001cn      & Ia         & K04c               \\
PS1-10w            & Ia         & CfAIR2              & \sn{}2005cf    & Ia        & CfAIR2; WV08; Pa07a        & \sn{}2001bt      & Ia         & K04c               \\
\sn{}2010Y            & Ia         & CfAIR2              & \sn{}2005bo    & Ia        & CfAIR2                     & \sn{}2001ba      & Ia         & K04b               \\
\sn{}2009na           & Ia         & CfAIR2              & \sn{}2005bl    & Ia-pec    & CfAIR2; WV08               & \sn{}2001ay      & Ia-pec     & K11                \\
\sn{}2009lf           & Ia         & CfAIR2              & \sn{}2005ao    & Ia        & CfAIR2; WV08               & \sn{}2000cx      & Ia-pec     & Ca03; So04; Cu02   \\
\sn{}2009le           & Ia         & CfAIR2              & \sn{}2004ef    & Ia        & C10                        & \sn{}2000ce      & Ia         & K01                \\
\sn{}2009kq           & Ia         & CfAIR2              & \sn{}2004ey    & Ia        & C10                        & \sn{}2000ca      & Ia         & K04b               \\
\sn{}2009kk           & Ia         & CfAIR2              & \sn{}2004gs    & Ia        & C10                        & \sn{}2000bk      & Ia         & K01                \\
\sn{}2009jr           & Ia         & CfAIR2              & \sn{}2004gu    & Ia-pec    & C10                        & \sn{}2000bh      & Ia         & K04b               \\
\sn{}2009im           & Ia         & CfAIR2              & \sn{}2005A     & Ia        & C10                        & \sn{}2000E       & Ia         & V03                \\
\sn{}2009ig           & Ia         & CfAIR2              & \sn{}2005M     & Ia        & C10                        & \sn{}1999gp      & Ia         & K01                \\
\sn{}2009fv           & Ia         & CfAIR2              & \sn{}2005ag    & Ia        & C10                        & \sn{}1999ek      & Ia         & K04c               \\
\sn{}2009fw           & Ia         & CfAIR2              & \sn{}2005al    & Ia        & C10                        & \sn{}1999ee      & Ia         & K04b               \\
\sn{}2009ds           & Ia         & CfAIR2              & \sn{}2005am    & Ia        & C10                        & \sn{}1999cp      & Ia         & K00                \\
\sn{}2009do           & Ia         & CfAIR2              & \sn{}2005hc    & Ia        & C10                        & \sn{}1999cl      & Ia         & K00                \\
\sn{}2009dc           & Ia-pec     & CfAIR2; T11; Y09    & \sn{}2005kc    & Ia        & C10                        & \sn{}1999by      & Ia-pec     & G04                \\
\sn{}2009bv           & Ia         & CfAIR2              & \sn{}2005ki    & Ia        & C10                        & \sn{}1999ac      & Ia-pec     & Ph06               \\
\sn{}2009an           & Ia         & CfAIR2              & \sn{}2006bh    & Ia        & C10                        & \sn{}1999aa      & Ia-pec     & K00                \\
\sn{}2009al           & Ia         & CfAIR2              & \sn{}2006eq    & Ia        & C10                        & \sn{}1998bu      & Ia         & H00; J99           \\
\sn{}2009ad           & Ia         & CfAIR2              & \sn{}2006gt    & Ia-pec    & C10                        & PTF09dlc    & Ia         & BN12               \\
\sn{}2009Y            & Ia         & CfAIR2              & \sn{}2006mr    & Ia-pec    & C10                        & PTF10hdv    & Ia         & BN12               \\
\sn{}2009D            & Ia         & CfAIR2              & \sn{}2006dd    & Ia        & S10                        & PTF10mwb    & Ia         & BN12               \\
\sn{}2008hy           & Ia         & CfAIR2              & \sn{}2005hj    & Ia        & S11                        & PTF10ndc    & Ia         & BN12               \\
\sn{}2008hv           & Ia         & CfAIR2; S11         & \sn{}2005ku    & Ia        & S11                        & PTF10nlg    & Ia         & BN12               \\
\sn{}2008hs           & Ia         & CfAIR2              & \sn{}2006bd    & Ia-pec    & S11                        & PTF10qyx    & Ia         & BN12               \\
\sn{}2008hm           & Ia         & CfAIR2              & \sn{}2006br    & Ia        & S11                        & PTF10tce    & Ia         & BN12               \\
\sn{}2008ha           & Iax        & CfAIR2; F09         & \sn{}2006bt    & Ia-pec    & S11                        & PTF10ufj    & Ia         & BN12               \\
\sn{}2008gl           & Ia         & CfAIR2              & \sn{}2006ej    & Ia        & S11                        & PTF10wnm    & Ia         & BN12               \\
\sn{}2008gb           & Ia         & CfAIR2              & \sn{}2006et    & Ia        & S11                        & PTF10wof    & Ia         & BN12               \\
\sn{}2008fx           & Ia         & CfAIR2              & \sn{}2006ev    & Ia        & S11                        & PTF10xyt    & Ia         & BN12               \\
\sn{}2008fv           & Ia         & CfAIR2; Bi12        & \sn{}2006gj    & Ia        & S11                        & \sn{}2011hr      & Ia         & W14                \\
\sn{}2008fr           & Ia         & CfAIR2              & \sn{}2006hb    & Ia        & S11                        & \sn{}2011gy      & Ia         & W14                \\
SNF20080522-011    & Ia         & CfAIR2              & \sn{}2006hx    & Ia        & S11                        & \sn{}2011hk      & Ia-pec     & W14                \\
SNF20080522-000    & Ia         & CfAIR2              & \sn{}2006is    & Ia        & S11                        & \sn{}2011fs      & Ia         & W14                \\
SNF20080514-002    & Ia         & CfAIR2              & \sn{}2006kf    & Ia        & S11                        & \sn{}2011gf      & Ia         & W14                \\
\sn{}2008af           & Ia         & CfAIR2              & \sn{}2006lu    & Ia        & S11                        & \sn{}2011hb      & Ia         & W14                \\
\sn{}2008ae           & Iax        & CfAIR2              & \sn{}2006ob    & Ia        & S11                        & \sn{}2011io      & Ia         & W14                \\
\sn{}2008Z            & Ia         & CfAIR2              & \sn{}2006os    & Ia        & S11                        & \sn{}2011iu      & Ia         & W14                \\
\sn{}2008C            & Ia         & CfAIR2; S11         & \sn{}2006ot    & Ia-pec    & S11                        & PTF11qri    & Ia         & W14                \\
\sn{}2008A            & Iax        & CfAIR2              & \sn{}2007A     & Ia        & S11                        & PTF11qmo    & Ia         & W14                \\
\sn{}2007sr           & Ia         & CfAIR2; S08         & \sn{}2007N     & Ia-pec    & S11                        & PTF11qzq    & Ia         & W14                \\
\sn{}2007rx           & Ia         & CfAIR2              & \sn{}2007af    & Ia        & S11                        & PTF11qpc    & Ia         & W14                \\
\sn{}2007qe           & Ia         & CfAIR2              & \sn{}2007ai    & Ia        & S11                        & \sn{}2011ha      & Ia         & W14                \\
\hline
\end{tabular}
\tablecomments{ \scriptsize
\\
(a) SN Spectroscopic Types: 
Ia = Normal \snIa{} including 91T-like, 86G-like, and spectroscopically normal objects; 
Iap = Peculiar \snIa{} including 91bg-like objects and extra-luminous, slow declining 06gz-like objects (\citealt{hicken07});
Iax = \snIax{} including 02cx-like objects distinct from peculiar SN Ia (\citealt{li03,foley13}). 
Spectroscopic type references for \F{} objects are in Tables~\ref{tab:general}-\ref{tab:general3}, and in the references below for non-\F{} objects with NIR photometry.  
SN with uncertain spectral types (\sn{}2011de, \sn{}2011aa, \sn{}2010iw) are denoted by a question mark ($?$) (see Table~\ref{tab:general} caption).
\\
(b) References for objects with at least 1 band of $YJHK_s$ photometry.  
CfAIR2: this paper; WV08: \citet{woodvasey08}; W14: \citet{weyant14}; S15: \citet{stritzinger15}; S14: \citet{stritzinger14}; F14b: \citet{foley14b}; Go14: \citet{goobar14}; Ca14: \citet{cartier14}; A14: \citet{amanullah14}; Si13: \citet{silverman13}; Ta12: \citet{taddia12}; M12: \citet{matheson12}; Bi12: \citet{biscardi12}; BN12: \citet{barone12}; T11: \citet{taubenberger11}; S11: \citet{stritzinger11}; K11: \citet{krisciunas11}; S10: \citet{stritzinger10}; C10: \citet{contreras10}; Y09: \citet{yamanaka09}; L09: \citet{leloudas09}; K09: \citet{krisciunas09}; F09: \citet{foley09b}; WX08: \citet{wangx08}; T08: \citet{taubenberger08}; S08: \citet{schweizer08}; P08: \citet{pignata08}; ER08: \citet{eliasrosa08}; S07: \citet{stritzinger07}; St07: \citet{stanishev07}; Ph07: \citet{phillips07}; Pa07b: \citet{pastorello07a}; Pa07a: \citet{pastorello07b}; K07: \citet{krisciunas07}; Ph06: \citet{phillips06}; ER06: \citet{eliasrosa06}; K05: \citet{krisciunas05}; So04: \citet{sollerman04}; K04c: \citet{krisciunas04c}; K04b: \citet{krisciunas04b}; G04: \citet{garnavich04}; B04: \citet{benetti04}; V03: \citet{valentini03}; K03: \citet{krisciunas03}; Ca03: \citet{candia03}; DP02: \citet{dipaola02}; Cu02: \citet{cuadra02}; K01: \citet{krisciunas01}; K00: \citet{krisciunas00}; H00: \citet{hernandez00}; J99: \citet{jha99}.
}
\label{tab:NIRcensus2} 
\end{center}
\end{table*}
\renewcommand{\arraystretch}{1}

\subsection{\PTL{} NIR Supernova Observations}
\label{sec:pobs}

Out of 121 total \snIa{} and \snIax{} observed from 2005-2011 by \PTL, \nIanogogo{} are not included in \F{}.  \F{} includes improved photometry for 20 of \nsnIaredo{} objects from \WV{}. For \sn{}2005cf, our photometry pipeline failed to produce a galaxy subtracted LC, so we include the \WV{} LC for \sn{}2005cf in \F{} and all applicable Figures or Tables.  These 20 objects include additional observations not published in \WV{}, processed homogeneously using upgraded mosaic and photometry pipelines (see \S\ref{sec:reduction}).  Table~\ref{tab:general} lists general properties of the \nsnIacfair{} \F{} \snIa{} and Table~\ref{tab:general3} lists these for the \nsnIaxcfair{} \F{} \snIax{}. 

Heliocentric galaxy redshifts are provided in Tables~\ref{tab:general}-\ref{tab:general3} and CMB frame redshifts are given in Table~\ref{tab:jhkdata} to ease construction of future Hubble diagrams including NIR \snIa{} data.\footnote{However, note that none of the redshifts in Tables~\ref{tab:general}-\ref{tab:general3} or Table~\ref{tab:jhkdata} have been corrected for local flow models. Objects with recession velocities $\lesssim 3000$ km s$^{-1}$ ($z \lesssim 0.01$) must have their redshifts corrected with local flow models or other distance information before being included in Hubble diagrams.} We obtained recession velocities from identified host galaxies as listed in the NASA/IPAC Extragalactic Database (NED). In cases where NED did not return a host galaxy or the host galaxy had no reported NED redshift, we either obtained redshift estimates from our own CfA optical spectra (\citealt{matheson08,blondin12}) or found redshifts reported in the literature. \fig{}~\ref{fig:zhist} shows a histogram of \F{} heliocentric galaxy redshifts $z_{\rm helio}$ for \nsnIanormtempl{} normal \snIa{} with \tbmax{} estimates accurate to within less than 10 days. 

\begin{figure}
\centering
\begin{tabular}{@{}c@{}}
\includegraphics[width=\scale\linewidth,angle=0]
{\colordir/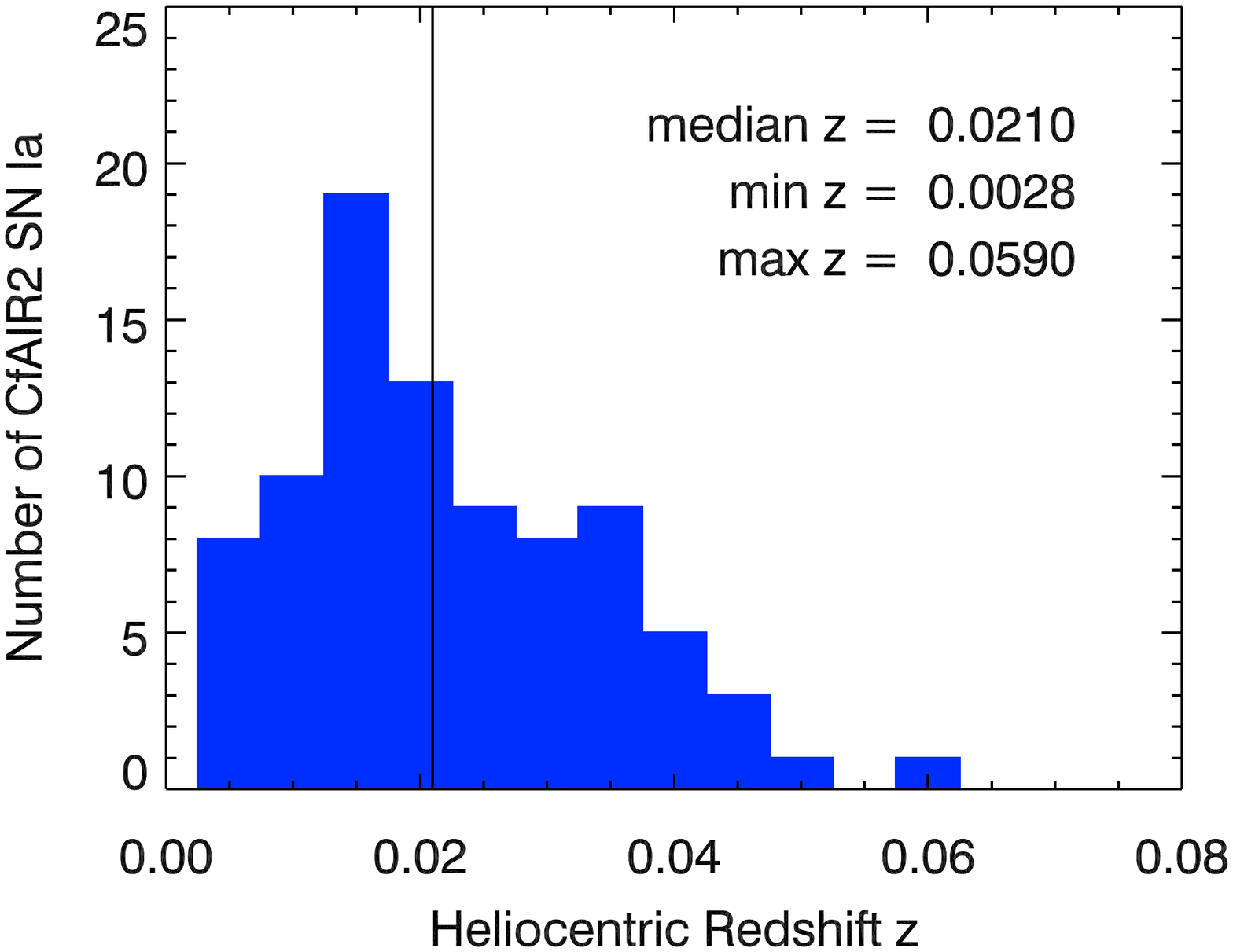}
\vspace{-0.3cm}
\end{tabular}
\caption[Histogram of \F{} Heliocentric Redshifts]{
Histogram of \F{} Heliocentric Redshifts
\\
\\
\scriptsize
(Color online) Histogram of heliocentric redshifts $z_{\rm helio}$ for \nsnIanormtempl{} spectroscopically normal \F{} \snIa{} from Table~\ref{tab:general} with \tbmax{} estimates accurate to within less than 10 days. Bin size $\Delta z = 0.005$.  Redshift statistics for the sample include: median (black vertical line, $0.0210$), minimum ($0.0028$), and maximum ($0.0590$). Heliocentric redshifts have not been corrected for any local flow models.
}
\label{fig:zhist}
\end{figure}

From 2005-2011, we also obtained extensive \PTL{} NIR observations of 
25 \snIbc{} \citep{bianco14}, and \nII{} \snII{} (to be presented elsewhere).
Table~\ref{tab:ptel_prev} references all previously published and in preparation papers using \PTL{} SN data, including multi-wavelength studies of individual objects (\citealt{tominaga05,kocevski07,modjaz09,wangx09,foley09b,sanders13,drout13,marion14a,margutti14,fransson14}) and NIR/optical LC compilations for SN of all types (e.g., \citealt{modjaz07}; \WV{}; F12; \citealt{bianco14}).
The most recent of these papers (\citealt{sanders13,marion14a,margutti14,fransson14,bianco14}) used the same mosaic and photometry pipelines also used to produce the CfAIR2 data for this paper (see \S\ref{sec:reduction}).  For completeness, we also include information on all other types of SN with published \PTL{} observations for both current and older pipelines.

\subsection{\PTL{} 1.3-m Specifications}
\label{sec:pairitel}

Dedicated in October 2004, \PTL{} uses the Two Micron All Sky Survey (2MASS; \citealt{skrutskie06}) northern telescope together with the 2MASS southern camera. \PTL{} is a fully automated robotic telescope with the sequence of observations controlled by an optimized queue-scheduling database \citep{bloom03,bloom06}.  Two dichroic mirrors allow simultaneous observing in \jhk{} (1.2, 1.6, and 2.2~$\mu$m, respectively; \citealt{cohen03,skrutskie06}) with three 256$\times$256 pixel HgCdTe NICMOS3 arrays.  Figure 1 of \WV{} shows a composite \jhk{} mosaicked image of SN 2006D (see \S\ref{sec:mosaics}).

Since the observations are conducted with the instrument that defined the 2MASS \jhk{} system, we use the 2MASS point source catalog \citep{cutri03} to establish photometric zero points. Typical 30-minute (1800-second) observations (including slew overhead) reach $10$-$\sigma$ sensitivity limits of $\sim18,17.5,$ and $17$ mag for point sources in \jhk{}, respectively (F12). For fainter objects, $10$-$\sigma$ point source sensitivities of $19.4$, $18.5$, and $18$ mag are achievable with 1.5 hours (5400-seconds) of dithered imaging \citep{bloom03} in \jhk{}, respectively. \PTL{} thus observes significantly deeper than 2MASS, which used a 7.8-second total exposure time to achieve 10-$\sigma$ point source sensitivities of 15.8, 15.1, 14.3 in \jhk{}, respectively (\citealt{skrutskie06}; see \S\ref{sec:phot}).

\subsection{Observing Strategy}
\label{sec:strategy}

Automation of \PTL{} made it possible to study SN with unprecedented temporal coverage in the NIR, by responding quickly to new SN and revisiting targets frequently (\citealt{bloom06}; \WV{}; F12).  \F{} followed up SN discovered by optical searches at $\delta \gtrsim -30$ degrees with $V \lesssim 18$ mag, with significant discovery contributions from both amateur and professional astronomers (see Tables~\ref{tab:general}-\ref{tab:general3}).  SN candidates with a favorable observation window and airmass $< 2.5$ from Mount Hopkins were considered for the \PTL{} observation queue.  We observed SN of all types but placed highest priority on the brightest \snIa{} discovered early or close to maximum brightness. SN candidates meeting these criteria were often added to the queue before spectroscopic typing to observe the early time LC. Since many optically discovered SN of all types brighter than $V < 18$ mag are spectroscopically typed by our group at the CfA\footnote{\singlespace \footnotesize \url{http://www.cfa.harvard.edu/supernova/OldRecentSN.html}} or other groups within 1-3 days of discovery, we rarely spent more than a few observations on objects we later deactivated after typing. All CfA supernovae are spectroscopically classified using the SuperNova IDentification code (SNID; \citealt{blondin07}). 

From 2005-2011, $\sim20$--$30$ SN per year were discovered that were bright enough to observe with the \PTL{} 1.3-m, with $\sim3$--$6$ available on any given night from Mount Hopkins.  Since we only perform follow-up NIR observations and are not conducting a NIR search to discover SN with \PTL{}, we suffer from all the heterogeneous sample selection effects and biases incurred by each of the independent discovery efforts.  A full analysis of the completeness of our sample is beyond the scope of this work.  
Overall, with $\sim30\%$ of the time on a robotic telescope available for supernova observations, effectively amounting to over 6 months on the sky, we observed over $2/3$ of the candidate SN that met our follow-up criteria.  We also observed galaxy template images (\snt{}) for each SN to enable host subtraction (see \S\ref{sec:host_sub}). 

\renewcommand{\arraystretch}{0.5}
\begin{table*}
\begin{center}
\caption[SN with \PTL{} Data]{SN with Published or Forthcoming \PTL{} Data\\}
\scriptsize
\begin{tabular}{llll}
\hline
Object or      & Type(s) & Reference & Comments \\
{\it Compilation} &             &                     &  \\
\hline	
\sn{}2005bf & Ic-Ib & \citealt{tominaga05} & Unusual core-collapse object \\
\sn{}2006aj & Ic-BL & \citealt{modjaz06,kocevski07} & Associated with GRB 060281\\
\sn{}2006jc & Ib/c  & \citealt{modjaz07} & Unusual core-collapse object; in M. Modjaz PhD Thesis \\
\sn{}2008D & Ib    & \citealt{modjaz09} & Associated with {\it Swift} X-ray transient XRT 080109 \\
\sn{}2005cf & Ia    & \citealt{wangx09} & Normal \snIa{}, significant multi-wavelength data \\
\sn{}2008ha & Iax    & \citealt{foley09b} & Extremely low luminosity supernova Iax \tablenotemark{a} \\
{\it WV08} & Ia,Ia-pec,Iax    & \citealt{woodvasey08} & Compilation of 20 \snIa{} and 1 \snIax{} NIR LCs \tablenotemark{a} \\
{\it F12} & Ia,Ia-pec,Iax    & \citealt{friedman12} & Compilation of \snIa{} and \snIax{} in A. Friedman PhD Thesis \tablenotemark{a}\\
{\it M07} & Ib,Ic    & \citealt{modjaz07} & Compilation of SN Ib and SN Ic in M. Modjaz PhD Thesis \tablenotemark{b} \\
PS1-12sk & Ibn    & \citealt{sanders13} & Pan-STARRS1 project observations \\
\sn{}2005ek & Ic    & \citealt{drout13} & Photometry from \citealt{modjaz07} PhD Thesis \tablenotemark{b} \\
\sn{}2011dh & IIb   & \citealt{marion14a} & SN in M51 \\
\sn{}2009ip & LBV   & \citealt{margutti14} & Luminous blue variable with outbursts. Not a supernova\\
\sn{}2010jl & IIn   & \citealt{fransson14} & Unusual core-collapse object \\
{\it B14} & Ib,Ic   & \citealt{bianco14} & Compilation of \PTL{} SN Ib and SN Ic \tablenotemark{b} \\
{\it \F{}} & Ia,Ia-pec,Iax     & \FRIED{} & This paper. Compilation of \PTL{} \snIa{}, SN Ia-pec, \snIax{} \tablenotemark{a}\\
\sn{}2012cg & Ia  & \HOWIEII{} & Bright Ia with multi-wavelength data \\ 
\hline
\end{tabular}
\tablecomments{ \scriptsize
\\ 
(a) Photometry in this paper supersedes \PTL{} LCs from \citealt{woodvasey08} (except \sn{}2005cf ), \sn{}2008ha LC in \citealt{foley09b}, F12.
\\
(b) B14 supersedes M. Modjaz PhD Thesis. 
\\
}
\label{tab:ptel_prev} 
\end{center}
\end{table*}
\renewcommand{\arraystretch}{1}

\section{Data Reduction}
\label{sec:reduction}

Since \WV{}, we have substantially upgraded our data reduction software, including both pipelines for combining the raw data into mosaics and for performing photometry on the mosaicked images. All \F{} data were processed homogeneously with a single mosaicking pipeline (hereafter \texttt{p3.6}) that adds and registers \PTL{} raw images into mosaics (\S\ref{sec:mosaics}).  The mosaics, and their associated noise and exposure maps, were then fed to a single photometry pipeline (hereafter \texttt{photpipe}), originally developed to handle optical data for the ESSENCE and SuperMACHO projects (\citealt{rest05,garg07,miknaitis07}) and modified to perform host galaxy subtraction and photometry  on the NIR mosaicked images (\S\ref{sec:host_sub}--\ref{sec:photpipe}). Earlier mosaic and \texttt{photpipe} versions have been used for previously published \PTL{} SN LCs (see Table~\ref{tab:ptel_prev}), with recent modifications by A. Friedman and W.M. Wood-Vasey to produce compilations of \snIa{} and \snIax{} (\F{}; this work) and SN Ib and SN Ic (\citealt{bianco14}). \texttt{Photpipe} now takes as input improved noise mosaics to estimate the noise in the mosaicked images (\S\ref{sec:sky}), registers the images to a common reference frame with \swarp{} \citep{bertin02}, subtracts host galaxy light at the SN position using reference images with \texttt{HOTPANTS} \citep{becker04,becker07}, and performs point-spread function (PSF) photometry using \dophot{}~\citep{schechter93}.  Photometry is extracted from either the unsubtracted or the subtracted images by forcing \dophot{} to measure the PSF-weighted flux of the object at a fixed position in pixel coordinates (see \S\ref{sec:host_sub}; F12).

In \S\ref{sec:mosaics}, we describe our \texttt{p3.6} mosaic pipeline.  In \S\ref{sec:sky}, we describe sky subtraction and our improved method to produce noise mosaics corresponding to the mosaicked images.   In \S\ref{sec:undersample}, we discuss the undersampling of the \PTL{} NIR camera. 
In \S\ref{sec:host_sub}--\ref{sec:nn2nnt} we detail the host galaxy subtraction process and describe our method for performing photometry on the subtracted or unsubtracted images. Major \texttt{photpipe} improvements are summarized in~\S\ref{sec:photpipe}. See F12 for additional details.

\subsection{Mosaics}
\label{sec:mosaics}
 
All \F{} images were processed into mosaics at the CfA using \texttt{p3.6} implemented in Python version 2.6.\footnote{\texttt{p1.0}-\texttt{p3.6} was developed at UC Berkeley and the CfA by J.S. Bloom, C. Blake, C. Klein, D. Starr, and A. Friedman.} F12 and references in Table~\ref{tab:ptel_prev} describe older mosaic pipelines. \citealt{klein14} provide a more detailed description of \texttt{p3.6} as used for \PTL{} observations of RR Lyrae stars. \figs{}~\ref{fig:galleryA}-\ref{fig:galleryC} show sample \texttt{p3.6} $J$-band mosaics for all 98 \F{} objects.

Including slew overhead for the entire dither pattern, typical exposure times range from 600 to 3600 seconds, yielding $\sim50$--$150$ raw images for mosaicking. Excluding slew overhead, effective exposure times are generally $\sim40-70\%$ of the time on the sky, yielding typical actual exposure times of $\sim 250$ to $\sim 2500$ seconds.
Raw images are obtained with standard double-correlated reads with the long exposure ($7.8$-second) minus short exposure ($51$-millisecond) frames in each filter treated as the ``raw'' frame input to \texttt{p3.6}. These raw $256\times256$ pixel images are of $\sim7.8$-second duration with a plate scale of 2 $\arcsec$/pixel and a 8.53\arcmin$\times$8.53\arcmin{} field of view (FOV).  To aid with reductions, the telescope is dithered after each set of three exposures with a step size $< 2$\arcmin{} based on a randomized dither pattern covering a typical $\sim$12\arcmin$\times$12\arcmin{} FOV. The three raw images observed at each dither position are then added into ``triplestacks" before mosaicking.  The \texttt{p3.6} pipeline processes all raw images by flat correction, dark current and sky subtraction, registration, and stacking to create final \jhk{} mosaics using \swarp{}~\citep{bertin02}.   Bad pixel masks are created dynamically and flat fields --- which are relatively stable --- were created from archival images. Since the short-timescale seeing also remains roughly constant in the several seconds of slew time between dithered images, we did not find it necessary to convolve the raw images to the seeing of a raw reference image before mosaicking. The seeing over long time periods (several months) remains relatively constant at 0.77--0.85\arcsec{} \footnote{\url{https://www.mmto.org/node/249}}. The raw images are resampled from a raw image scale of 2\arcsec/pixel into final mosaics with 1\arcsec/pixel sampling with \swarp{}~\citep{bertin02}.  The typical FWHM in the final \PTL{} mosaics is $\sim 2.5$--$3.0$\arcsec{}, consistent with the average image quality obtained by 2MASS \citep{bloom03,skrutskie06}.  
 
The desired telescope pointing center for all dithered images is set to the SN RA and DEC coordinates from the optical discovery images.  Unfortunately, as a result of various software and/or mechanical issues --- for example problems with the RA drive --- the \PTL{} 1.3-m telescope pointing accuracy can vary by $\sim1-30$ arcminutes from night to night.  Catastrophic pointing errors can result in the SN being absent in all of the raw images and missing in the $\sim 12\arcmin \times 12$\arcmin{} mosaic FOV.  More often, non-fatal pointing errors result in the SN being absent or off-center in some, but not all, raw images.  In \texttt{p2.0} used for \WV{}, the mosaic center was constrained to be the SN coordinates and the mosaic size in pixels was fixed.  This resulted in a significant fraction of failed or low S/N mosaics using an insufficient number of raw images.  For \texttt{p3.0}-\texttt{p3.6}, the constraint fixing the SN at the mosaic center was relaxed and the mosaic center was allowed to be the center of all imaging. This resulted in $\sim15\%$ more mosaic solutions than \texttt{p2.0}.  Mosaics that failed processing at intermediate \texttt{photpipe} stages were excluded from the LC automatically.  Some mosaics that succeeded to the end of \texttt{photpipe} were excluded based on visual inspection or by identifying outlier LC points during post processing.

\afterpage{
\renewcommand{\scale}{0.9}
\begin{figure*}
\centering
\begin{tabular}{@{}c@{}}

\includegraphics[width=\scale\linewidth,angle=0]
{\colordir/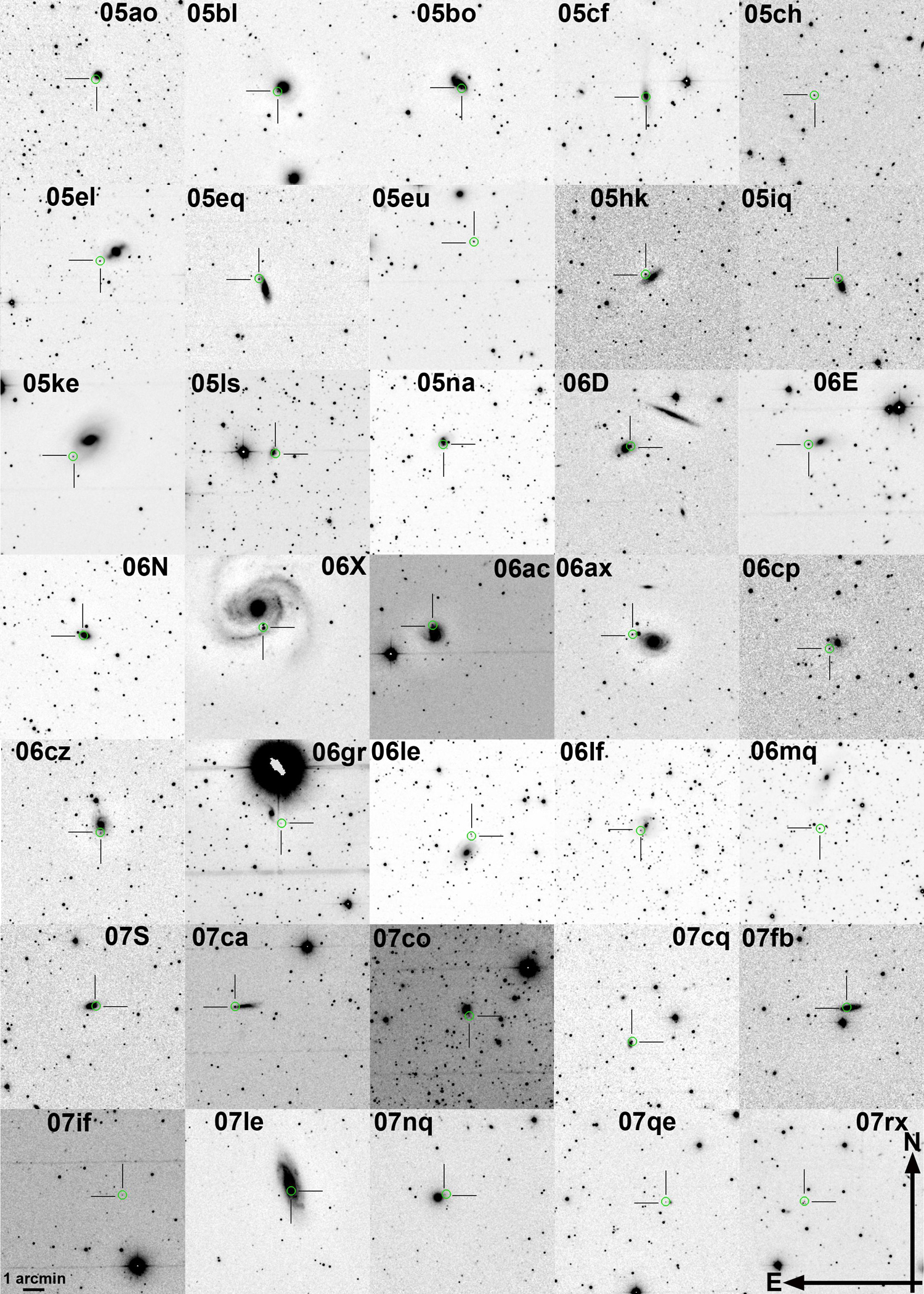}  
\end{tabular}
\caption[$J$-band mosaics for \F{} \snIa{} (Gallery 1)]
{Gallery of 35 \PTL{} $J$-band Mosaics 
\\
\\
\scriptsize
(Color online)
A subset of 35 \PTL{} $J$-band Mosaics from the set of \nsnIacfair{} \F{} \snIa{} and \nsnIaxcfair{} \snIax{} observed with \PTL{} from 2005-2011.  \snIa{} are marked by green circles (color online) and crosshairs. SN names are of the shortened form {\bf 06X} = \sn{}2006X. North and East axes for all mosaics are indicated in the lower right corner of the figure.
}
\label{fig:galleryA}
\end{figure*}

\renewcommand{\scale}{0.9}
\begin{figure*}
\centering
\begin{tabular}{@{}c@{}}

\includegraphics[width=\scale\linewidth,angle=0]
{\colordir/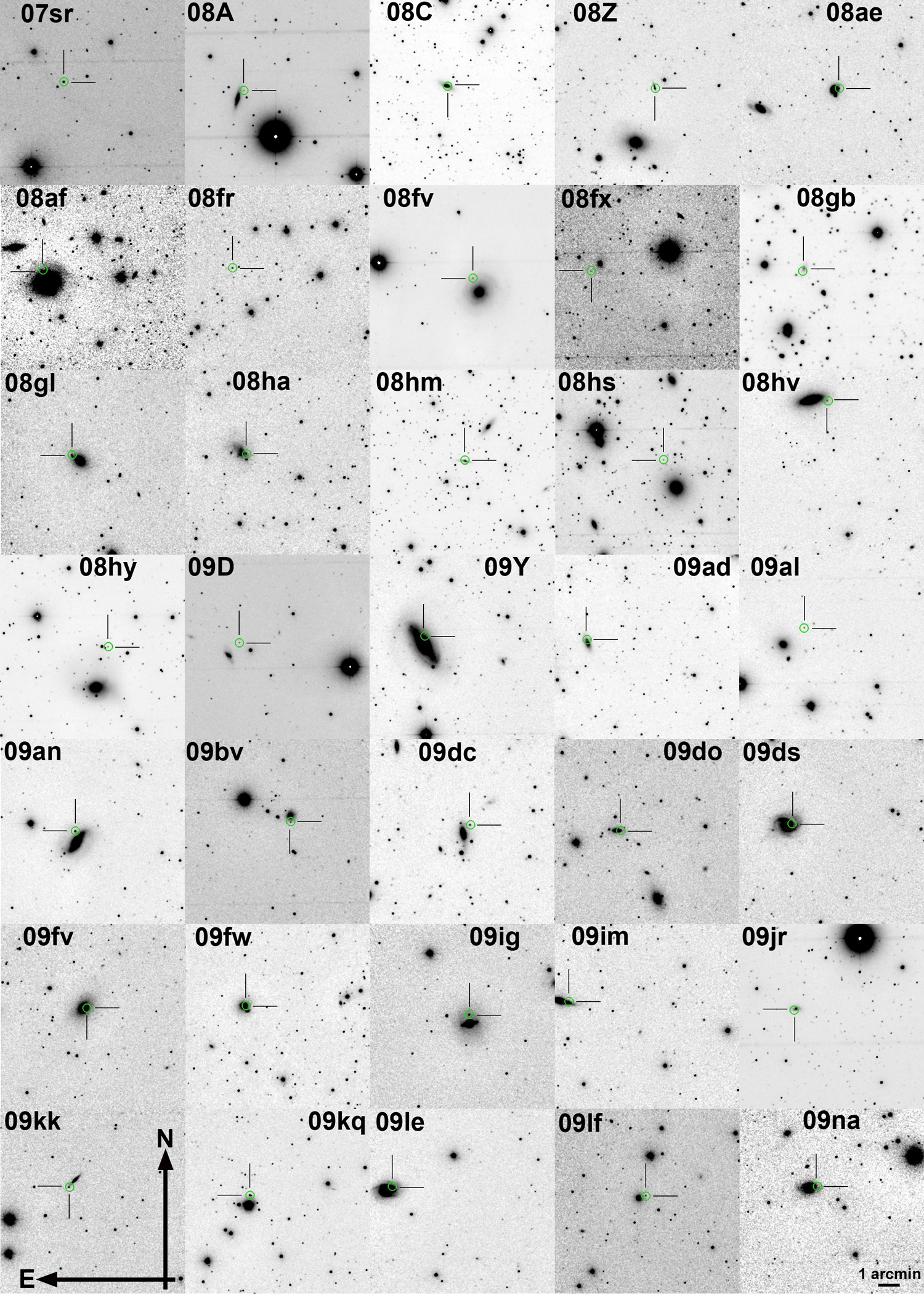}  
\end{tabular}
\caption[$J$-band mosaics for \F{} \snIa{} (Gallery 2)]
{Gallery of 35 \PTL{} $J$-band Mosaics 
\\
\\
\scriptsize
(Color online)
A subset of 35 \PTL{} $J$-band Mosaics from the set of \nsnIacfair{} \F{} \snIa{} and \nsnIaxcfair{} \snIax{} observed with \PTL{} from 2005-2011.  \snIa{} are marked by green circles (color online) and crosshairs. SN names are of the shortened form {\bf 09an} = \sn{}2009an. North and East axes for all mosaics are indicated in the lower left corner of the figure.
}
\label{fig:galleryB}
\end{figure*}

\renewcommand{\scale}{1}
\begin{figure*}
\centering
\begin{tabular}{@{}c@{}}

\includegraphics[width=\scale\linewidth,angle=0]
{\colordir/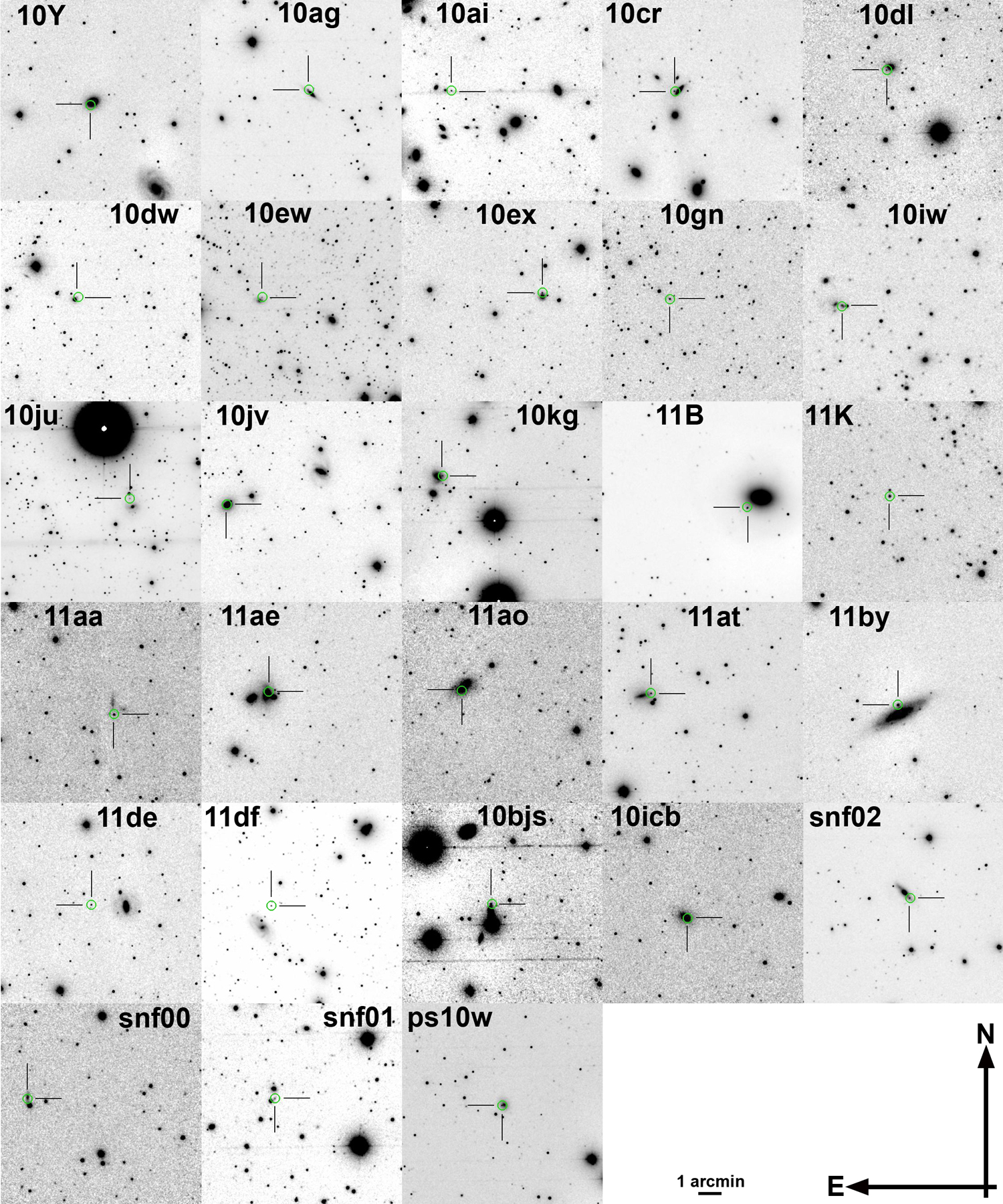}  
\end{tabular}
\caption[$J$-band mosaics for \F{} \snIa{} (Gallery 3)]
{Gallery of 28 \PTL{} $J$-band Mosaics 
\\
\\
\scriptsize
(Color online)
A subset of 28 \PTL{} $J$-band Mosaics from the set of \nsnIacfair{} \F{} \snIa{} and \nsnIaxcfair{} \snIax{} observed with \PTL{} from 2005-2011.  \snIa{} are marked by green circles (color online) and crosshairs. SN names are of the shortened form {\bf 06X} = \sn{}2006X. North and East axes for all mosaics are indicated in the lower right corner of the figure.  Non-IAUC SN Names include: {\bf 10bjs}=PTF10bjs, {\bf 10icb}=PTF10icb, {\bf snf02}=SNF20080514-002, {\bf snf00}=SNF20080522-000, {\bf snf01}=SNF20080522-011, {\bf ps10w}=PS1-10w.
}
\label{fig:galleryC}
\end{figure*}
}

\subsection{Sky Subtraction and Noise Maps}
\label{sec:sky}

\renewcommand{\scale}{1}
\begin{figure}[h]
\centering
\begin{tabular}{@{}c@{}}
\includegraphics[width=\scale\linewidth,angle=0]
{\colordir/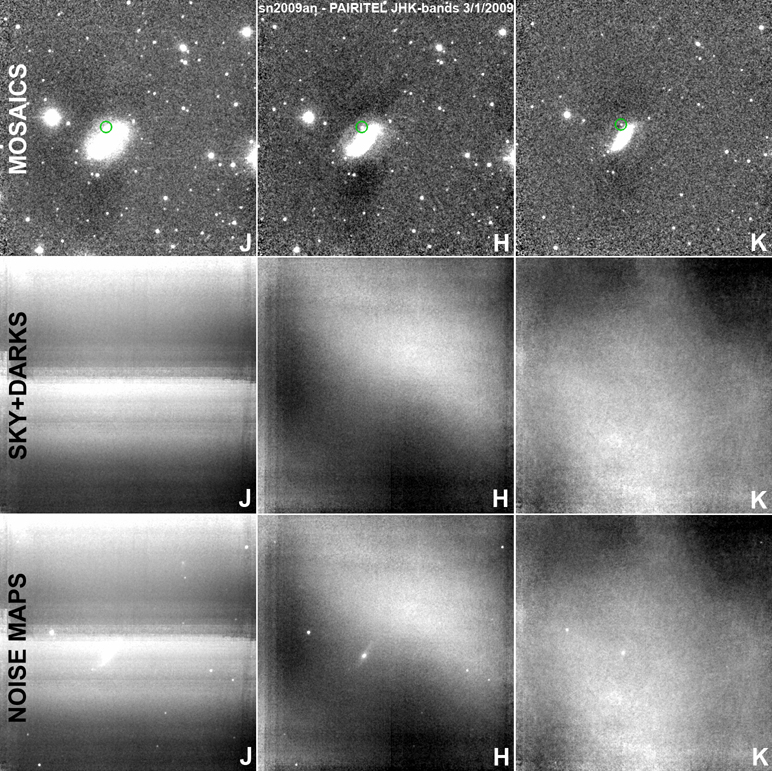}
\end{tabular}
\caption[Skarks and Noise Mosaics]{
\PTL{} Source, Skark, \& Noise Mosaics
\\
\\
\scriptsize
(Color online)
Mosaics ({\bf first row}), Skark Mosaics ({\bf second row}), and Noise Mosaics ({\bf third row}) for the \PTL{} \jhk{} images of \sn{}2009an from 3/1/2009.  The SN is marked with green circles (color online).
Images are displayed 
in SAOimage ds9 with zscale scaling, in grayscale with counts increasing from black to white.  The skark images contain the number of sky + dark current + bias counts (skark counts) subtracted from each mosaic pixel.  Median skark counts for these images were $\sim800$, $6700,$ and $19600$ counts in \jhk{}, respectively, reflecting the sky noise increase towards longer NIR wavelengths which is worst in $K_s$-band. The large scale patterns in the skark mosaics come from arcminute scale spatial variations in the sky brightness of the raw frames, and both thermal dark current and amplifier glow, which peak at the corners of each detector quadrant, and which both contribute Poisson noise.  The skark mosaics also show signatures of the relatively stable electronic bias shading patterns in each quadrant of the raw \jhk{} detectors, which differ by bandpass. All of these contributions get smeared out over the mosaic dither pattern. Noise mosaics use source counts from the mosaic, skark counts from the skark mosaics, and noise from other sources (see \S7.1 of F12 for assumptions used to estimate the noise per pixel). The large scale patterns in the $J$-band skark and noise mosaics are dominated by the cumulative detector noise contributions, including thermal dark current, shading, and amplifier glow. By contrast, the $H$ and $K_s$ skark and noise mosaics are dominated by sky counts and sky noise, respectively, which combine with the various detector imprints and spatiotemporal sky variation across the dither pattern to form the large scale patterns in those bandpasses.
}
\label{fig:skark_noise}
\end{figure}
\renewcommand{\scale}{1}

The \PTL{} camera has no cold shutter, so dark current cannot be measured independently, and background frames include both sky and dark photons (``skark''). Fortunately, the thermal dark current counts across the raw frames, are {\it negligible} in \jhk{} for the NICMOS3 arrays on timescales comparable to the individual, raw, $7.8$-second exposures (\citealt{skrutskie06}). Furthermore, the dark current rate does not detectably vary across the 1.5 hours of the maximum dither pattern used in these observations. Background frames also include an electronic bias, characterized by shading in each of the four raw image quadrants which produces no noise, and amplifier glow, which peaks at the corners of the quadrants, and which, like thermal dark current, does produce Poisson noise.  These intrinsic detector and sky noise contributions get smeared out over the mosaic dither pattern, producing characteristic patterns in the skark mosaics and mosaic noise maps (see \fig{}~\ref{fig:skark_noise}).\footnote{The shading is an electronic bias which technically produces no noise.  Shading was subtracted out as part of the skark counts for each corresponding raw image.  However, the shading was included as a generic background contribution along with thermal dark current, amplifier glow, and sky counts, and thus effectively contributes to the noise mosaics in \fig{}~\ref{fig:skark_noise}.} 

\PTL{} SN observations did not include on-off pointings alternating between the source and a nearby sky field, so skark frames were created for each raw image in the mosaic by applying a pixel-by-pixel average through the stack of a time series of unregistered raw frames, after removing the highest and lowest pixel values in the stack. The stack used a time window of 5 minutes before and after each raw image.  This approximation assumes that the sky is constant on timescales less than $10$ minutes. For reference, typical dithered image sequences have effective exposure times of $10$--$30$ minutes. \fig{}~\ref{fig:skark_noise} shows that for $J$-band, where the sky counts are small compared to the various sources of detector noise, the skark and noise mosaics are dominated by the cumulative effect of the intrinsic detector features over the entire dither pattern, including dark current, shading, and amplifier glow.\footnote{For further information on these features of NICMOS arrays, also used on the Hubble Space Telescope, see {\tiny \url{http://documents.stsci.edu/hst/nicmos/documents/handbooks/v10/c07\_detectors4.html} or \url{http://www.stsci.edu/hst/nicmos/documents/handbooks/DataHandbookv8/nic\_ch4.8.3.html}}.} By contrast, the $H$ and $K_s$-band skark and noise mosaics in \fig{}~\ref{fig:skark_noise} are dominated by sky counts and sky noise, respectively, which combine with the various detector imprints and spatiotemporal sky variation to produce the large scale patterns smeared across the dither pattern.

Although the telescope is dithered ($< 2$\arcmin) after three exposures at the same dither position, for host galaxies with large angular size $\gtrsim 2-5\arcmin$ (in the 8.53\arcmin\ raw image FOV), host galaxy flux contamination introduces additional systematic uncertainty by biasing skark count estimates toward larger values, leading to {\it over-subtraction} of sky light in those pixels (F12). Still, the relatively large \PTL{} 8.53\arcmin{} FOV combined with a dither step size comparable or greater than the $\sim1$-$2\arcmin{}$ angular size of typical galaxies at $z \sim 0.02$ allows us to safely estimate the sky from the raw frames in most cases.  This observing strategy also gives us more time on target compared to on-off pointing. While our approach can lead to systematic sky over-subtraction for SN and stars near larger galaxies, by testing the radial dependence of \PTL{} photometry of 2MASS stars within $3\arcmin{}$ of the SN (and close to the host galaxy), we estimate this systematic error to be negligible compared to our photometric errors, biasing SN photometry fainter by $\lesssim 0.01$ mag in $JH$ and $\lesssim 0.02$ in $K_s$ (F12). By comparison, mean photometric errors for each of the highest S/N LC points from the set of SN in \F{} are $\sim0.03$, $0.05$, and $0.12$ mag in \jhk{}, respectively, (with larger mean statistical errors for all LC points of $\sim0.09$, $0.12$, and $0.18$ mag in \jhk{}, respectively). We thus choose to ignore systematic errors from sky over-subtraction in this work.

Since three raw frames are taken at each dither position and co-added into triplestacks before mosaicking, \texttt{p3.6} now also constructs ``tripleskarks",  by co-adding the three associated skark frames taken at each dither position.  To remove the estimated background counts, \texttt{p3.6} now subtracts the associated tripleskark from each triplestack before creating final mosaics and new skark and noise mosaics (see \fig{}~\ref{fig:skark_noise}). Since the estimated skark noise can vary by $\sim10$--$100$\% across individual skark mosaics, modeling the noise in each pixel provides more reliable {\it differential} noise estimates at the positions of all 2MASS stars and the SN, although our {\it absolute} noise estimate is still underestimated since the noise mosaics do not model all sources of uncertainty (see \S7.1 of F12). To account for this, we also use 2MASS star photometry to empirically calculate inevitable noise underestimates, and correct for them in SN photometry on subtracted or unsubtracted images (see F12; \S\ref{sec:phot}). 
 
\subsection{The \PTL{} NIR Camera is Undersampled}
\label{sec:undersample}

The \PTL{} Infrared camera is undersampled because the $2\arcsec$ detector pixels are larger than the atmospheric seeing disk at FLWO.  This means we can not fully sample the point spread function (PSF) of the detected image.  To achieve some sub-pixel sampling, \PTL{} implements a randomized dither pattern. 
While dithering can help recover some of the image information lost from undersampling, large pixels with dithered imaging cannot fully replace a fully sampled imaging system \citep{lauer99,fruchter02,rowe11}, and in practice, dithering does not always reliably produce the desired sub-pixel sampling.  When we subtract host galaxy light, which requires PSF matching SN and \snt{} mosaics, undersampling leads to uncertainty in photometry for individual subtractions that can underestimate or overestimate the flux at the SN position. We correct for this by averaging many subtractions, and removing bad subtractions, when producing \F{} LCs (see \S\ref{sec:host_sub}-\ref{sec:nn2nnt}).

\subsection{Host Galaxy Subtraction}
\label{sec:host_sub}
 
We obtain \snt{} images after the SN has faded below detection for the \PTL{} Infrared camera, typically $\gtrsim 6$--$12$ months after the last SN observation.  We use \snt{} images to subtract the underlying host galaxy light at the SN position for each SN image that meets our image quality standards (see~\S\ref{sec:dophot_iso}-\ref{sec:galsub_phot}).  To limit the effects of variable observational conditions, sensitivity to individual template observations of poor quality, and to minimize the photometric uncertainty 
from individual subtractions, we try to obtain at least $\nt=2$, and as many as $\nt=\jmaxsntemp$ \snt{} images that satisfy our image quality requirements (see~\S\ref{sec:nn2nnt}). In practice, we obtained medians of $\nt=\jmedsntemp$, $\hmedsntemp$, and $\kmedsntemp$ usable \snt{} images in \jhk{}, respectively (\fig{}~\ref{fig:n_sntemp_hist}). In cases with only $\nt=1$ \snt{} image, galaxy-subtracted LCs are deemed acceptable only for bright, well isolated SN that are consistent with the unsubtracted LCs (see~\S\ref{sec:dophot_iso}, \S\ref{sec:galsub_syserr}).

\subsection{Forced \dophot{} on Unsubtracted Images}
\label{sec:dophot_iso}

Forced \dophot{} photometry~\citep{schechter93} at a fixed position was performed on the unsubtracted SN images as an initial step for all \PTL{} SN.  Forced \dophot{} LCs on unsubtracted images provide an excellent approximation to the final galaxy-subtracted LCs for SN that were clearly separated from their host galaxy (F12). Approximately $30\%$ of SN of all types observed by \PTL{} are well isolated from the host galaxy and bright enough so that the measured galaxy flux at the SN position is $\lesssim 10\%$ of the SN flux at peak brightness. 
We use \nwelliso{} of these bright, well isolated SN to perform internal consistency checks to test for 
errors incurred from host galaxy subtraction (see \S\ref{sec:phot_sys}; F12).

\subsection{Forced \dophot{} on Difference Images}
\label{sec:galsub_phot}

We perform galaxy subtraction on all \F{} objects to reduce the data with a homogeneous method.\footnote{Only \sn{}2008A (and the \sn{}2005cf LC retained from \WV) use forced \dophot{} and no host subtraction. \nnt{} failed for \sn{}2008A as a result of poor quality \snt{} images (see \S\ref{sec:nn2nnt}).} 
We used subtraction-based photometry following \citet{miknaitis07}. 
The SN flux in the difference images is measured with forced \dophot{} photometry at fixed pixel coordinates, determined by averaging SN centroids from $J$-band or CfA optical $V$-band difference images with photometric detections of the object that had a S/N $>5$.  SN centroids are typically accurate to within $\lesssim0.2$\arcsec{}. Tests show no systematic LC bias for forced \dophot{} photometry as a result of SN astrometry errors if the SN centroid is accurate to within $\lesssim 0.5$\arcsec{} (F12). The RA and DEC values in Tables~\ref{tab:general}-\ref{tab:general3} show best fit SN centroid coordinates.  These are typically more accurate than optical discovery coordinates from IAU/CBET notices, which may only be accurate to within $\lesssim1-2$\arcsec{}. Forced \dophot{} photometry at this fixed position in the difference images employs the \dophot{} PSF calculated from standard stars in the un-convolved image.  For the difference images the calibrated zero point from the template is used, with suitable correction for the convolution of the \snt{} image as detailed by \citet{miknaitis07}.

\subsection{Averaging Subtractions: NNT Method}
\label{sec:nn2nnt}

\renewcommand{\scale}{1}
\begin{figure}
\centering
\begin{tabular}{@{}c@{}}
\includegraphics[width=\scale\linewidth,angle=0]
{\colordir/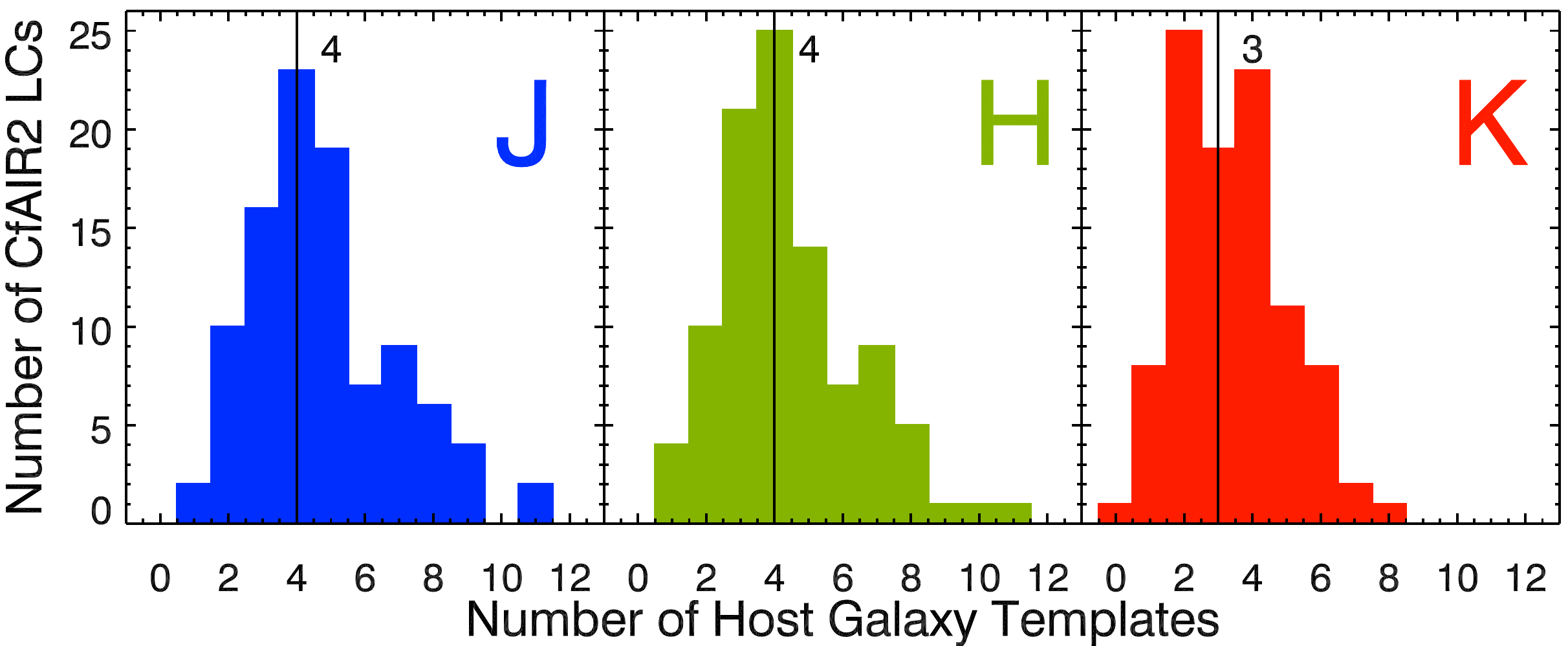}
\end{tabular}
\caption[Histograms of \jhk{} \snt{}]{
Histograms of \jhk{} \snt{} Subtractions
\\
\\
\scriptsize
(Color online) Histogram of the number of host galaxy template images $\nt$ in each bandpass used for each SN. $\nt$ is the maximum number of \snt{} subtractions used over all nights per LC and bandpass.  Some subtractions fail during \texttt{photpipe} or are rejected as bad subtractions on individual nights during post-processing.  We generally obtain $> \nt$ host galaxy images, but some images fail the mosaicking pipeline (especially in $K_s$-band) prior to \texttt{photpipe}. We tried to obtain at least $\nt=2$, and as many as $\nt=\jmaxsntemp$ usable \snt{} images, with medians of $\nt=\jmedsntemp$, $\hmedsntemp$, and $\kmedsntemp$ \snt{} images in \jhk{}, respectively. For some SN, only $\nt=1$ template images were usable and \sn{}2008A had no usable \snt{} images.
}
\label{fig:n_sntemp_hist}
\end{figure}
\renewcommand{\scale}{1}

We use \nnt{}, an alternative galaxy subtraction method for \F{}, which uses fewer individual subtractions than the NN2 method  \citep{barris05} used in \WV{}. With \nnt{}, for each of the $N_{\rm SN}$ mosaicked SN images, we subtract each of the usable $\nt$ \snt{} images, yielding at most $N_{\rm NNT} = N_{\rm SN} \times \nt$ individual subtractions. 
\nnt{} yields $\nt$ realizations of the LC which can be combined into a final galaxy-subtracted LC with a night-by-night weighted flux average after robust \threesig{} rejection and manual checks to exclude individual bad subtractions.\footnote{Weighted mean flux values on each night are weighted by the corrected \dophot{} uncertainties. A S/N $> 1$ cut is employed for individual subtractions before \nnt{}.  A S/N $> 3$ cut is employed for final LC points.  $\nt$ can differ nightly and by bandpass and is often smallest in $K_s$. See~\S\ref{sec:phot_precision},~\S\ref{sec:galsub_syserr}, Table~\ref{tab:nnt_err}, and Appendix \ref{sec:nnt_math}.}  SN or SNTEMP images that failed our image quality requirements were also excluded from \nnt{} via automatic \texttt{photpipe} tests and manual checks, yielding fewer bad subtractions than the purely automated process used in \WV{}.  

By obtaining $1 \lesssim \nt \le \jmaxsntemp$ usable SNTEMP images, including additional observations since \WV{}, most \F{} \snIa{} have $\nt \gtrsim 4$ SNTEMP images suitable for galaxy subtraction (see \fig{}~\ref{fig:n_sntemp_hist}). \nnt{} allowed us to exclude individual bad subtractions, average over variance across subtractions from different templates, and produce \F{} \snIa{} LCs with more accurate flux measurements
compared to NN2 for \WV{}.  We discuss the 
statistical and systematic 
uncertainty incurred from \nnt{} host galaxy subtraction in~\S\ref{sec:phot_sys}.  \F{} \nnt{} LCs also show better agreement with CSP photometry for the same objects compared to \WV{} (see \S\ref{sec:csp_comp}).\footnote{Some fainter \snIa{} LCs which used NN2 in \WV{} showed significant systematic deviations from the published CSP photometry for the same objects. These discrepancies exceeded deviations expected from small bandpass differences without S-corrections (\citealt{contreras10}; M. Phillips --- private communication).}

\subsection{Photpipe Improvements}
\label{sec:photpipe}

Since \WV{}, we have implemented several improvements to \texttt{photpipe}. 
\texttt{Photpipe} now takes \texttt{p3.6} mosaics as input (see \S\ref{sec:mosaics}). To use SN that are not in the \texttt{p3.6} mosaic center,  \texttt{photpipe} uses larger radius photometric catalogs and improved image masks (see F12). In \WV{}, our ``skark" noise estimate was assumed to be constant throughout the mosaic (see \S\ref{sec:sky}). Figure~\ref{fig:skark_noise} shows this is a bad approximation. Instead, \texttt{p3.6} noise mosaics are used by \texttt{photpipe} and fed as inputs to \dophot{} \citep{schechter93}, our point source photometry module, and \texttt{HOTPANTS} \citep{becker04,becker07}, our difference imaging module (see \S\ref{sec:host_sub}), leading to improved image subtraction. See F12 for details on the computational implementation of \texttt{photpipe} and \texttt{p3.6}.

As a result of improvements discussed throughout \S\ref{sec:reduction}, \F{} supersedes \WV{} photometry for 20 out of \nsnIaredo{} LCs (excluding \sn{}2005cf). \F{} and \WV{} photometry agree best for the brightest, well isolated, SN with little galaxy light at the SN position. Fainter SN that required significant host galaxy subtraction show the most disagreement between \F{} and \WV{} due mainly to the differences between NN2 and \nnt{} (see \S4.3.1 of F12). Problems with \WV{} NN2 photometry are most evident in the set of \nwvcsp{} \WV{} SN also observed by the CSP, which are discussed in \S\ref{sec:csp_comp}.  
The improved agreement between \F{} and CSP (see \S\ref{sec:disc}) gives evidence that \F{} photometry is superior to \WV{}.

Although individual LCs show differences between \F{} and \WV{} data, we do not expect the revised photometry to significantly affect the overall conclusions of \WV{}. Preliminary analysis, which will be presented elsewhere, will derive mean NIR LC templates and mean absolute magnitudes using only normal \F{} \snIa{} and compare these to mean templates derived using only 18 normal \PTL{} \snIa{} from \WV{}.

\section{Photometric Calibration And Verification}
\label{sec:phot}

We now discuss the methods used to calibrate \PTL{} photometry and test the calibration, including internal consistency checks and comparison with external data sets with NIR photometry for the same objects.  In \S\ref{sec:standard}, we present \PTL{} photometry for 2MASS stars which we use to test for systematic problems with \PTL{} \dophot{} photometry. In \S\ref{sec:phot_sys}, we investigate potential systematic photometry errors from host galaxy subtraction.  In \S\ref{sec:csp_comp}, we compute approximate color terms describing offsets between \PTL{} and CSP $J$ and $H$ bandpasses using 2MASS field stars observed by both groups. In \S\ref{sec:csp_comp1}, we compare \F{} data to an overlapping subset of CSP \snIa{} photometry, demonstrating overall agreement between the data sets. Throughout, we refer to F12 for additional details.

\subsection{Photometric Calibration}
\label{sec:standard}

We organize \S\ref{sec:standard} as follows. In \S\ref{sec:ptel_2mass_star_lcs}, we present \PTL{} mean photometric measurements and uncertainties for all 2MASS stars for \nIa{} out of 121 \snIa{} and \snIax{} fields observed from 2005-2011.  In \S\ref{sec:phot_precision}, we test whether \dophot{} is correctly estimating photometric uncertainties for \PTL{} point sources. In \S\ref{sec:phot_accuracy}, we assess whether \PTL{} \dophot{} photometry globally agrees with 2MASS star photometry. Overall, \S\ref{sec:phot_precision}-\ref{sec:phot_accuracy} test the precision and accuracy of \dophot{} photometry on unsubtracted \PTL{} images. We find no significant systematic differences with 2MASS. 

\subsubsection{\PTL{} Photometry of 2MASS Standard Stars}
\label{sec:ptel_2mass_star_lcs}

\renewcommand{\arraystretch}{0.5}
\renewcommand{\tabcolsep}{2pt}
\begin{table*}
\begin{center}
\caption[\PTL{} Photometric Catalog of 2MASS Standard Stars]{\PTL{} \jhk{} Photometry of 2MASS Standard Stars in \snIa{} Fields} 
\scriptsize
\begin{tabular}{@{}ccccccccccccccccccc@{}}
\hline
 SN & Star & $\alpha$(2000) & $\delta$(2000) & $N_{J}$ & $m^{\rm \ptel}_{J}$ & $\sigma^{\rm \ptel}_{m_{J}}$ & $m^{\rm \twomass}_{J}$ & $\sigma_{m^{\rm \twomass}_{J}}$ & $N_{H}$ & $m^{\rm \ptel}_{H}$ & $\sigma^{\rm \ptel}_{m_{H}}$ & $m^{\rm \twomass}_{H}$ & $\sigma_{m^{\rm \twomass}_{H}}$ & $N_{K}$ & $m^{\rm \ptel}_{K}$ & $\sigma^{\rm \ptel}_{m_{K}}$ & $m^{\rm \twomass}_{K}$ & $\sigma_{m^{\rm \twomass}_{K}}$ \\
  \tablenotemark{a} &  \tablenotemark{b} & \tablenotemark{c} & \tablenotemark{c} &  
  \tablenotemark{d} & [mag]\tablenotemark{e} & [mag]\tablenotemark{f} & [mag]\tablenotemark{g} & [mag]\tablenotemark{g} & 
  \tablenotemark{d} & [mag]\tablenotemark{e} & [mag]\tablenotemark{f} & [mag]\tablenotemark{g} & [mag]\tablenotemark{g} & 
  \tablenotemark{d} & [mag]\tablenotemark{e} & [mag]\tablenotemark{f} & [mag]\tablenotemark{g} & [mag]\tablenotemark{g} \\
\hline
\sn{}2005ak & 01 & 14:40:18.45 & +03:30:55.44 & 34 & 16.549 & 0.007 & 16.504 & 0.159 & 35 & 15.940 & 0.009 & 16.024 & 0.183 & 30 & 15.675 & 0.012 & 15.251 & 0.173 \\ 
\sn{}2005ak & 02 & 14:40:18.56 & +03:34:12.76 & 34 & 15.918 & 0.006 & 15.858 & 0.097 & 33 & 15.230 & 0.008 & 15.230 & 0.105 & 33 & 15.024 & 0.010 & 15.075 & 0.148 \\ 
\sn{}2005ak & 03 & 14:40:19.41 & +03:30:22.95 & 34 & 15.112 & 0.006 & 15.118 & 0.056 & 35 & 14.768 & 0.007 & 14.822 & 0.085 & 33 & 14.686 & 0.008 & 14.814 & 0.112 \\ 
\sn{}2005ak & 04 & 14:40:20.77 & +03:27:36.99 & 34 & 16.404 & 0.006 & 16.430 & 0.150 & 35 & 15.793 & 0.009 & 16.057 & 0.219 & 34 & 15.549 & 0.012 & 15.326 & 0.197 \\ 
\sn{}2005ak & 05 & 14:40:20.94 & +03:33:41.82 & 33 & 15.013 & 0.006 & 15.071 & 0.049 & 34 & 14.408 & 0.007 & 14.511 & 0.071 & 34 & 14.301 & 0.007 & 14.285 & 0.074 \\ 
\sn{}2005ak & 06 & 14:40:22.26 & +03:31:18.61 & 33 & 17.032 & 0.007 & 16.521 & 0.147 & 33 & 16.386 & 0.010 & 16.101 & 0.215 & 29 & 16.153 & 0.014 & 15.598 & 0.255 \\ 
\sn{}2005ak & 07 & 14:40:22.58 & +03:32:56.39 & 35 & 15.637 & 0.006 & 15.665 & 0.066 & 35 & 15.001 & 0.007 & 15.133 & 0.089 & 34 & 14.765 & 0.008 & 14.946 & 0.148 \\ 
\sn{}2005ak & 08 & 14:40:26.00 & +03:31:41.52 & 34 & 13.255 & 0.005 & 13.233 & 0.024 & 35 & 12.617 & 0.006 & 12.608 & 0.030 & 35 & 12.406 & 0.007 & 12.404 & 0.032 \\ 
\sn{}2005ak & 09 & 14:40:26.55 & +03:30:58.65 & 34 & 14.780 & 0.006 & 14.762 & 0.037 & 35 & 14.212 & 0.007 & 14.121 & 0.035 & 35 & 13.967 & 0.007 & 14.003 & 0.071 \\ 
\sn{}2005ak & 10 & 14:40:29.45 & +03:32:34.68 & 35 & 16.402 & 0.006 & 16.596 & 0.163 & 35 & 15.757 & 0.008 & 15.736 & 0.152 & 32 & 15.571 & 0.011 & 15.228 & 0.173 \\ 
\sn{}2005ak & 11 & 14:40:29.89 & +03:28:05.44 & 33 & 14.455 & 0.006 & 14.444 & 0.038 & 34 & 14.160 & 0.006 & 14.114 & 0.035 & 33 & 14.055 & 0.007 & 14.095 & 0.072 \\ 
\sn{}2005ak & 12 & 14:40:30.02 & +03:30:15.93 & 34 & 15.424 & 0.005 & 15.319 & 0.072 & 33 & 14.958 & 0.007 & 15.021 & 0.090 & 35 & 14.793 & 0.008 & 14.624 & 0.123 \\ 
\sn{}2005ak & 13 & 14:40:31.33 & +03:28:33.93 & 24 & 15.472 & 0.010 & 15.589 & 0.082 & 28 & 14.814 & 0.011 & 15.169 & 0.100 & 31 & 14.488 & 0.010 & 14.898 & 0.150 \\ 
\sn{}2005ak & 14 & 14:40:31.52 & +03:32:31.31 & 36 & 14.373 & 0.005 & 14.367 & 0.036 & 36 & 14.171 & 0.007 & 14.212 & 0.042 & 36 & 14.145 & 0.007 & 14.277 & 0.086 \\ 
\sn{}2005ak & 15 & 14:40:31.74 & +03:29:10.30 & 35 & 15.420 & 0.006 & 15.304 & 0.056 & 34 & 14.804 & 0.007 & 14.823 & 0.070 & 35 & 14.574 & 0.008 & 14.704 & 0.116 \\ 
\sn{}2005ak & 16 & 14:40:32.31 & +03:31:13.54 & 34 & 16.087 & 0.006 & 15.902 & 0.090 & 36 & 15.501 & 0.008 & 15.476 & 0.132 &  \nodata  &  \nodata  &  \nodata  &  \nodata  &  \nodata  \\ 
\sn{}2005ak & 17 & 14:40:32.43 & +03:33:34.39 & 28 & 14.766 & 0.010 & 14.756 & 0.056 & 26 & 14.069 & 0.012 & 14.143 & 0.085 & 29 & 13.836 & 0.011 & 13.802 & 0.070 \\
\hline
\end{tabular}
\tablecomments{ \tiny
\\
\scriptsize
({\bf A full machine-readable Table is available online in the electronic version of this paper. A portion is shown here for guidance}).
\\
\scriptsize
{\bf (a)} Tables like the above sample are provided online for \nIa{} out of 121 \snIa{} and \snIax{} fields observed with \PTL{} from 2005-2011 (\sn{}2005ak-\sn{}2011df), including \nIanogo{} \snIa{} without \F{} photometry (e.g., \sn{}2005ak above). Tables include weighted mean \PTL{} photometry and uncertainties for all 2MASS stars in each \snIa{} field.
3 \snIa{} are not included in Table~\ref{tab:ptel_catalog} as a result of unresolved software errors: \sn{}2008fv, \sn{}2008hs (in \F{}), and \sn{}2011ay (not in \F{}). 
\\
{\bf (b)} Superscripts ${\rm \ptel}$ and ${\rm \twomass}$ denote \PTL{} and 2MASS, respectively. Missing data is denoted by \ldots.  
\\
{\bf (c)} RA ($\alpha$) and DEC ($\delta$) for Epoch 2000 in sexagesimal coordinates. 
\\
{\bf (d)} $N_{X}$ is the number of \PTL{} SN images in band $X={J,H,K}$ with this standard star used to measure $m^{\rm \ptel}_{X}$ and $\sigma^{\rm \ptel}_{m_{X}}$.
\\
{\bf (e)} \PTL{} apparent brightness in magnitudes $m^{\rm \ptel}_{X}$ is computed as the weighted mean \PTL{} magnitude over all $N_{X}$ SN images with that 2MASS star.
\\
{\bf (f)} \PTL{} magnitude uncertainty $\sigma^{\rm \ptel}_{m_{X}}$ is computed as the error on the weighted mean of the $N_{X}$ measurements, each of which have already been corrected for \dophot{} uncertainty estimates as described in \S\ref{sec:phot_precision} and F12.  
(see \S7.3 of F12).
\\
{\bf (f)} The 2MASS magnitudes $m^{\rm \twomass}_{X}$ and uncertainties $\sigma^{\rm \twomass}_{m_{X}}$ for each star are from the 2MASS point source catalog \citep{cutri03}.
}
\label{tab:ptel_catalog}
\end{center}
\end{table*}
\renewcommand{\tabcolsep}{6pt}
\renewcommand{\arraystretch}{1}

For 121 \PTL{} SN fields observed from 2005-2011, including 23 objects not in \F{}, we performed \dophot{} photometry on all 2MASS stars to measure the photometric zero point for each image. In a typical 12\arcmin$\times$12\arcmin{} \texttt{p3.6} mosaic FOV, there were between 6 and 92 2MASS stars in each filter (see \figs{}~\ref{fig:galleryA}-\ref{fig:galleryC}).
While the exact coverage for a mosaic during a given night varies (see \S\ref{sec:mosaics}), the majority of the 2MASS stars are covered by each observation of a given SN field.
Fewer 2MASS stars are detected by \dophot{} as wavelength increases from $J$ to $H$ to $K_s$.  For all \snIa{} or \snIax{} fields with at least \nminstar{} mosaic images, the mean number of 2MASS stars was $39$, $38$, and $34$ in \jhk{}, respectively (see Table 4.1 of F12). 

We interpret the error on the weighted mean of the \PTL{} photometric measurements to be the uncertainty in the measurement of the mean \PTL{} magnitude for that 2MASS star (see~\S\ref{sec:phot_precision} and \S7.3 of F12 for mathematical details). Table~\ref{tab:ptel_catalog} presents weighted mean \PTL{} photometric measurements and uncertainties for all 2MASS stars in \nIa{} SN fields observed by \PTL{}.
A global comparison of \PTL{} and 2MASS star measurements is presented in \S\ref{sec:phot_precision}--\ref{sec:phot_accuracy}.

\subsubsection{Photometric Precision}
\label{sec:phot_precision}

We assess the repeatability of \dophot{} measurements of 2MASS stars to quantify the photometric precision of \PTL{}. This tests whether we have correctly estimated our {\it uncertainties} for point sources measured on individual nights.  Although a small fraction of 2MASS stars are variable \citep{plavchan08,quillen14}, by averaging over $\gtrsim 4000$ 2MASS stars for each filter (see Table~\ref{tab:ptel_catalog}) and removing outlier points, we do not expect this to significantly affect our results. Assuming 2MASS stars have constant brightness, the measured scatter indicates if the \PTL{} \dophot{} uncertainties are under or overestimated. Because we do not model all known sources of uncertainty in computing our noise mosaics (see~\S\ref{sec:sky} and \S7.1 of F12), we expect to underestimate our photometric errors.  Empirical tests using \dophot{} photometry of 2MASS stars in the unsubtracted images confirm we are underestimating our photometric magnitude uncertainties by factors of $\sim1.5$--$3$, depending on the brightness of the point source and the filter (F12).  We then multiply the uncorrected \dophot{} magnitude uncertainties ($\sgphot$) for individual points in the \snIa{} LCs by this empirically measured, magnitude-dependent correction factor $C$. Corrected \dophot{} magnitude uncertainties are given by $\sgphotcorr = C \times \sgphot$ 
(see \S4 of F12).

\subsubsection{Photometric Accuracy}
\label{sec:phot_accuracy}

We test whether \PTL{} and 2MASS star photometry are consistent within the estimated uncertainties {\it after} correcting the \PTL{} \dophot{} uncertainties as discussed in \S\ref{sec:phot_precision}. This tests the photometric accuracy of \PTL{} to identify any statistically significant systematic offsets from 2MASS. 
We expect mean \PTL{} and 2MASS photometry to agree when averaged over many stars {\it by construction}, so this is a self-consistency check to rule out any glaring systematic problems with \PTL{} \dophot{} photometry.  For these tests, we measure the difference between the weighted mean \PTL{} magnitudes for each star and the 2MASS catalog magnitudes in Table~\ref{tab:ptel_catalog}. Because \PTL{} photometry goes deeper than 2MASS for each image
and the weighted mean \PTL{} magnitude of each 2MASS star is determined from measurements over many nights, we do not expect the 2MASS catalog magnitude and the weighted mean \PTL{} magnitude to be strictly equal for all standard stars.  We expect greatest agreement for the brightest 2MASS stars with decreasing agreement and increased scatter as the 2MASS catalog brightness decreases, consistent with measurements drawn from a distribution with Gaussian uncertainties. See \S4 of F12.

Aggregated \PTL{}-2MASS residuals for all 2MASS stars in 121 \PTL{} SN fields yield weighted mean residuals of $\meanrestpj \pm \meanrestpjerr$, $\meanrestph \pm \meanrestpherr$, and $\meanrestpk \pm \meanrestpkerr$ in \jhk{}, respectively (uncertainties are standard errors of the mean).  Thus, when averaging over thousands of stars observed over a 6-year span from 2005-2011, \PTL{} and 2MASS agree to within a few thousandths of a magnitude in $\jhk{}$, with evidence for a small, but statistically significant \PTL{}-2MASS offsets of $\sim 0.001$, $0.001$, and $-0.006$ mag in \jhk{}, respectively, at the $\sim2$--$3 \sigma$ level. If we correct for the slight underestimate of our uncertainties in the \PTL{}-2MASS residuals, we find that $\sim$68\%, $\sim$95\%, and $\sim$99\% of the standard stars have \PTL{}-2MASS residuals consistent within 0 to 1, 2, and $3$-$\sigma$ respectively, as expected with correctly estimated Gaussian errors (see \S7.4 of F12).

\subsection{Photometry Systematics}
\label{sec:phot_sys}

In \S\ref{sec:phot_sys}, we discuss internal consistency tests to assess other potential statistical and systematic errors with the photometry. In \S\ref{sec:galsub_err}--\ref{sec:embedded_galsub}, we evaluate our most important systematic and statistical uncertainty from the \nnt{} host galaxy subtraction process, both for bright, well isolated objects and for objects superposed on the nucleus or spiral arms of host galaxies. See \S4 of F12 for discussions of systematic errors from sky subtraction and astrometric errors in the best fit SN centroid position.


\subsubsection{Galaxy Subtraction: Statistical \& Systematic Errors}
\label{sec:galsub_err}

When subtracting SN and \snt{} images observed under different seeing conditions, undersampling of the \PTL{} NIR camera introduces uncertainties into both the estimates of the PSF and convolution kernel solution when attempting to transform the SN or \snt{} image to the PSF of the other. This leads to flux being added or subtracted from photometry on individual subtractions.  While \nnt{} attempts to correct for this by averaging over many subtractions, there is always remaining uncertainty as a result of undersampling (see \S\ref{sec:reduction}).  

For an individual night of photometry, we conservatively estimate the statistical uncertainty from \nnt{}, $\sgnnt$, as the error weighted standard deviation of the input flux measurements, weighted by the corrected \dophot{} flux uncertainties for each of the $\nt$ subtractions (for details see \S\ref{sec:reduction} and Appendix \ref{sec:nnt_math}). For cases where only $\nt=1$ or $2$ subtractions survive both the pipeline's cuts and any manual rejection, \nnt{} flux estimates can be biased high or low and either the weighted standard deviation can not be computed or it is not a reliable estimate of the statistical uncertainty. To ensure accurate photometric uncertainties for these cases --- at the expense of reduced photometric precision --- we adopt a conservative systematic error floor of $0.25$ mag or $0.175$ mag for $\nt=1$ and $\nt=2$, respectively.  Final galaxy subtracted uncertainties $\sgnntcorr$ are computed as in Table~\ref{tab:nnt_err}, which includes a final signal-to-noise cut of S/N $>3$. 
Thus, when a given LC point has an uncertainty larger than its neighbors, either only 1 or 2 good subtractions were used or the scatter amongst the surviving 3+ subtractions was large.

\renewcommand{\arraystretch}{0.5}
\begin{table}
\begin{center}
\caption[Computing NNT Errors]{Computing NNT Errors  \\}
\scriptsize
\begin{tabular}{cccc}
\hline
$\nt$      & $\sgnntcorr$ mag error                              & S/N   & Note \\
\hline
1             &         max$( 0.25 {\rm \ mag}, \ \sgnnt )$     & $3 < {\rm S/N} < \sim4.2$   &  \tablenotemark{a}     \\
2             &         max$( 0.175 {\rm \ mag}, \ \sgnnt )$    & $3 < {\rm S/N} < \sim5.5$   &       \\
3+           &          $\sgnnt$                                                 & $3 < {\rm S/N}$                    & \tablenotemark{b}      \\
\hline
\end{tabular}
\tablecomments{ \scriptsize 
\\
(a) If $\nt=1$, $\sgnnt = \sgphotcorr$, the corrected \dophot{} error for a single subtraction.
\\
(b) A S/N $>1$ cut is used before \nnt{} averaging.  A S/N $> 3$ cut is placed on the final \nnt{} LC points.
}
\label{tab:nnt_err}
\end{center}
\end{table}
\renewcommand{\arraystretch}{1}

In \S\ref{sec:galsub_syserr}-\ref{sec:embedded_galsub}, both for bright, well isolated objects and SN superposed on the host galaxy, \nnt{} produces no net systematic bias given $\nt \gtrsim 3$--$4$ usable host galaxy templates.  For fainter objects, SN superposed on the host galaxy nucleus, or SN with insufficient high quality \snt{} images, the additional uncertainty from host galaxy subtraction can yield many LC points that are excluded based on S/N cuts, outlier rejection, or final quality checks, sometimes yielding LCs of insufficient quality for publication or cosmological analysis. 

\subsubsection{Galaxy Subtraction for Bright, well isolated Objects}
\label{sec:galsub_syserr}

To test if \nnt{} biases the photometry, we first use SN that are well isolated from their host galaxy nuclei. In these cases,  photometry on the unsubtracted images gives a good approximation to the final galaxy subtracted LC at most phases, providing an internal consistency check of \nnt{}.  We use bright SN for which the host galaxy flux at the SN position is a small fraction of the SN flux in the $[-10,50]$ day phase range,  including \nwelliso{} bright and/or well isolated SN of all types (see \S4 of F12). We test if the weighted mean residuals of the unsubtracted and subtracted LCs are consistent with zero to within the standard deviation of the residuals in this phase range, which are each only $\sim 0.001$--$0.002$ mag, depending on the filter. After removing $3$-$\sigma$ outliers and S/N $< 3$ points, the weighted means of the aggregated residuals for all \nwelliso{} SN are consistent with 0 by this measure, with weighted means and standard deviations of the residuals of \galsubwmeanj $\pm$ \galsubewmeanj, \galsubwmeanh $\pm$ \galsubewmeanh, and \galsubwmeank $\pm$ \galsubewmeank{} magnitudes in \jhk{}, respectively.  At least for bright, well isolated objects with sufficient host galaxy templates, \nnt{} does not introduce a net bias in the photometry.

\subsubsection{Galaxy Subtraction for Superposed SN}
\label{sec:embedded_galsub}

For SN superposed on the host galaxy, we can not make the same comparison in the absence of a suitable unsubtracted reference LC.  In these cases, we test the subtraction process by performing forced \dophot{} \nnt{} photometry on the galaxy subtracted difference images at positions near the host galaxy. 
We perform forced photometry on a $3\times3$ grid of positions with evenly spaced increments of $15\arcsec = 15$ pixels centered around the SN position. At least some of these 9 grid positions are likely to be superposed on the galaxy.
If the subtraction process is working correctly (no net bias), the difference image LCs should have a weighted mean of zero flux at all grid positions {\it except} for the central position with the SN, albeit with larger scatter for grid positions superposed on the galaxy (see \S4 of F12). 

We performed this test for all SN fields. The standard deviation of the difference image flux values for each LC is used to estimate the 
uncertainty in the measured flux at each grid position.\footnote{The scatter also increases towards longer wavelength since the signal-to-noise ratio decreases from $J$ to $H$ to $K$ as a result of the presence of additional contaminating sky noise (see \S\ref{sec:sky}).} For all \F{} objects, grid positions offset from the SN showed weighted mean flux consistent with zero to within 1--3 standard deviations.  Highly embedded SN fainter than $J \sim 18-19$ mag at the brightest LC point are often too faint for \PTL{}, and \nnt{} can yield LCs with inaccurate flux values that are not suitable for publication. However, if $\nt \gtrsim 3-4$ host galaxy template images are obtained for sufficiently bright SN which reach $J \lesssim 18$ mag, \nnt{} galaxy subtraction yields a net bias of $\lesssim 0.01$ mag even at positions clearly superposed on host galaxies.

\subsubsection{NNT vs. Forced \dophot{} Errors}
\label{sec:nnt_v_fdo}

\renewcommand{\scale}{0.5}
\begin{figure*}
\centering
\begin{tabular}{@{}c@{}c@{}}

\includegraphics[width=\scale\linewidth,angle=0]
{\colordir/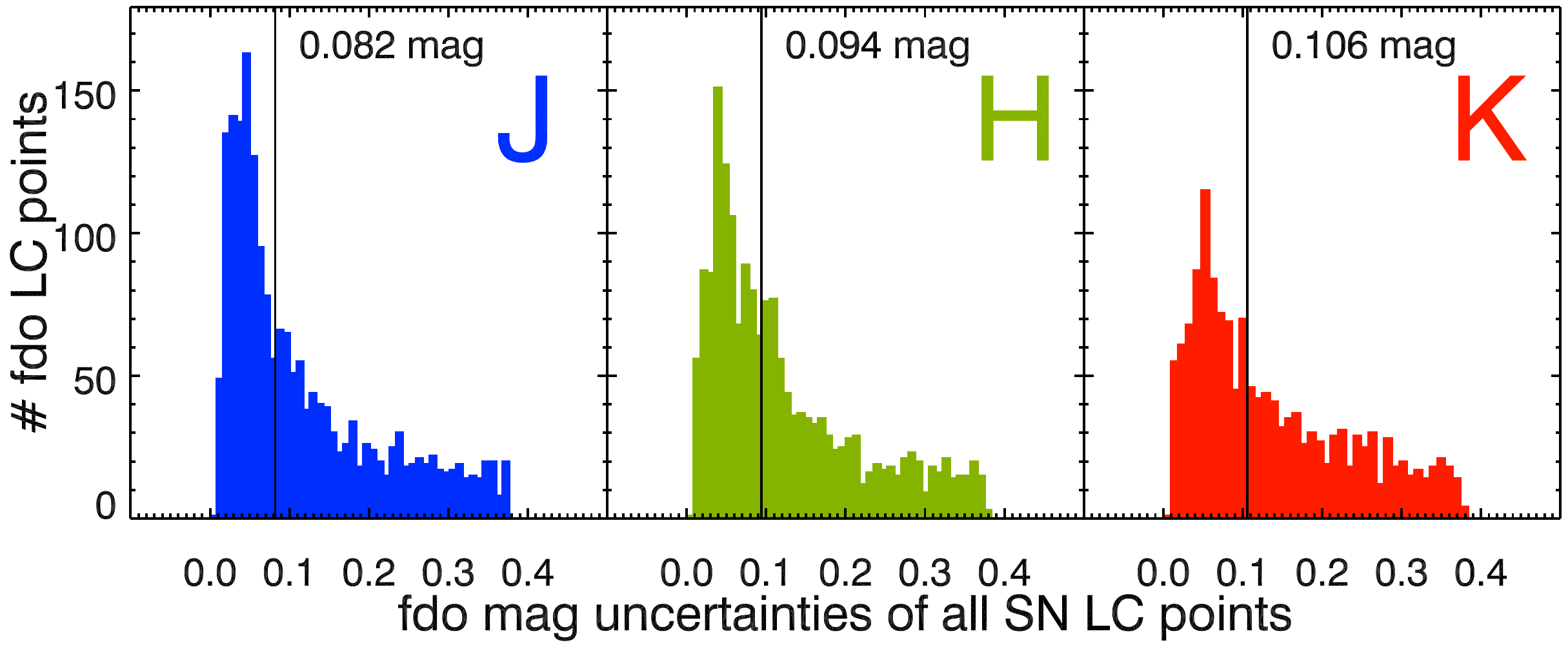} &

\includegraphics[width=\scale\linewidth,angle=0]
{\colordir/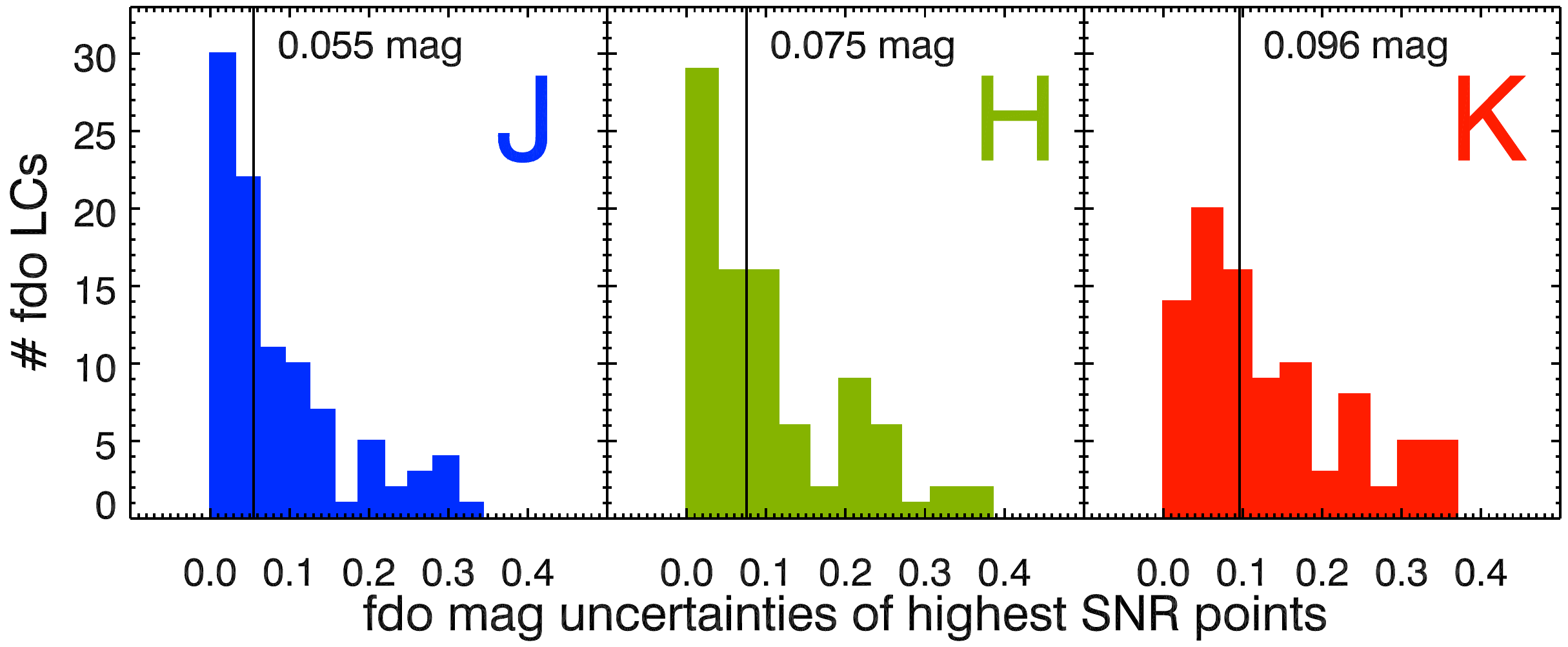} \\

\includegraphics[width=\scale\linewidth,angle=0]
{\colordir/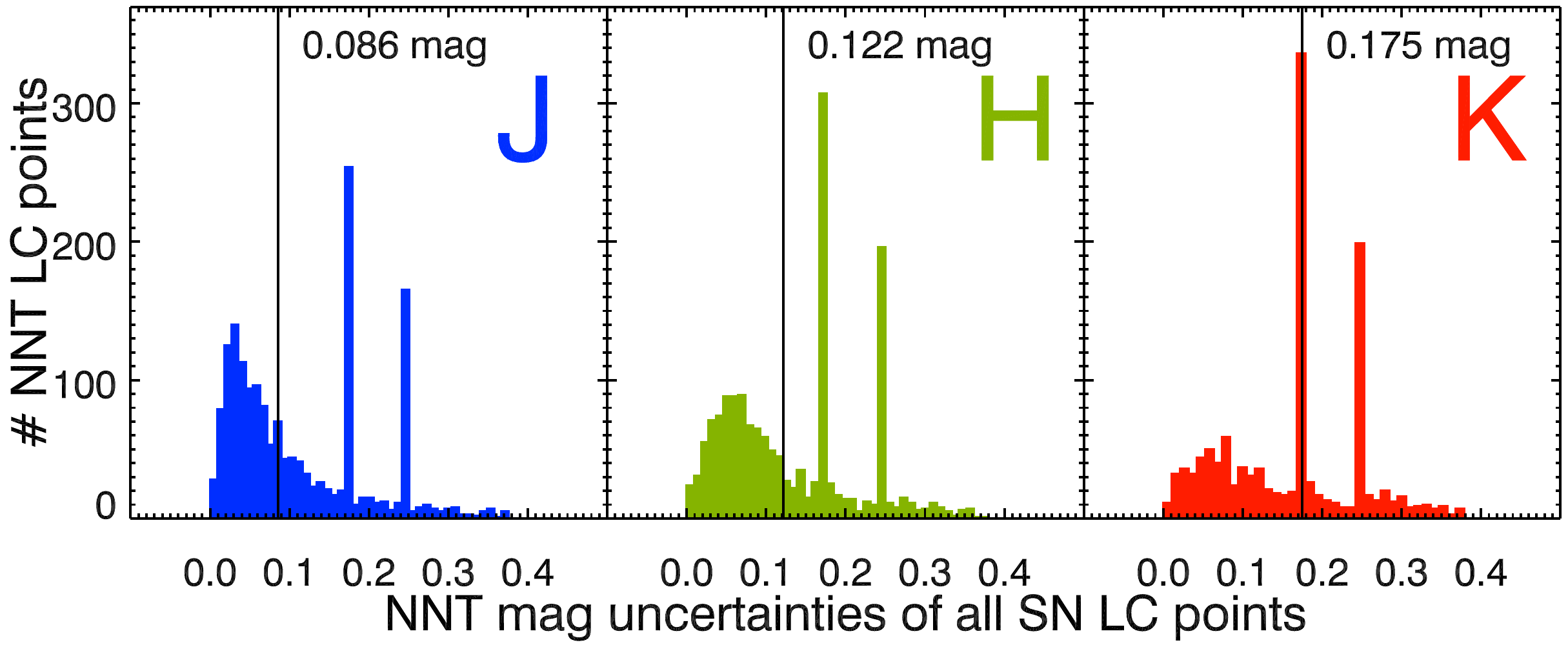} &

\includegraphics[width=\scale\linewidth,angle=0]
{\colordir/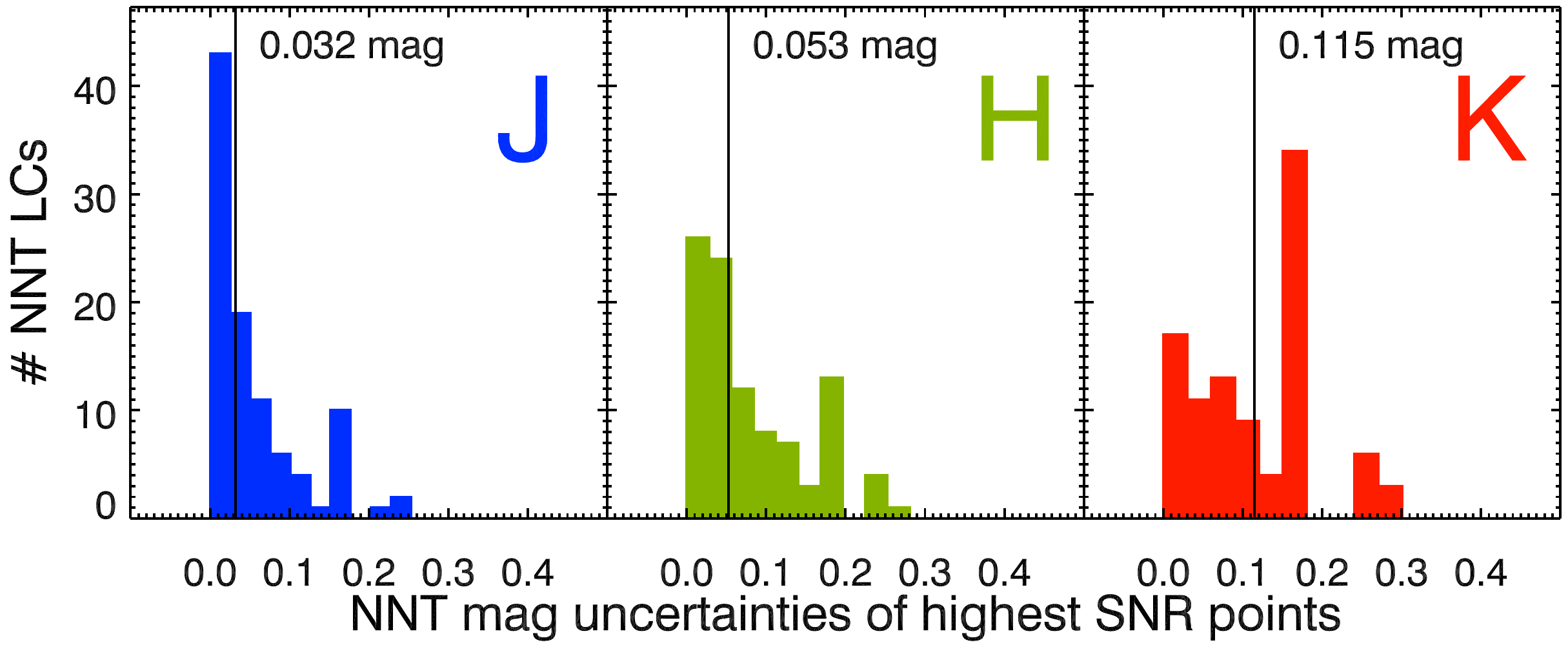} \\

\end{tabular}
\caption[Forced \dophot{} and \nnt{} Errors]{Forced \dophot{} and \nnt{} Errors
\\
\\
\scriptsize
(Color online) Magnitude uncertainty histograms for ({\bf Row 1}) forced \dophot{} photometry (fdo) on unsubtracted images and ({\bf Row 2}) host galaxy subtracted photometry (\nnt). Median values are indicated with vertical lines and plot annotations. Left columns show errors for all \F{} LC points. Right columns show errors for only the highest S/N points for each \F{}  LC. Spikes at $0.25$ and $0.175$ mag (lower left figure), and at $0.175$ mag (lower right figure) reflect the conservative systematic error floor imposed for cases with $\nt=1$ or $2$ usable subtractions (see Table~\ref{tab:nnt_err}). The highest S/N LC points have median uncertainties of $\sim \jmedmagerrlcpeak$, $\hmedmagerrlcpeak$, and $\kmedmagerrlcpeak$ mag in \jhk{}, respectively (lower right plot). Even in these cases, the systematic error floor skews histograms toward larger median errors; for $\jhk$, there are $\sim 10-35$ LCs with only $\nt=2$ usable subtractions, leading to spikes at $0.175$ mag.  All \F{} \nnt{} LC points have median uncertainties of $\jmedmagerr$, $\hmedmagerr$, and $\kmedmagerr$ mag in \jhk{}, respectively (lower left plot). \nnt{} errors are generally comparable to or less than forced \dophot{} errors on unsubtracted images provided $\nt \gtrsim 3-4$. This again reflects the systematic error floor for $\nt=1$ or $2$. For the highest S/N points for each LC, the median \nnt{} photometric precision is smaller than forced \dophot{} for $J$ and $H$, but not in $K_s$, again as a result of the systematic error floor (see right column figures).  
\vspace{0.5cm}
}
\label{fig:fdo_fdi_errs}
\end{figure*}

\nnt{} can lead to larger reported errors ($\sgnnt$) compared to corrected \dophot{} point source photometry without galaxy subtraction ($\sgphotcorr$) for cases with $\nt \lesssim 2-3$, due primarily to our imposed systematic error floor for these cases (see Table~\ref{tab:nnt_err}).  However, for cases with $\nt \gtrsim 3-4$ templates, $\sgnnt \lesssim \sgphotcorr$ and \nnt{} performs as well or better than \dophot{} without host subtraction as a result of the effective division by $\sim \sqrt{\nt}$ inside the error weighted standard deviation used to compute $\sgnnt$ (see Appendix~\ref{sec:nnt_math}).  \fig{}~\ref{fig:fdo_fdi_errs} shows median magnitude uncertainties for both the highest S/N LC points for each SN and for all LC points for both forced \dophot{} and \nnt{} photometry.  The spikes in the \nnt{} error distributions are artifacts of our systematic error floor chosen for cases with $\nt=1$--$2$ \snt{} images.

\subsection{Comparing \PTL{} and CSP Photometry}
\label{sec:csp_comp}

Comparing \PTL{} \F{} \nnt{} LCs with published CSP photometry for the same \snIa{} provides an important external consistency check.  Although CfA and CSP observatories with NIR detectors are in the northern and southern hemispheres, respectively, an overlapping subset of \nptelcsp{} \F{} objects in the declination range $ -24.94410 < \delta < 25.70778$ were observed in \jhk{} by both groups (see Table~\ref{tab:ptel_csp_overlap} and \figs{}~\ref{fig:f12_csp}-\ref{fig:f12_csp3}).\footnote{ 
The latitudes and longitudes of the \PTL{} and CSP observatories are (FLWO: 31.6811$^{\circ}$N, 110.8783$^{\circ}$W) and (LCO: 29.0146$^{\circ}$S, 70.6926$^{\circ}$W), respectively. \PTL{} observes objects with $\delta \gtrsim -30^{\circ}$.}
Similar to Tables~\ref{tab:general}-\ref{tab:general3} of this paper, Table 1 of \citealt{contreras10} (hereafter C10) and Table 1 of \citealt{stritzinger11} (hereafter S11) present general properties of 35 and 50 \snIa{} observed by the CSP, respectively. Some CSP \snIa{} had only optical observations and no NIR data.\footnote{All \PTL{} and CSP \snIa{} with NIR overlap are included in \F{} except \sn{}2006is (CSP NIR data in S11) and \sn{}2005mc (CSP optical data in C10), which had poor quality \PTL{} LCs. Two other \snIa{} (\sn{}2005bo, \sn{}2005bl) have \PTL{} \jhk{} observations in \F{} and CSP optical observations but no CSP NIR data (\sn{}2005bl: \citealt{taubenberger08}; \sn{}2005bo: C10), and are not included in the \PTL{} and CSP NIR comparison set. \sn{}2005bl was also included in \WV{}.}
  The \nptelcsp{} CSP NIR objects independently observed by \PTL{} include \nptelcspnormal{} normal \snIa{}, 1 peculiar, fast-decining object, \nptelcspsuperch{} overluminous, slowly-declining objects, and 1 \snIax{}. Of these, \nwvcsp{} had data published in \WV{} and \nfcspnew{} are new to \F{}. See Table~\ref{tab:ptel_csp_overlap}.
  
\subsubsection{CSP - \PTL{} Offsets and Color Terms}
\label{sec:ptel_csp_stars}
\citet{cohen03} and \citet{skrutskie06} describe the 2MASS \jhk{} filter system while \citet{carpenter01} and \citet{leggett06} provide color transformations from other widely used photometric systems to 2MASS.  The \PTL{}/2MASS \jhk{} bandpasses are very similar to the CSP \jhk{} filters, so it is a reasonable approximation to compare the LCs directly, without first attempting to transform the CSP data to the 2MASS system.  However, to justify this approximation, following C10, we investigate whether there exist non-negligible zero point offsets or color terms between \PTL{} and CSP NIR filters using 2MASS stars in fields observed by both groups. While C10 compared CSP measurements of 2MASS stars to the 2MASS point source catalog \citep{cutri03}, here we also compare CSP and \PTL{} measurements of 2MASS stars from Table~\ref{tab:ptel_catalog} to derive zero point estimates and color terms to approximately transform CSP natural system data to the 2MASS system. Although \PTL{} is on the 2MASS natural system, \PTL{} observations are deeper than 2MASS, so \PTL{} measurements of 2MASS stars are more appropriate than 2MASS catalog data for estimating differences  between \PTL{} and CSP photometry. 

\subsubsection{Zero Point Offsets from 2MASS Star Photometry}
\label{sec:ptel_csp_stars2}

C10 used CSP photometric measurements of 984 $J$ and $H$-band 2MASS stars in their SN fields, finding these mean zero point offsets between the CSP Swope 1.0-m natural system and the 2MASS $J$ and $H$ filters:
\begin{eqnarray}
\label{eq:contreras_zp}
J_{\rm CSP}  - J_{\twomass} & = & 0.010 \pm 0.003 \ {\rm mag} \\ \nonumber
H_{\rm CSP} - H_{\twomass} & = & 0.043 \pm 0.003 \ {\rm mag} 
\end{eqnarray}
\noindent C10 did not derive zero point offsets in $K_s$ because they had only 41 CSP 2MASS star observations in $K_s$.

For 19 objects observed by both \PTL{} and CSP (including \sn{}2006is, which is not in \F), we obtained CSP standard star photometry for the local sequences for 16 objects from the literature (C10; S11; \citealt{taubenberger11}) and 3 additional objects from the CSP (M. Stritzinger --- private communication; see \S4.33 of F12). In these 19 SN fields, we used 269, 264, and 24 2MASS stars observed both by \PTL{} and CSP in \jhk{}, respectively, limited to the color range $0.2 < \jhcsp < 0.7$ mag also used by C10.  We compute CSP - \PTL{} residuals for each 2MASS star in \jhk{} and interpret the weighted mean residuals and the error on the weighted mean as our estimate of the zero point offset and uncertainty between the CSP natural system ($JH$ Swope, $K_s$ duPont) and the \PTL{}/2MASS \jhk{} system.  Although column 6 of Table~\ref{tab:ptel_catalog} reports uncertainties on the weighted mean \PTL{} magnitudes of 2MASS stars as the error on the weighted mean, we follow the method reported by the CSP here and instead use the RMS 
to estimate our local sequence uncertainties (C10; S11), which yield larger, more conservative error estimates.

\renewcommand{\scale}{1}
\begin{figure}[h]
\centering
\begin{tabular}{@{}c@{}}

\includegraphics[width=\scale\linewidth,angle=0]
{\colordir/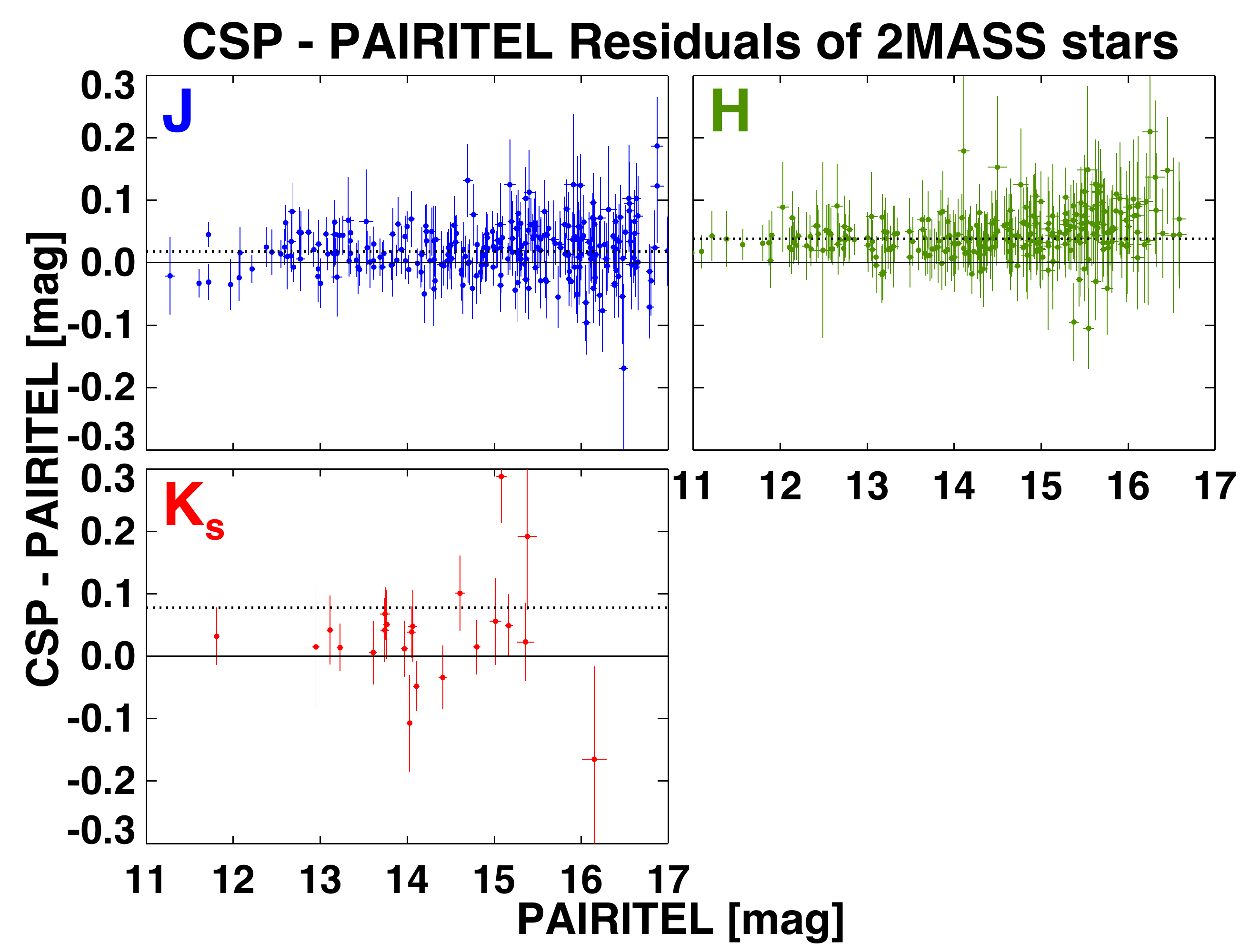} \\

\end{tabular}
\caption
[Zero Point Offsets of \PTL{} and CSP \jhk{} Filters]
{\PTL{} and CSP \jhk{} Offsets
\\
\\
\scriptsize
(Color online) For 19 NIR SN fields, we use 269, 264, and 24 2MASS stars observed by both \PTL{} and the CSP in \jhk{}, respectively, in the color range $0.2 < \jhcsp < 0.7$ mag also used by C10. Plots show CSP - \PTL{} \jhk{} magnitude residuals on the y-axis 
versus the \PTL{} star magnitude on the x-axis.  Errors on the residuals are the quadrature sum of the quoted CSP errors and the \PTL{} errors on the weighted mean magnitude of 2MASS stars, given by the RMS errors for \PTL{} (not shown in Table~\ref{tab:ptel_catalog}; see \S\ref{sec:ptel_csp_stars}).  
The weighted mean zero-point offsets (dotted lines) in each panel
are the values given in \eq{}~\ref{eq:csp_ptel_2mass_stars_zp2}.
}
\label{fig:ptel_csp_stars_zeropoint}
\end{figure}

Using the RMS error for \PTL{} measurements of 2MASS stars, 
we find zero point offsets of:
\begin{eqnarray}
\label{eq:csp_ptel_2mass_stars_zp2}
J_{\rm CSP}  - J_{\rm \ptel} & = & 0.018 \pm 0.002 \ {\rm mag} \\  \nonumber
H_{\rm CSP} - H_{\rm \ptel} & = & 0.038 \pm 0.003 \ {\rm mag} \\  \nonumber
{K_s}_{\rm CSP} - {K_s}_{\rm \ptel}  & = & 0.077 \pm 0.011 \ {\rm mag}  
\end{eqnarray}
\noindent The \jhk{} CSP - \PTL{} zero point offsets from \eq{}~\ref{eq:csp_ptel_2mass_stars_zp2} are also shown in \fig{}~\ref{fig:ptel_csp_stars_zeropoint} and agree with those from C10 in \eq{}~\ref{eq:contreras_zp} to within 2-$\sigma$ in $J$ and 1-$\sigma$ in $H$.  While C10 used $\sim3$--$4$ times as many 2MASS stars, \eq{}~\ref{eq:contreras_zp} technically estimates the offsets between CSP and 2MASS, not the offsets between CSP and \PTL{} given by \eq{}~\ref{eq:csp_ptel_2mass_stars_zp2}.  Since we are most interested in the latter, and since we do not consider the slight differences between \eq{}~\ref{eq:contreras_zp} and \eq{}~\ref{eq:csp_ptel_2mass_stars_zp2} to be significant, we simply use our own offsets from \eq{}~\ref{eq:csp_ptel_2mass_stars_zp2} as needed.  We do not consider the zero point offset for $K_s$ in \eq{}~\ref{eq:csp_ptel_2mass_stars_zp2} to be reliable, since it is based on only 24 2MASS stars measured by both groups.

\subsubsection{CSP - \PTL{} Color Terms}
\label{sec:ptel_csp_color_terms}

Considering only 2MASS stars in the color range $0.2 < \jhcsp< 0.7$ mag, C10 obtained the following linear fits for the $JH$ bands:
\begin{eqnarray}
\label{eq:contreras_colorterms} 
J_{\rm CSP}  - J_{\twomass} & = & (-0.045 \pm 0.008)  \times \jhcsp \\
 & + &(0.035 \pm 0.067) \ {\rm mag} \nonumber \\  
H_{\rm CSP} - H_{\twomass} & = & (0.005 \pm 0.006) \times \jhcsp \nonumber \\
& + & (0.038 \pm 0.080) \ {\rm mag} \nonumber 
\end{eqnarray}
\noindent  C10 thus find some evidence for a small color term slope in $J$, a negligible color term in $H$, and do not attempt to derive any color terms involving $K_s$.

\renewcommand{\scale}{1}
\begin{figure}[h]
\centering
\begin{tabular}{@{}c@{}}

\includegraphics[width=\scale\linewidth,angle=0]
{\colordir/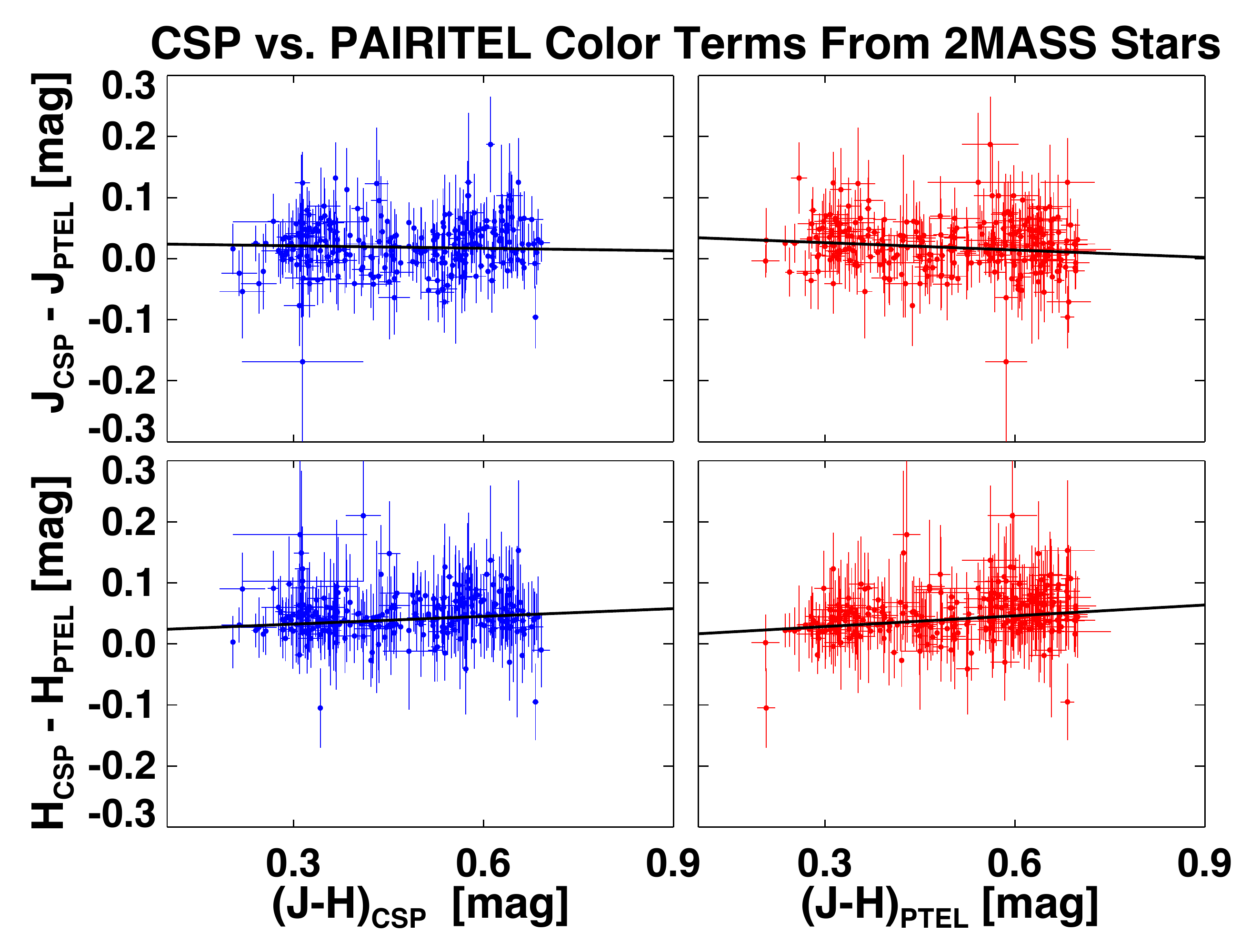} \\

\end{tabular}
\caption
[\PTL{} and CSP $J-H$ Color Terms]
{\PTL{} and CSP $J-H$ Color Terms
\\
\\
\scriptsize
(Color online) Linear fits for $JH$ color terms using 2MASS stars observed by \PTL{} and CSP, given by \eq{}~\ref{eq:friedman_colorterms}.  Following C10, we include only stars in the color range $0.2 < \jhcsp < 0.7$ mag, yielding 263 2MASS stars with $\jhcsp$ data (blue, left panels) and 259 stars with $(J-H)_{\ptel}$ data (red, right panels).  Error bars assume RMS errors for \PTL{} (not shown in Table~\ref{tab:ptel_catalog}; see \S\ref{sec:ptel_csp_stars}). Linear fits have $\chi^2/$doF$=\chi^2_{\nu} < 1$ ($\chi^2_{\nu} = 0.79, 0.35$, left panels and $\chi^2_{\nu} = 0.79, 0.33$, right panels, both top to bottom).
}
\label{fig:ptel_csp_stars_colorterms}
\end{figure}

Following C10, we test for linear color terms between CSP and \PTL{} filters using 263 2MASS stars with both $J$ and $H$ band data. 
We use the \citealt{carpenter01} color terms for $K_s$.\footnote{\singlespace \footnotesize 
\citealt{carpenter01} find these fits for the Las Campanas Observatory (LCO) $K_s$ band using the Persson Standard stars: \\
{\tiny ${K_s}_{\rm CSP} - {K_s}_{\twomass}  =  (-0.015 \pm 0.004) \times (J-K_s)_{\rm CSP}+ (0.002 \pm 0.004) \ {\rm mag}$}.\\
 The \citealt{carpenter01} color transformations have been updated at \url{http://www.astro.caltech.edu/\~{}jmc/2mass/v3/transformations/} as of 2003. \citealt{carpenter01} find a fairly small color term for $K_s$ (the CSP $K_s$ filter is on the 2.5-m duPont telescope at LCO).
}
We find the following $JH$ linear color term fits using the RMS error for the \PTL{} uncertainties of 2MASS stars (also see \fig{}~\ref{fig:ptel_csp_stars_colorterms}):
\begin{eqnarray}
\label{eq:friedman_colorterms} 
J_{\rm CSP}  - J_{\rm \ptel} & = &(-0.014 \pm 0.017) \times \jhcsp  \\
& + & (0.025 \pm 0.009) \ {\rm mag} \nonumber \\  
H_{\rm CSP} - H_{\rm \ptel} & = & (0.042 \pm 0.022) \times \jhcsp  \nonumber \\
& + & (0.020 \pm 0.011) \ {\rm mag}  \nonumber 
\end{eqnarray}
\noindent Linear color term fits yield $\chi^2_{\nu} < 1$, indicating that while the fits are good, the errors are slightly overestimated by using the RMS.  For all panels in \fig{}~\ref{fig:ptel_csp_stars_colorterms}, the probability that a correct model would give the observed $\chi^2_{\nu}$ is $\sim 1$. $JH$ color term fits from \eq{}~\ref{eq:friedman_colorterms} and from C10 in \eq{}~\ref{eq:contreras_colorterms} agree in the slopes at 2-$\sigma$ and the intercepts at 1-$\sigma$. Both fits also yield the same signs for the slopes and indicate at most small $JH$ color terms.

Again, although the C10 fits used $\sim3$--$4$ times as many 2MASS stars, we consider the color terms from either \eqs{}~\ref{eq:contreras_colorterms} or \ref{eq:friedman_colorterms} to be equally reliable. For SN LCs with sufficient sampling to compute reliable colors, applying either set of color terms produced comparable results, since both color terms are small.  In summary, either set of color terms (or no color terms) are reasonable choices to approximately put CSP data on \PTL{}/2MASS system. Still, to compare CSP and \F{} data on the same footing, for the analysis in \S\ref{sec:csp_comp1}, we apply our own $JH$ color terms from \eq{}~\ref{eq:friedman_colorterms} and $K_s$ color terms from \citealt{carpenter01} as needed.

\subsection{Comparing \F{} and CSP LCs}
\label{sec:csp_comp1}

Because \F{} and CSP observations were generally performed at slightly different phases, it is usually not possible to compute direct LC data differences.  We thus require a smooth model fit to interpolate from to compute residuals, which we apply to all \nptelcsp{} overlap objects.\footnote{Model fits to joint \F{}+CSP data all use cubic splines, with some LCs using simple linear fits at late epochs $\gtrsim 30$ days. All fits are boxcar-smoothed with a 5 day moving window. These steps avoid spline over-fitting.  All fits to normal \snIa{} use the \WV{} normal \snIa{} template LC to inform the fit for missing data, with data given greater weight than the template to account for intrinsic variation of the NIR LC shapes. Re-fitting the mean template LC using spectroscopically normal \F{} \snIa{} yielded very similar results to the \WV{} template, so we did not find it necessary to construct a new mean template LC for the purposes of these LC fits. This will be presented elsewhere.
Fits to peculiar \snIa{} or \snIax{} are direct fits to data only.}
The purpose of these model fits is not to estimate LC shape parameters, but merely to provide a baseline with which to compute residuals.
\figs{}~\ref{fig:f12_csp}-\ref{fig:f12_csp3} overplot all 18 example \F{} and CSP \snIa{} LCs for comparison. 
Applying either set of color terms from \S\ref{sec:ptel_csp_color_terms} (or no color terms) had a negligible effect on the CSP LCs, model fits, and weighted mean residuals for the CSP-\F{} data in Table~\ref{tab:ptel_csp_overlap}.
 
\renewcommand{\arraystretch}{0.5}
\begin{table}
\begin{center}
\caption[\nptelcsp{} NIR \snIa{} with \PTL{}/CSP Overlap]{\nptelcsp{} NIR \snIa{} Observed by \PTL{} \&  CSP}
\tiny
\begin{tabular}{@{}l@{}l@{}r@{}r@{}r@{}c@{}c@{}}
\hline
\multicolumn{1}{c}{SN}                         & \multicolumn{1}{c}{Type}                      & \multicolumn{1}{c}{$\Delta J$ [mag]}                           &  \multicolumn{1}{c}{$\Delta H$ [mag]}                         & \multicolumn{1}{c}{$\Delta K_s$ [mag]} & \multicolumn{1}{c}{Agree?} & \multicolumn{1}{c}{CSP}\\
\multicolumn{1}{c}{\tablenotemark{a}}  & \multicolumn{1}{c}{\tablenotemark{b}}   & \multicolumn{1}{c}{\tablenotemark{c}}   &  \multicolumn{1}{c}{\tablenotemark{c}} & \multicolumn{1}{c}{\tablenotemark{c}}  &  \multicolumn{1}{c}{\tablenotemark{d}}    & \multicolumn{1}{c}{Refs\tablenotemark{e}}  \\
\hline
\sn{}2005el    \ &  Ia     \ &  $  0.032 \pm 0.026 $    \ &  $  0.042 \pm 0.018 $    \ &  $  0.078 \pm 0.024 $             \ &  234    \ &  1    \\
\sn{}2005eq    \ &  Ia     \ &  $ -0.010 \pm 0.030 $    \ &  $ -0.003 \pm 0.024 $    \ &  $ -0.034 \pm 0.030 $             \ &  112    \ &  1    \\
\sn{}2005hk    \ &  Iax    \ &  $ -0.031 \pm 0.027 $    \ &  $ -0.012 \pm 0.028 $    \ &  $  0.050 \pm 0.048 $             \ &  212    \ &  3    \\
\sn{}2005iq    \ &  Ia     \ &  $ -0.025 \pm 0.029 $    \ &  $  0.080 \pm 0.060 $    \ &  $ -0.077 \pm 0.045 $             \ &  122    \ &  1    \\
\sn{}2005ke    \ &  Iap    \ &  $ -0.001 \pm 0.014 $    \ &  $ -0.001 \pm 0.014 $    \ &  $  0.010 \pm 0.020 $             \ &  111    \ &  1    \\
\sn{}2005na    \ &  Ia     \ &  $ -0.059 \pm 0.030 $    \ &  $ -0.000 \pm 0.023 $    \ &   \multicolumn{1}{c}{\nodata}     \ &  21     \ &  1    \\
\sn{}2006D     \ &  Ia     \ &  $  0.003 \pm 0.011 $    \ &  $ -0.006 \pm 0.014 $    \ &  $  0.000 \pm 0.010 $             \ &  111    \ &  1    \\
\sn{}2006X     \ &  Ia     \ &  $  0.009 \pm 0.018 $    \ &  $  0.006 \pm 0.011 $    \ &  $ -0.007 \pm 0.010 $             \ &  111    \ &  1    \\
\sn{}2006ax    \ &  Ia     \ &  $ -0.026 \pm 0.014 $    \ &  $  0.003 \pm 0.005 $    \ &  $  0.007 \pm 0.018 $             \ &  211    \ &  1    \\
\hline
\sn{}2007S     \ &  Ia     \ &  $  0.029 \pm 0.023 $    \ &  $  0.015 \pm 0.020 $    \ &  $  0.006 \pm 0.024 $             \ &  211    \ &  2    \\
\sn{}2007ca    \ &  Ia     \ &  $  0.004 \pm 0.012 $    \ &  $  0.036 \pm 0.025 $    \ &   \multicolumn{1}{c}{\nodata}     \ &  12     \ &  2    \\
\sn{}2007if    \ &  Iap    \ &  $  0.058 \pm 0.033 $    \ &  $  0.053 \pm 0.038 $    \ &   \multicolumn{1}{c}{\nodata}     \ &  22     \ &  2    \\
\sn{}2007le    \ &  Ia     \ &  $  0.015 \pm 0.013 $    \ &  $  0.006 \pm 0.008 $    \ &   \multicolumn{1}{c}{\nodata}     \ &  21     \ &  2    \\
\sn{}2007nq    \ &  Ia     \ &  $  0.004 \pm 0.020 $    \ &  $  0.000 \pm 0.054 $    \ &   \multicolumn{1}{c}{\nodata}     \ &  11     \ &  2    \\
\sn{}2007sr    \ &  Ia     \ &  $  0.022 \pm 0.017 $    \ &  $  0.017 \pm 0.012 $    \ &   \multicolumn{1}{c}{\nodata}     \ &  22     \ &  4    \\
\sn{}2008C     \ &  Ia     \ &  $ -0.004 \pm 0.018 $    \ &  $ -0.001 \pm 0.018 $    \ &   \multicolumn{1}{c}{\nodata}     \ &  11     \ &  2    \\
\sn{}2008hv    \ &  Ia     \ &  $  0.024 \pm 0.024 $    \ &  $  0.011 \pm 0.020 $    \ &   \multicolumn{1}{c}{\nodata}     \ &  21     \ &  2    \\
\sn{}2009dc    \ &  Iap    \ &  $ -0.004 \pm 0.019 $    \ &  $ -0.006 \pm 0.015 $    \ &  $ -0.002 \pm 0.019 $             \ &  111    \ &  5    \\
\hline
\end{tabular}
\tablecomments{ \tiny
\\
{\bf (a)} All SN LCs use \nnt{} galaxy subtraction (see \S\ref{sec:nn2nnt}). 
The horizontal line in the middle of the table divides the 9 \PTL{} SN with \F{} data which supersedes \WV{} data (top: \sn{}2005el-\sn{}2006ax) from the 9 SN with \PTL{} data new to this work (bottom: \sn{}2007S-\sn{}2009dc).
\\
{\bf (b)}  Ia: spectroscopically normal. Iap: peculiar, under-luminous (\sn{}2005ke), peculiar over-luminous (\sn{}2007if, \sn{}2009dc). Iax: 02cx-like (\sn{}2005hk).
\\
{\bf (c)} Weighted mean CSP - \F{} residuals and 1-$\sigma$ errors, estimated by the error weighted standard deviation of the residuals divided by 3. $K_s$-band data not available for some CSP \snIa{}.
\\
{\bf (d)} Do CSP - \F{} weighted mean residuals agree within 1, 2, or $\ge$ 3-$\sigma$ for \jhk{}, respectively?  For example, 132 would mean the NIR LCs agree in $J$ within 1-$\sigma$, $H$ within $\ge$ 3-$\sigma$, and $K_s$ within 2-$\sigma$. All 18 LCs in $JH$ and all 8 in $K_s$ agree within at least 3-$\sigma$ by this metric (Except for sn2005el, $K_s$, which agrees at 4-$\sigma$).
\\ 
{\bf (e)} CSP References: (1) \citealt{contreras10}, (2) \citealt{stritzinger11}, (3) \citealt{phillips07}, (4) \citealt{schweizer08}, (5) \citealt{taubenberger11} 
}
\label{tab:ptel_csp_overlap}
\end{center}
\end{table}
\renewcommand{\arraystretch}{1}

For all \F{} and color-term-corrected CSP LC points at similar phases, the scatter in the residuals arises from both statistical photometric uncertainties and systematic uncertainties as a result of imperfect model fits, which can dominate, especially at late times. For individual \snIa{}, we compute the weighted mean of the residuals about the joint model fit in the phase range $[-10,60]$ days where the model fit is generally valid. 
To include systematic uncertainty from the joint model fit, we conservatively estimated the 1-$\sigma$ uncertainty on the weighted mean CSP - \F{} residual as the error weighted standard deviation of the residuals, which we then divided by a factor of 3 to avoid overestimating the uncertainty.
We then compute whether the mean CSP - \F{} residuals are consistent with zero to within 1, 2 or $\ge$ 3-$\sigma$ in the selected phase range. We find that nearly all \F{} and color term corrected CSP \snIa{} LCs
(18 $JH$ and 8 $K_s$ LCs) are consistent to within 3-$\sigma$ by this metric.\footnote{Except for \sn{}2005el in $K_s$, which agrees at 4-$\sigma$.} See Figs.~\ref{fig:f12_csp}-\ref{fig:f12_csp3} and Table~\ref{tab:ptel_csp_overlap}. 
 
While this method is useful to compare entire LCs, we note that some CSP and \F{} LCs in specific bands do show significant $\sim 0.1-0.4$ mag deviations for individual data points at similar phases or ranges of data points over smaller phase ranges, beyond what can be explained from poor model fits alone.  For example, these discrepancies were noted: \sn{}2005iq, $H$, $<0$ days; \sn{}2005na, $H$, $20$-$40$ days; \sn{}2007if, $JH$, $20$-$30$ days; \sn{}2008hv, $J$, $>40$ days; \sn{}2006D, $H$, $>40$ days; \sn{}2005el, $JH$, $>40$ days; \sn{}2007sr, $H$, $10$-$20$ days. Nevertheless, many of these differences come from $\sim$1-2, individual outlier \F{} data points, and most of the LCs show broad agreement by the above metric across a broad range of phases. See Figs.~\ref{fig:f12_csp}-\ref{fig:f12_csp3}.

\renewcommand{\scale}{0.5}
\begin{figure*}
\centering
\begin{tabular}{@{}c@{}c@{}}

\includegraphics[width=\scale\linewidth,angle=0]
{\colordir/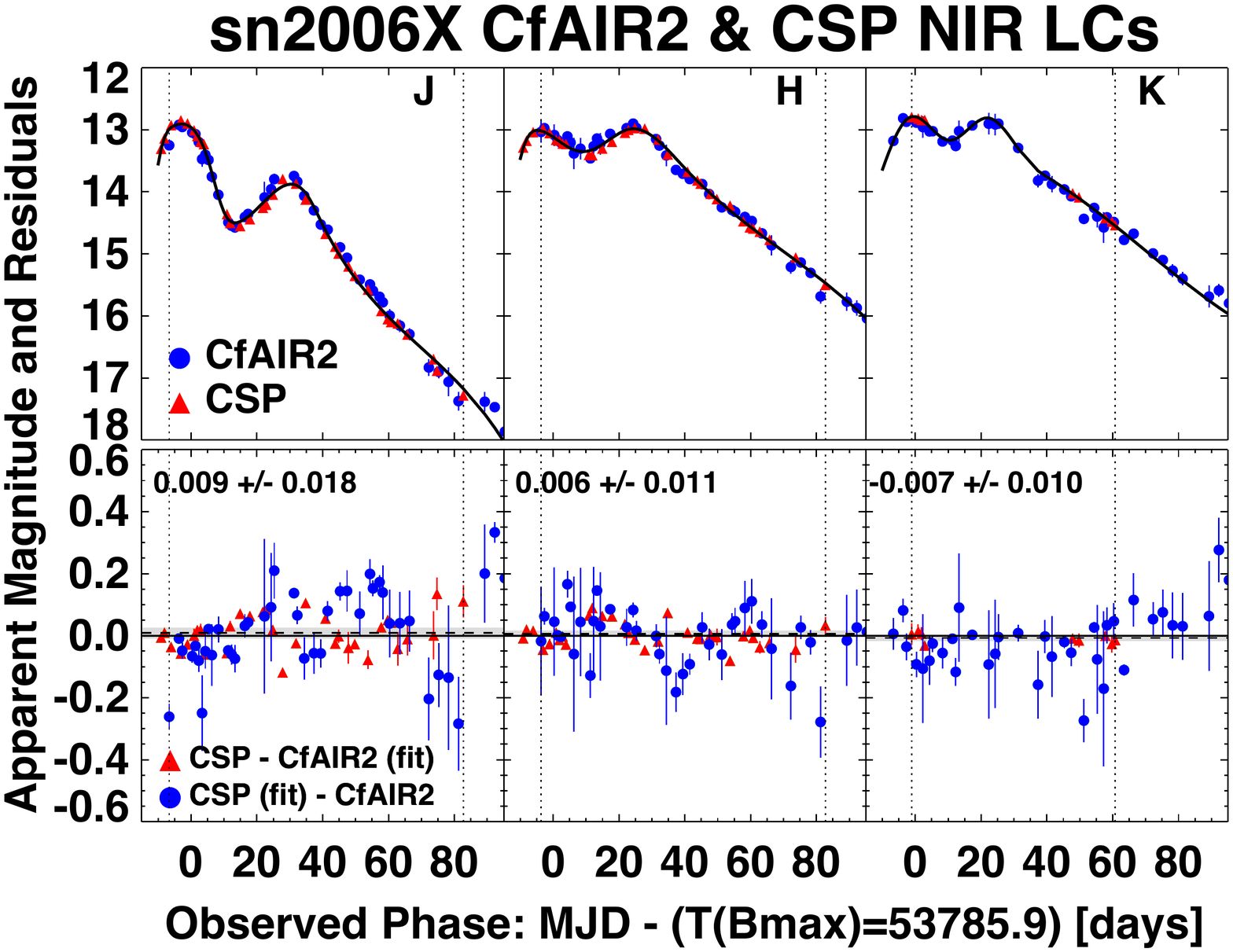} &

\includegraphics[width=\scale\linewidth,angle=0]
{\colordir/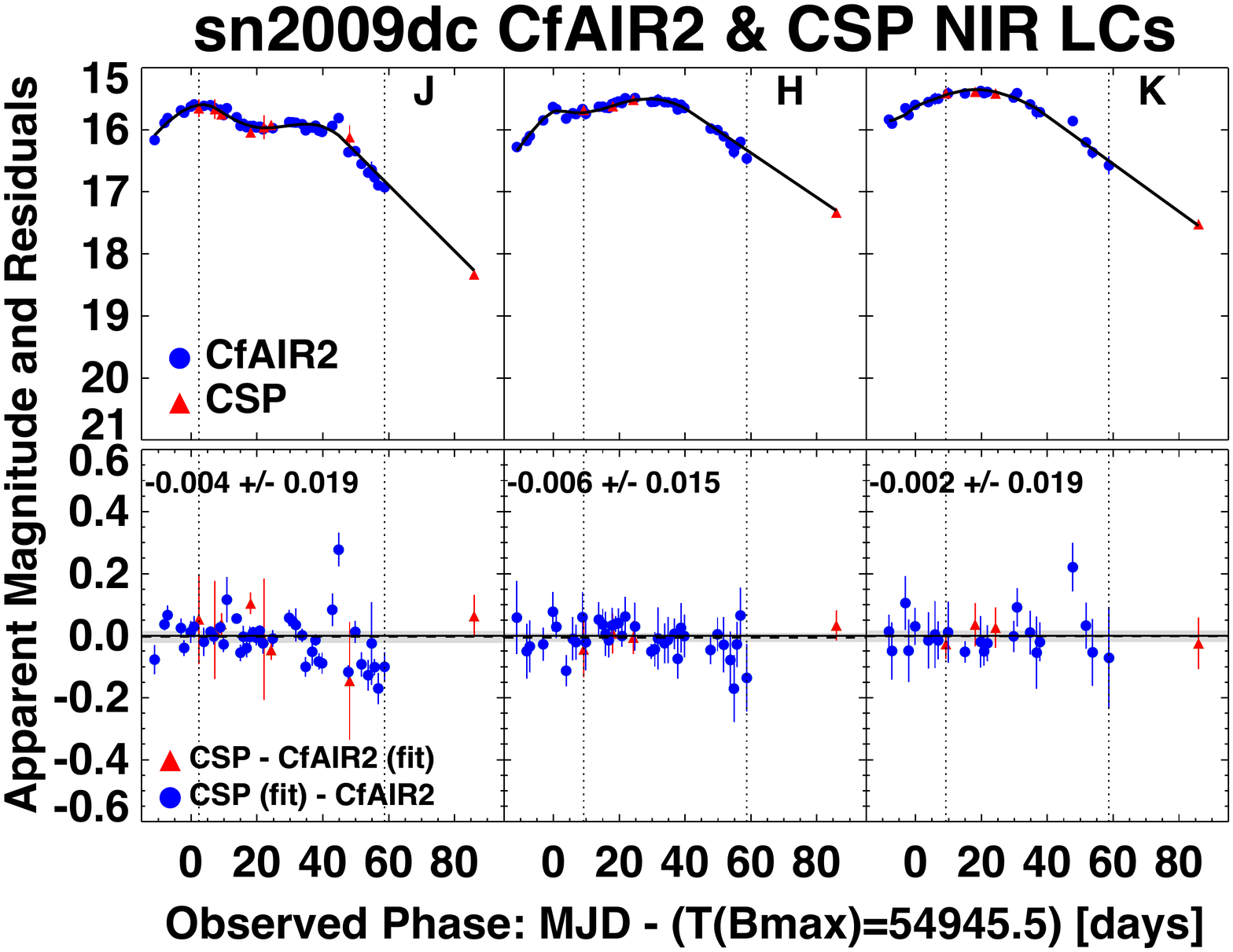} \\

\includegraphics[width=\scale\linewidth,angle=0]
{\colordir/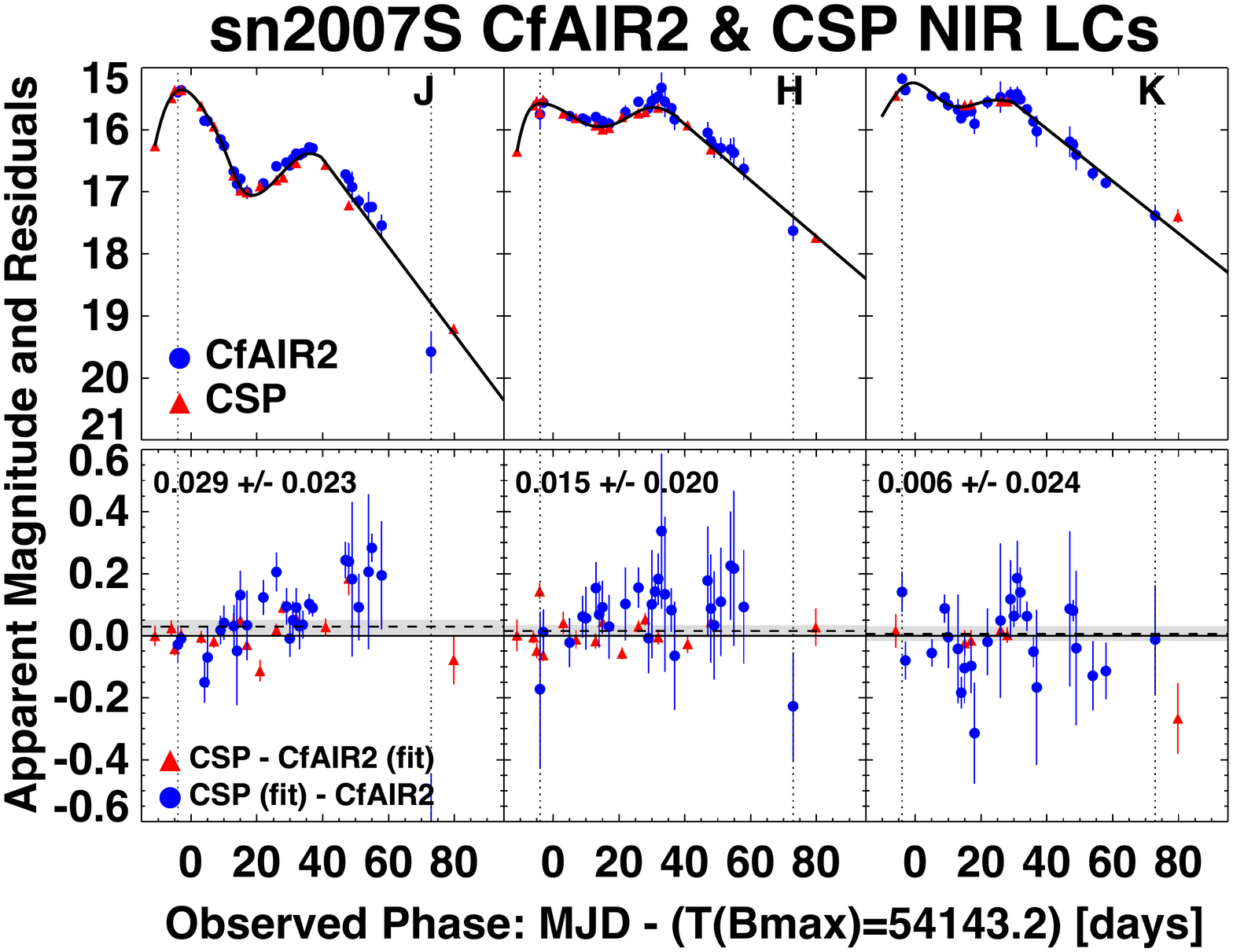} &

\includegraphics[width=\scale\linewidth,angle=0]
{\colordir/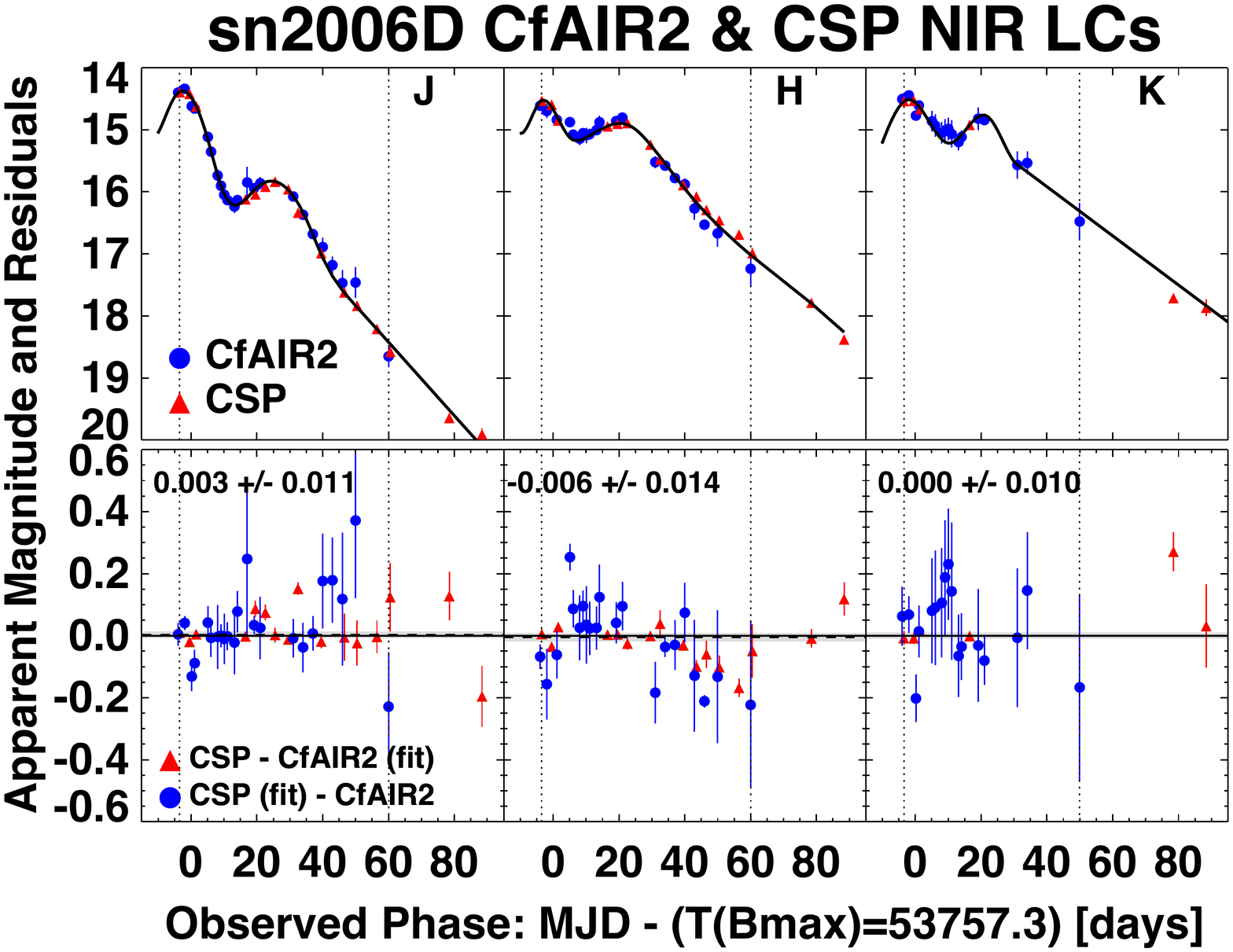} \\

\includegraphics[width=\scale\linewidth,angle=0]
{\colordir/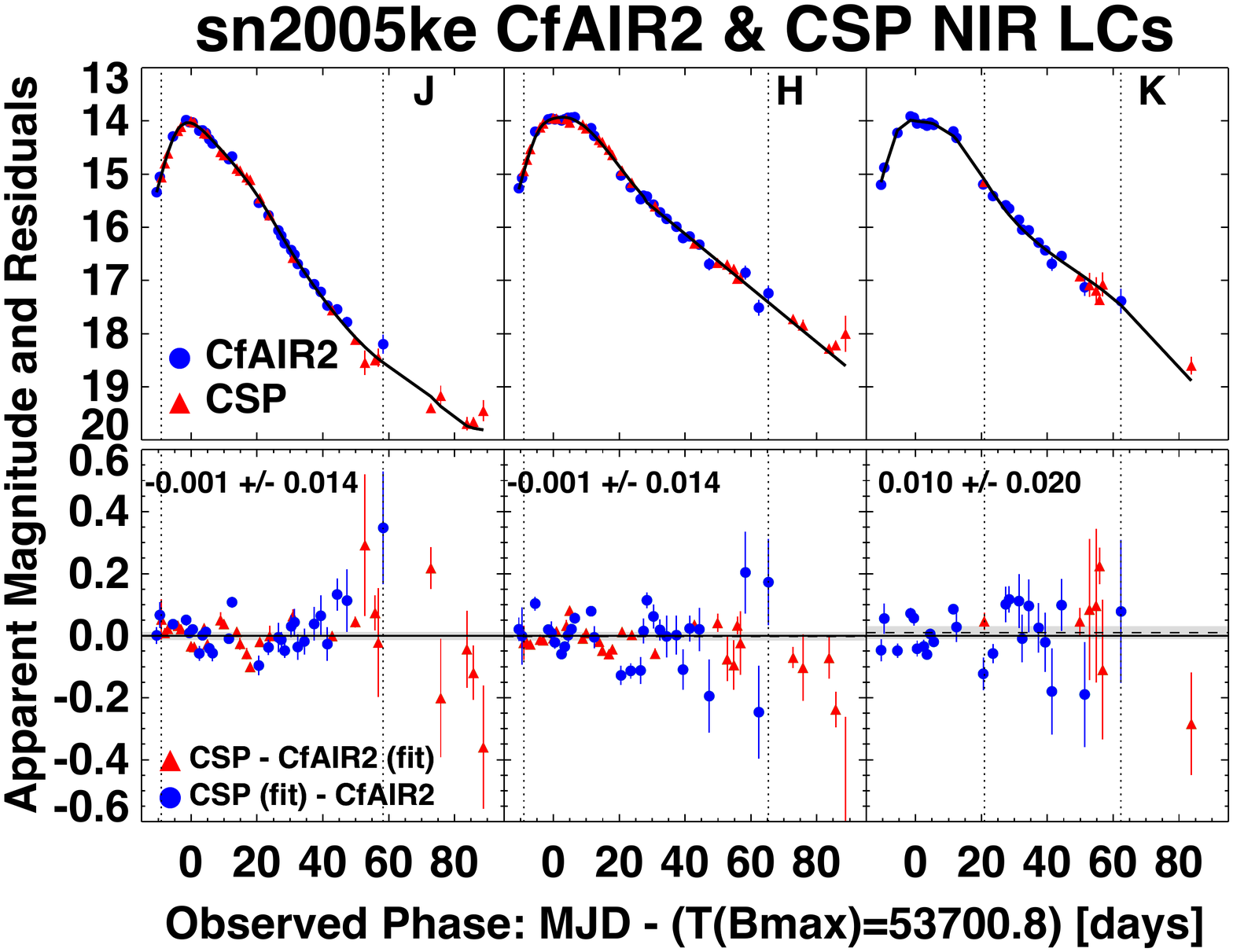} &

\includegraphics[width=\scale\linewidth,angle=0]
{\colordir/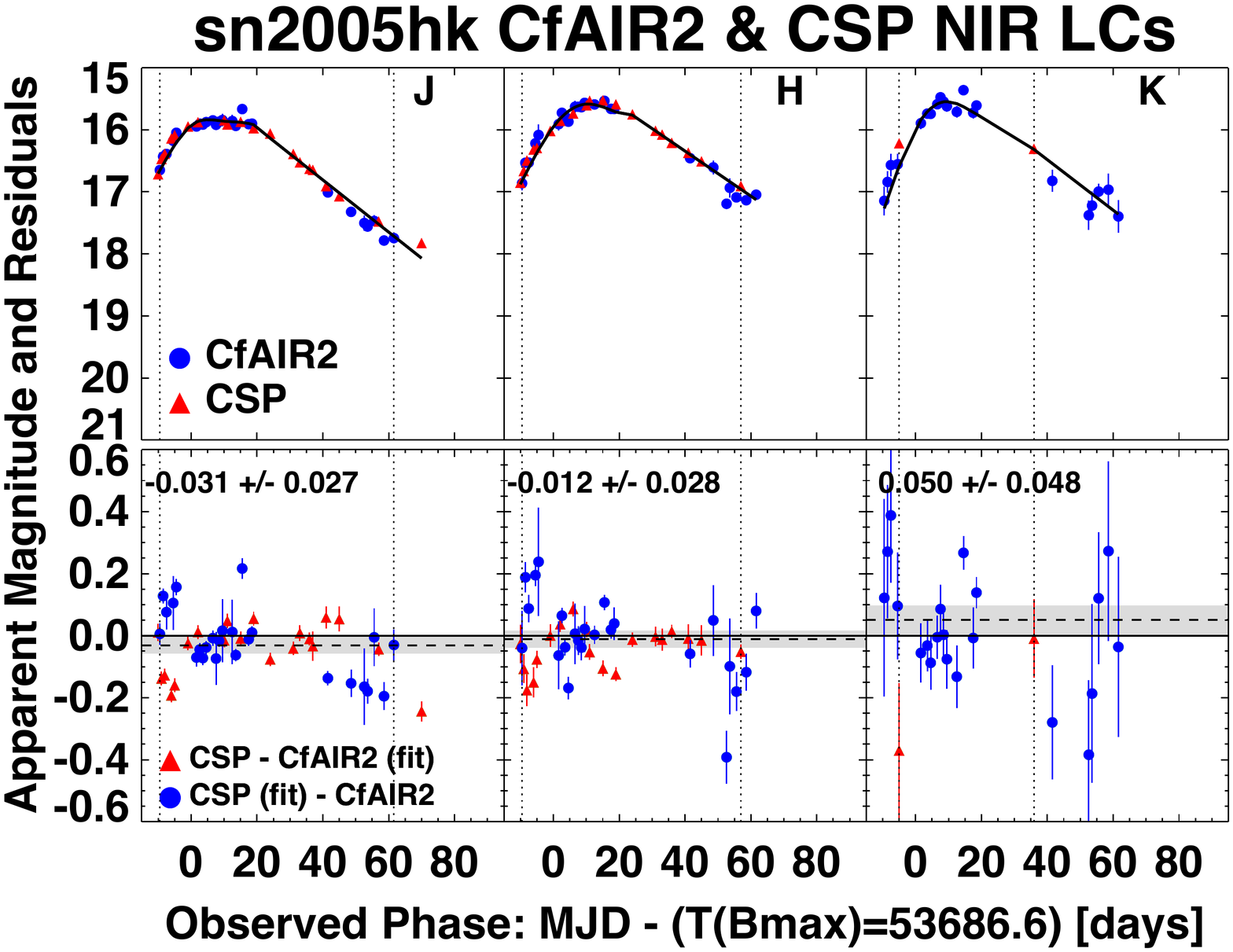} \\

\end{tabular}
\caption[\F{}/CSP Photometry Comparison]{
Comparing \F{} to CSP Photometry 
\\
\\
\tiny
(Color online) 
{\bf (Top panels)} Plot shows 6 example NIR \snIa{} LCs out of the \nptelcsp{} \F{} objects observed by both \PTL{} and CSP.
\jhk{} \snIa{} LCs are shown from PAIRITEL \F{} galaxy subtracted photometry (blue circles) and CSP LCs (red triangles) after applying color terms from \eq{}~\ref{eq:friedman_colorterms} of this paper (see \S\ref{sec:ptel_csp_color_terms}).  Vertical dotted lines show regions of temporal overlap for both LCs. The black line is a cubic spline model fit to the joint PAIRITEL+CSP data with a simple linear fit applied $\gtrsim 30$-$40$ days in specific cases.  For normal \snIa{}, the \WV{} mean template LC is used to help fit for missing data (not for Ia-pec or Iax: \sn{}2009dc, \sn{}2005ke, \sn{}2005hk).   
\\
\\
{\bf (Bottom panels)} CSP - \F{} residuals are computed as either (CSP data minus \F{} joint model fit) or (CSP joint model fit - \F{} data) for each epoch, using the same plot symbols as above for differences computed using CSP or \F{} data. While the CSP (fit) - \F{} residuals (blue circles) are above the zero residual line when the corresponding \F{} data point has a larger magnitude value than the joint model fit in the top row panels, since we are computing CSP - \F{} residuals, the CSP - \F{} (fit) (red triangles) residuals behave in the opposite sense. For example, when the CSP data has a larger magnitude than the joint model fit in the top row panels, the corresponding residual lies {\it below} the zero residual line.
Weighted mean residuals and 1-$\sigma$ uncertainties for CSP - \F{} data in the phase range $[-10,60]$ days, as listed in Table~\ref{tab:ptel_csp_overlap}, are also shown in the upper left corner of each panel and indicated by the dashed line and the gray strip, respectively.
}
\label{fig:f12_csp}
\end{figure*}

\renewcommand{\scale}{0.5}
\begin{figure*}
\centering
\begin{tabular}{@{}c@{}c@{}}

\includegraphics[width=\scale\linewidth,angle=0]
{\colordir/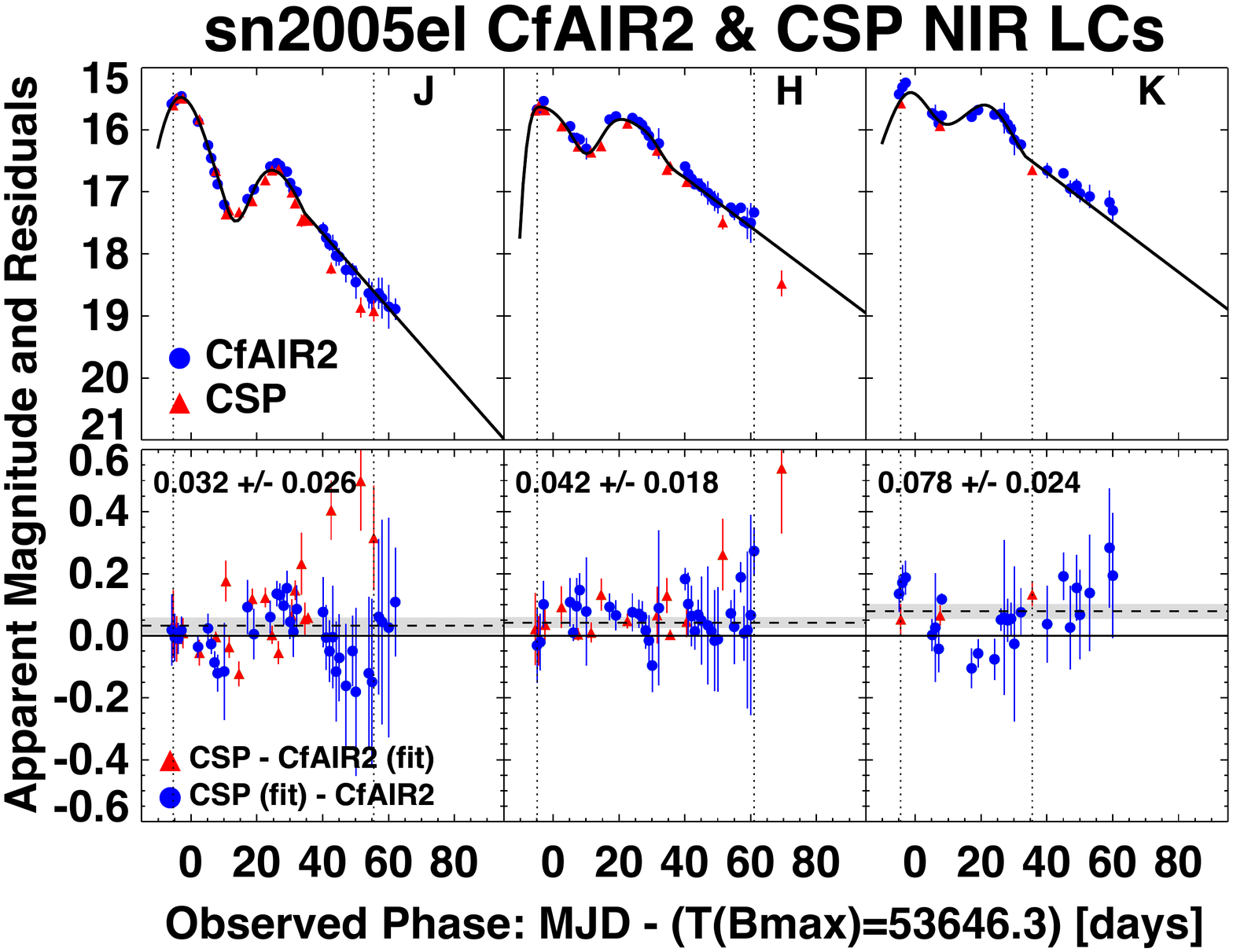} &

\includegraphics[width=\scale\linewidth,angle=0]
{\colordir/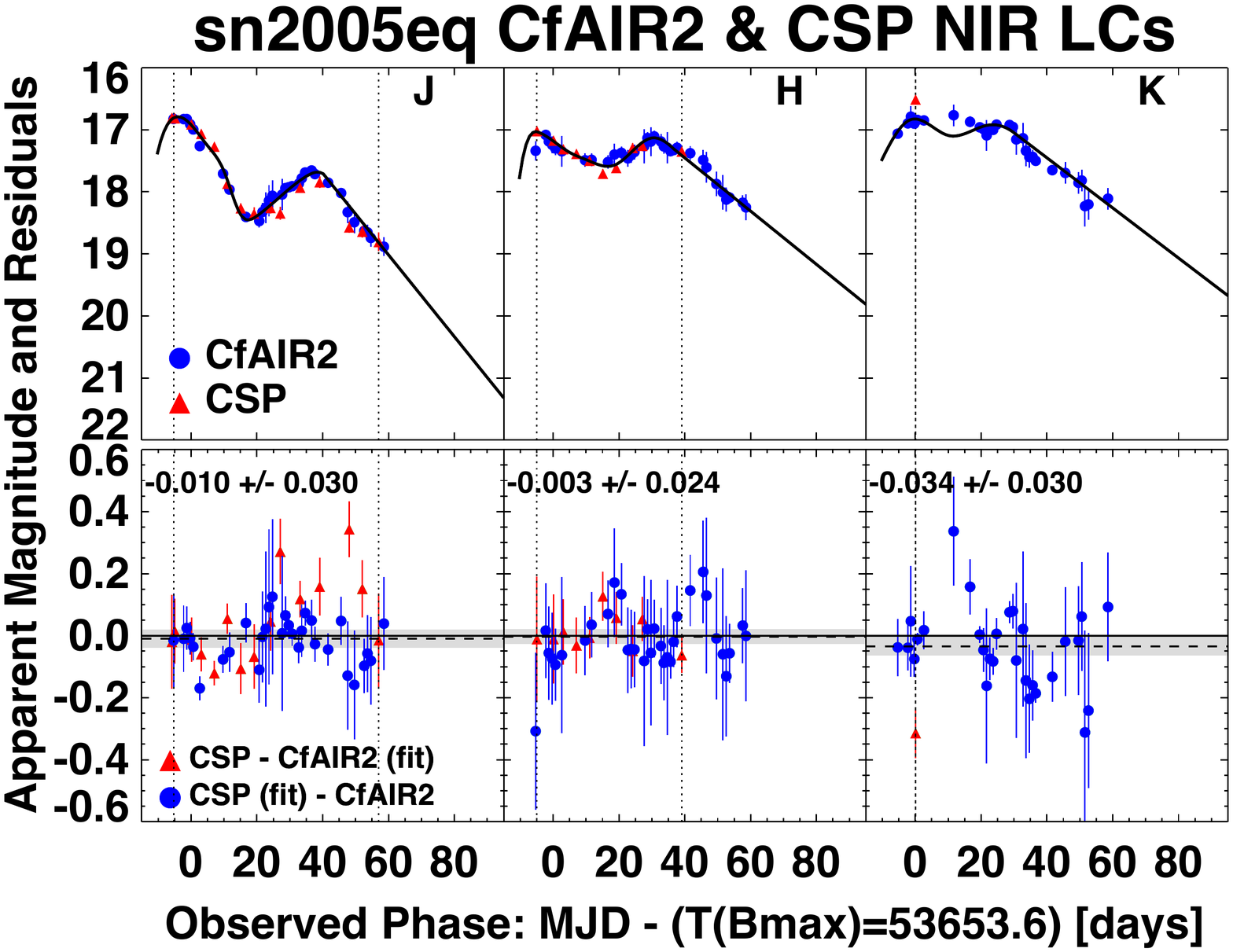} \\

\includegraphics[width=\scale\linewidth,angle=0]
{\colordir/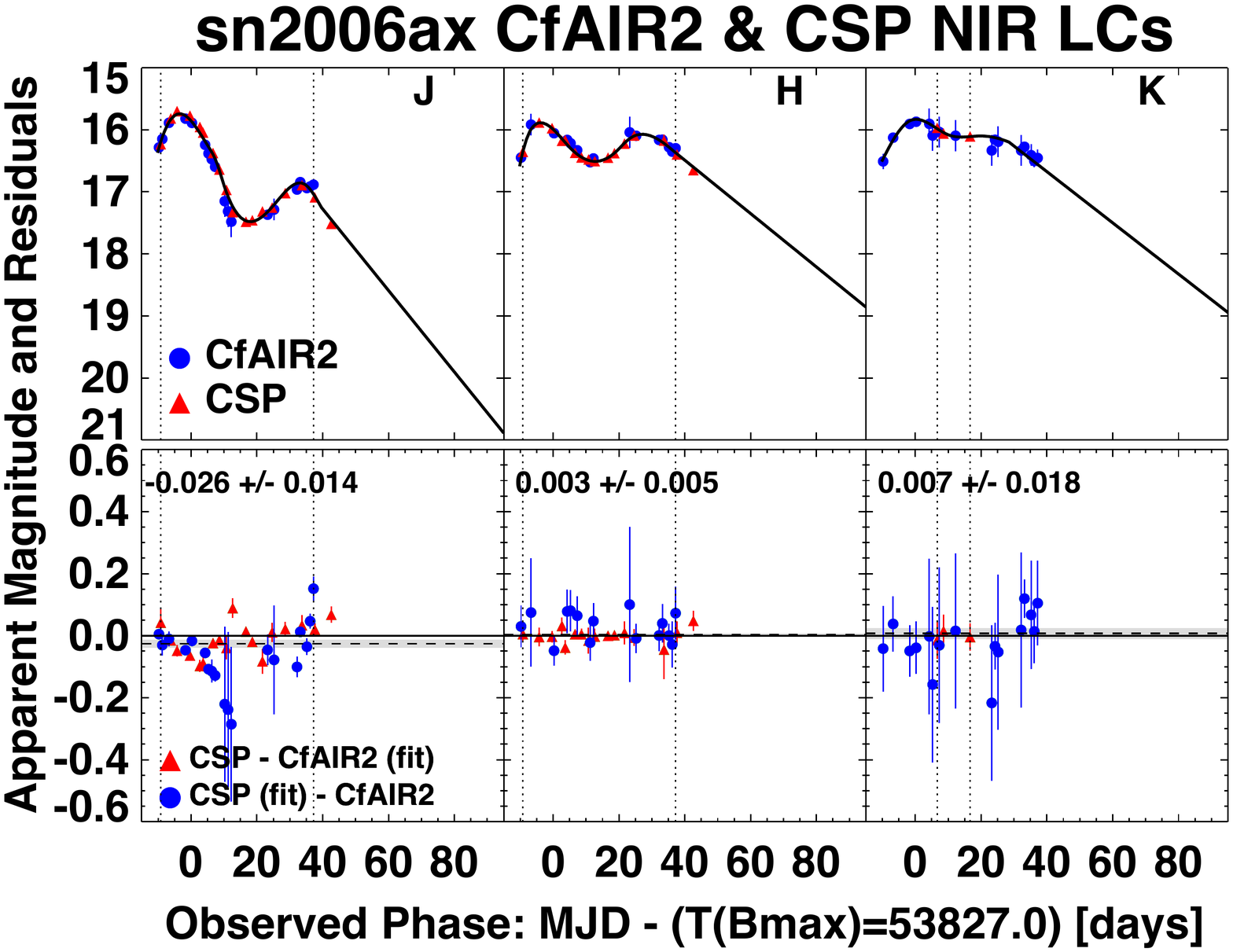} &

\includegraphics[width=\scale\linewidth,angle=0]
{\colordir/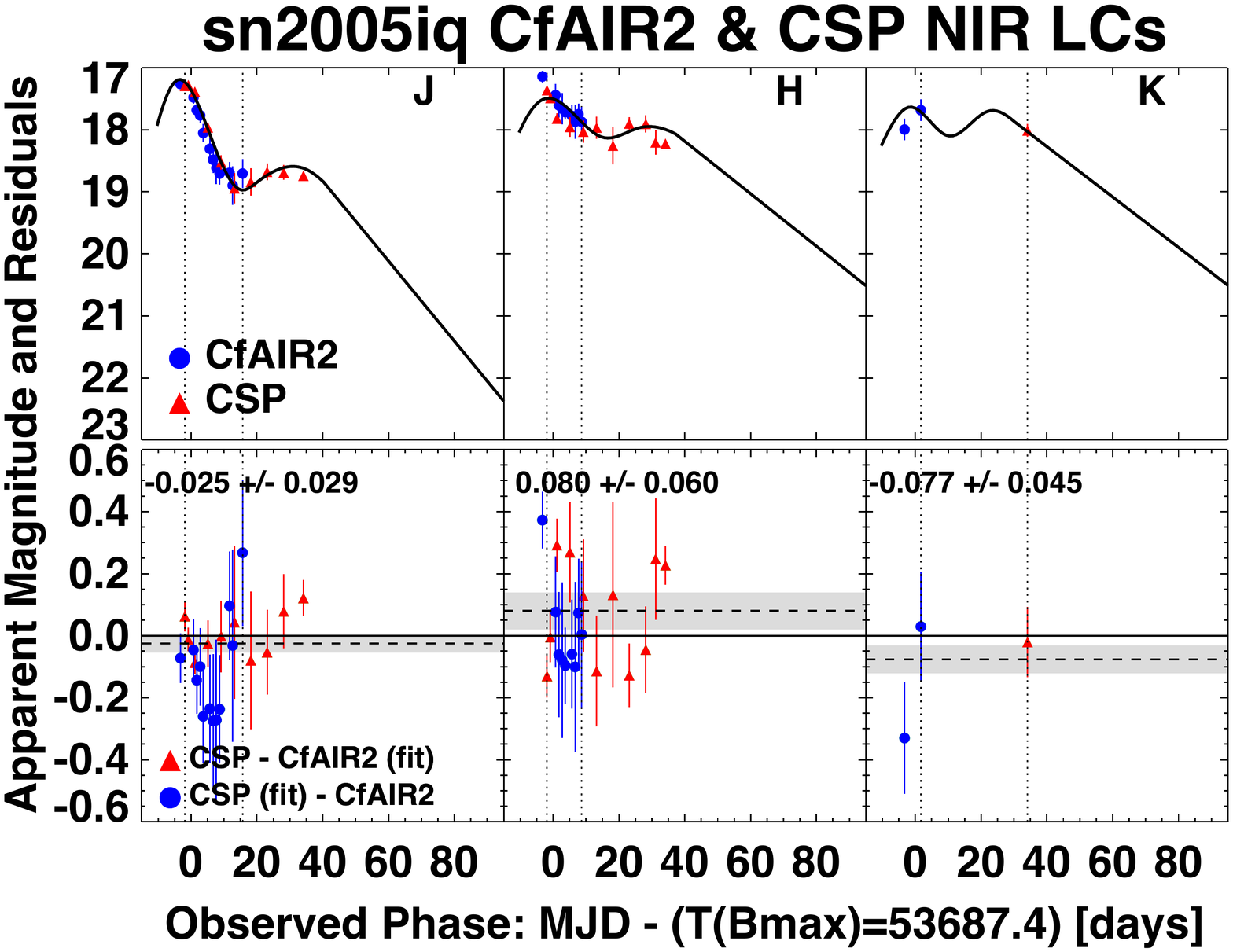} \\

\includegraphics[width=\scale\linewidth,angle=0]
{\colordir/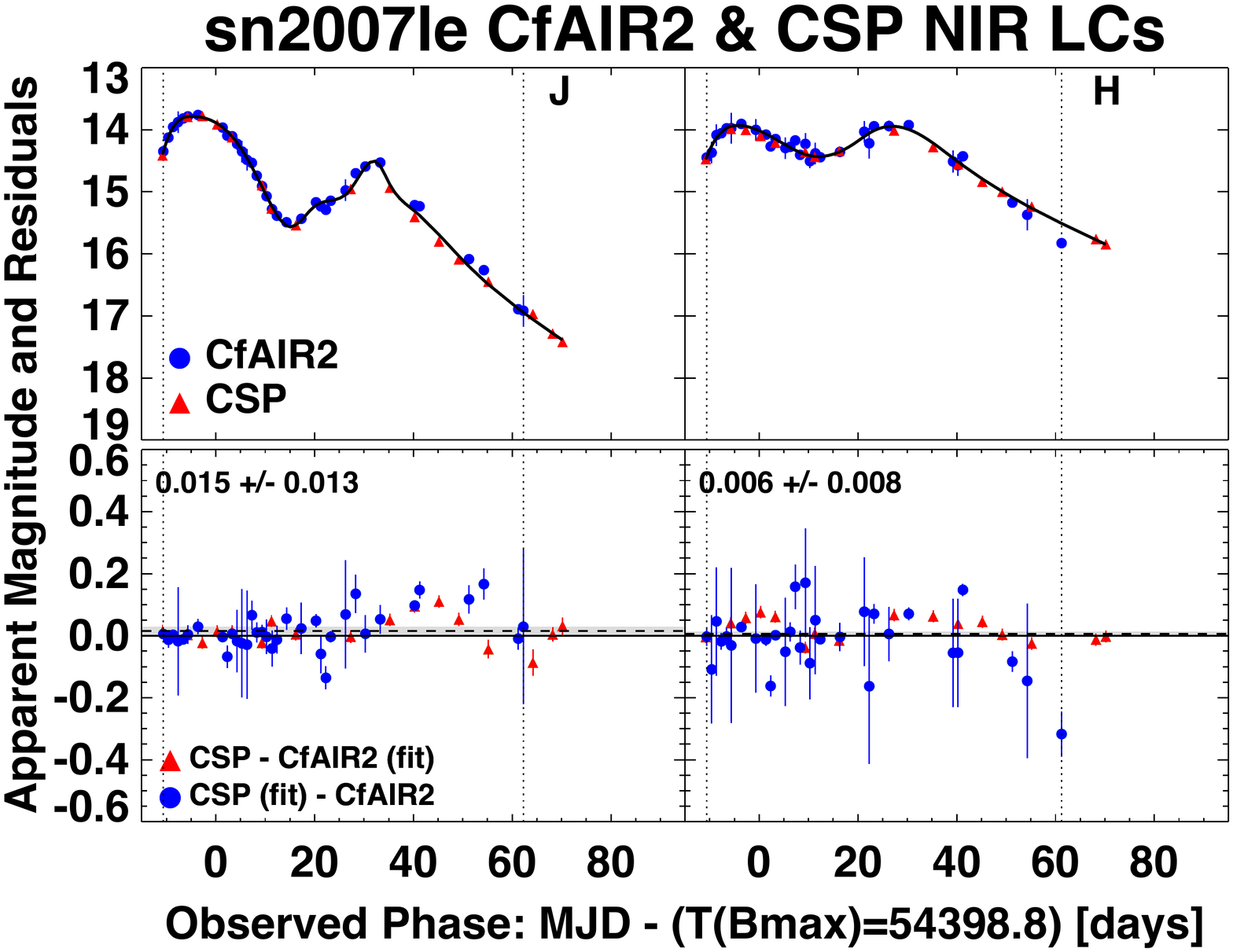} &

\includegraphics[width=\scale\linewidth,angle=0]
{\colordir/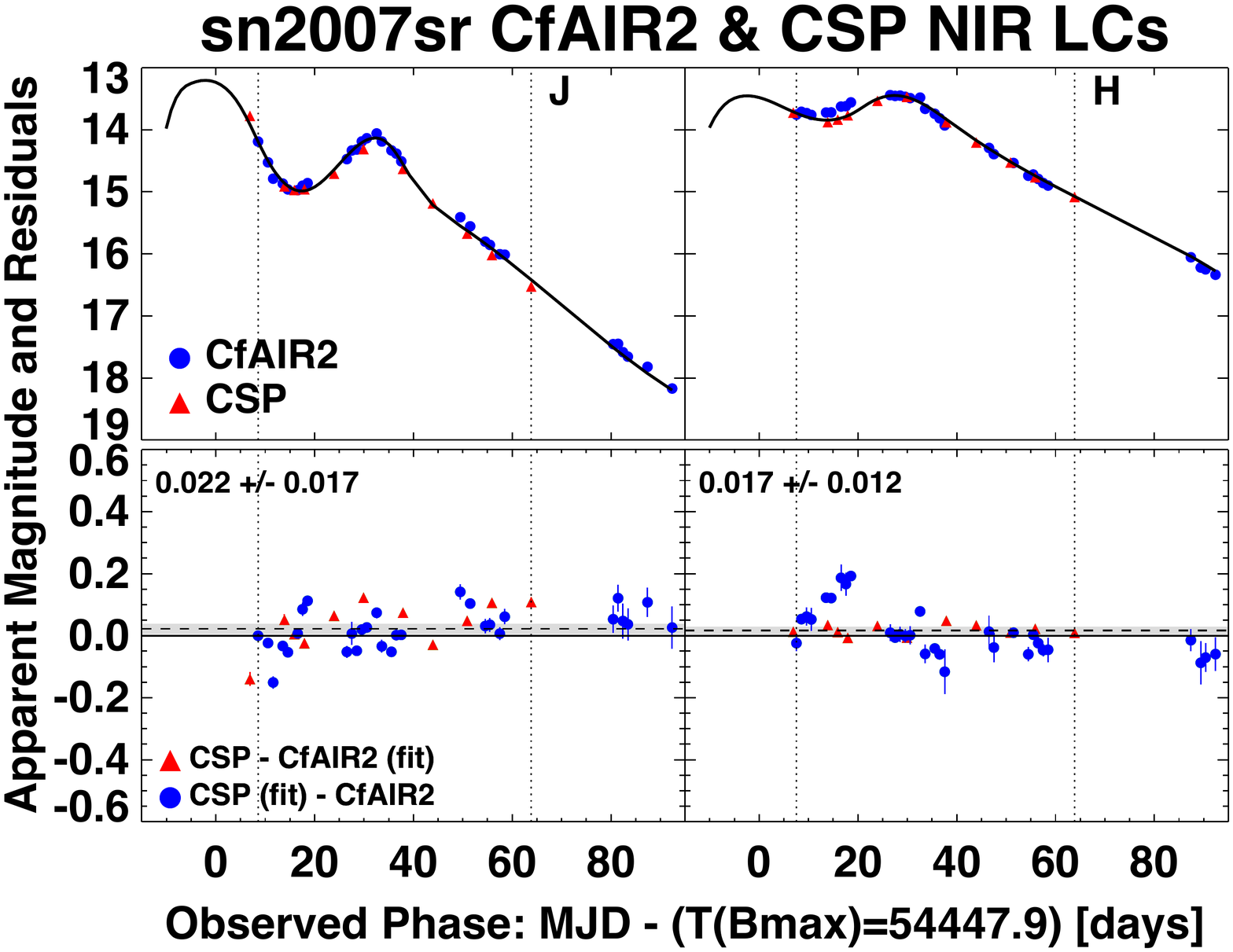} \\

\end{tabular}
\caption[\F{}/CSP Photometry Comparison]{
Comparing \F{} to CSP Photometry 
\\
\\
\tiny
(Color online) 
{\bf (Top panels)} Plot shows 6 example NIR \snIa{} LCs out of the \nptelcsp{} \F{} objects observed by both \PTL{} and CSP.
\jhk{} \snIa{} LCs are shown from PAIRITEL \F{} galaxy subtracted photometry (blue circles) and CSP LCs (red triangles) after applying color terms from \eq{}~\ref{eq:friedman_colorterms} of this paper (see \S\ref{sec:ptel_csp_color_terms}).  Vertical dotted lines show regions of temporal overlap for both LCs. The black line is a cubic spline model fit to the joint PAIRITEL+CSP data with a simple linear fit applied $\gtrsim 30$-$40$ days in specific cases.  For normal \snIa{}, the \WV{} mean template LC is used to help fit for missing data.   
CSP $K_s$-band is missing for some \snIa{} (e.g., \sn{}2007le and \sn{}2007sr).  
\\
\\
{\bf (Bottom panels)} CSP - \F{} residuals are computed as either (CSP data minus \F{} joint model fit) or (CSP joint model fit - \F{} data) for each epoch, using the same plot symbols as above for differences computed using CSP or \F{} data. While the CSP (fit) - \F{} residuals (blue circles) are above the zero residual line when the corresponding \F{} data point has a larger magnitude value than the joint model fit in the top row panels, since we are computing CSP - \F{} residuals, the CSP - \F{} (fit) (red triangles) residuals behave in the opposite sense. For example, when the CSP data has a larger magnitude than the joint model fit in the top row panels, the corresponding residual lies {\it below} the zero residual line.
Weighted mean residuals and 1-$\sigma$ uncertainties for CSP - \F{} data in the phase range $[-10,60]$ days, as listed in Table~\ref{tab:ptel_csp_overlap}, are also shown in the upper left corner of each panel and indicated by the dashed line and the gray strip, respectively.
}
\label{fig:f12_csp2}
\end{figure*}

\renewcommand{\scale}{0.5}
\begin{figure*}
\centering
\begin{tabular}{@{}c@{}c@{}}

\includegraphics[width=\scale\linewidth,angle=0]
{\colordir/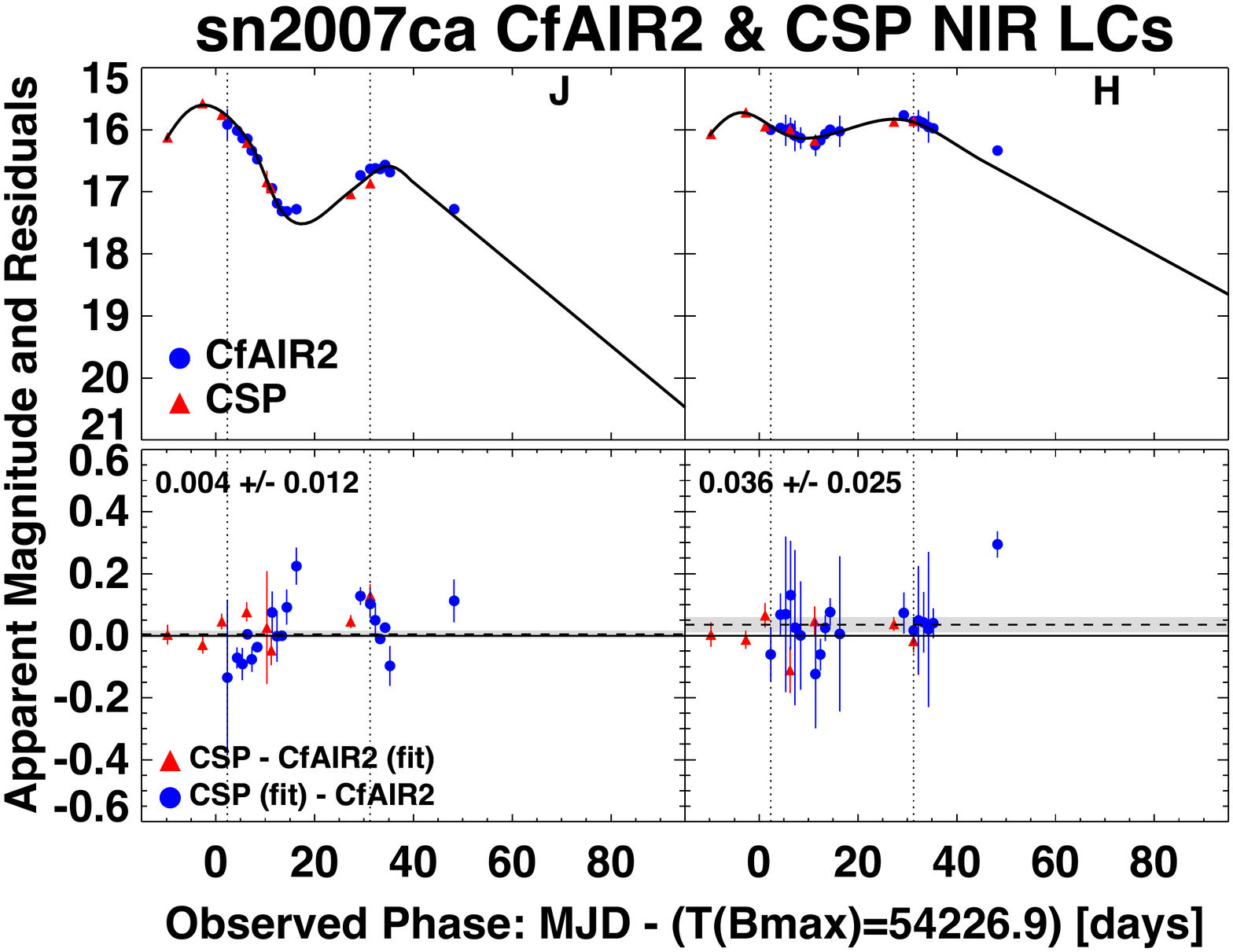} &

\includegraphics[width=\scale\linewidth,angle=0]
{\colordir/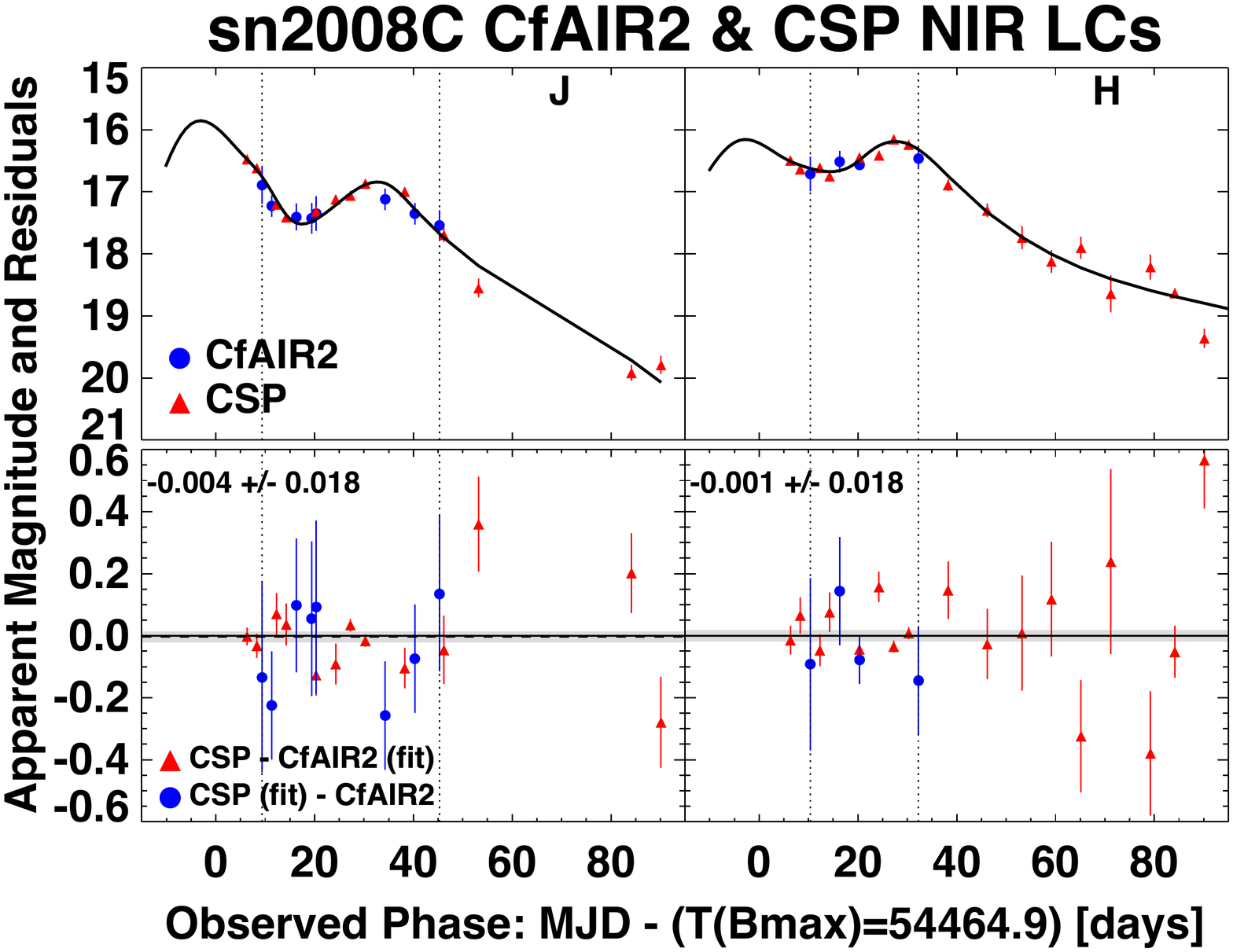} \\

\includegraphics[width=\scale\linewidth,angle=0]
{\colordir/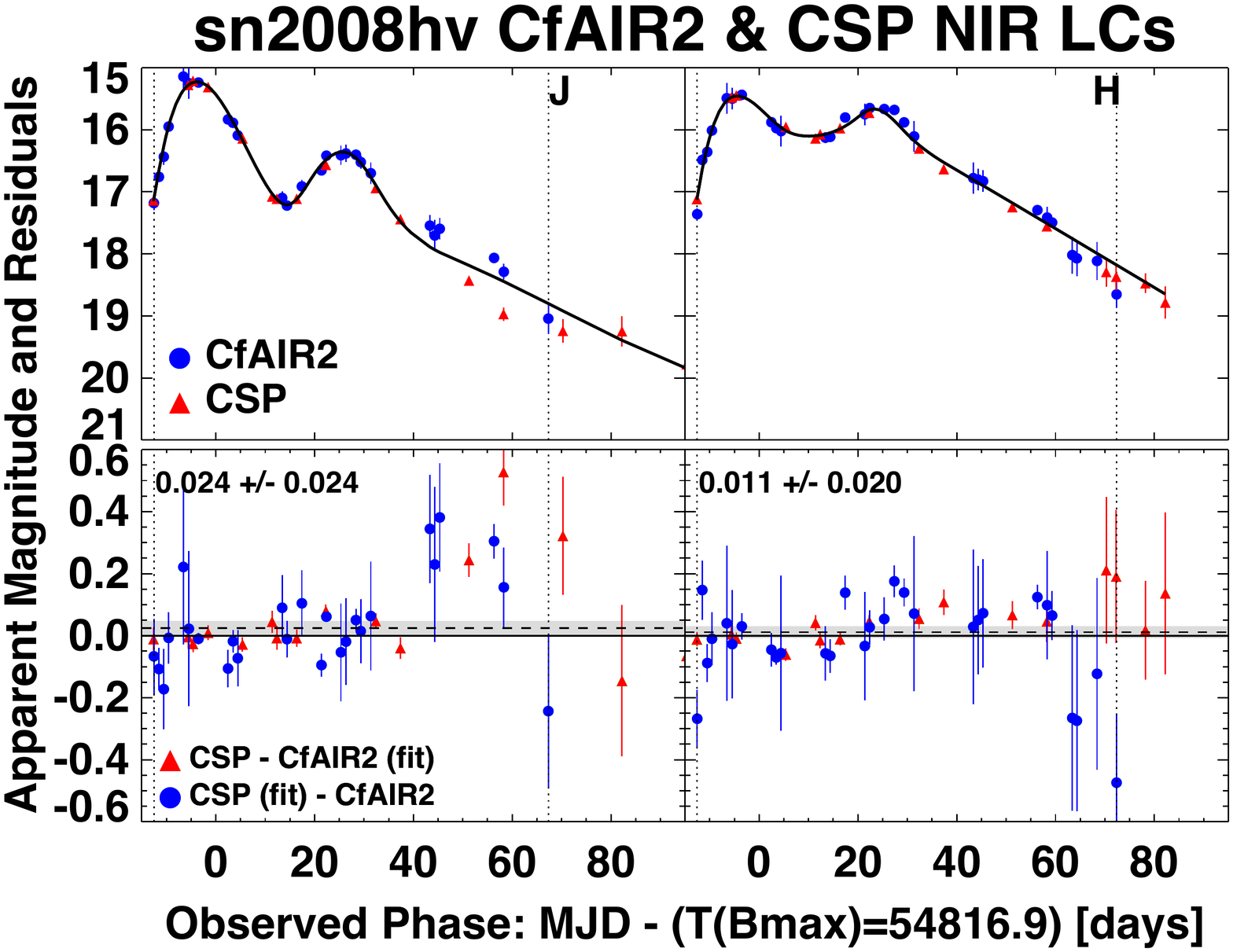} &

\includegraphics[width=\scale\linewidth,angle=0]
{\colordir/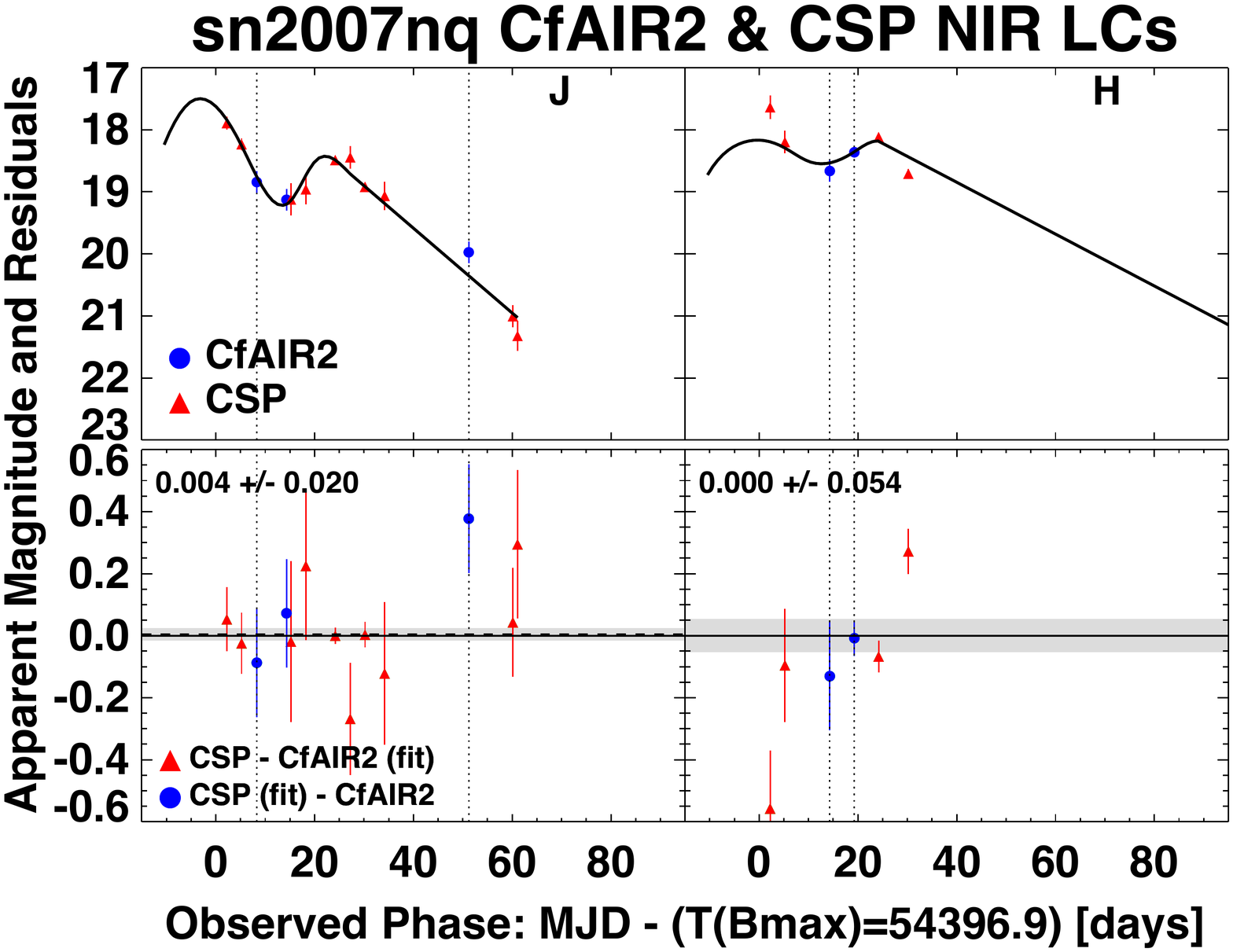} \\

\includegraphics[width=\scale\linewidth,angle=0]
{\colordir/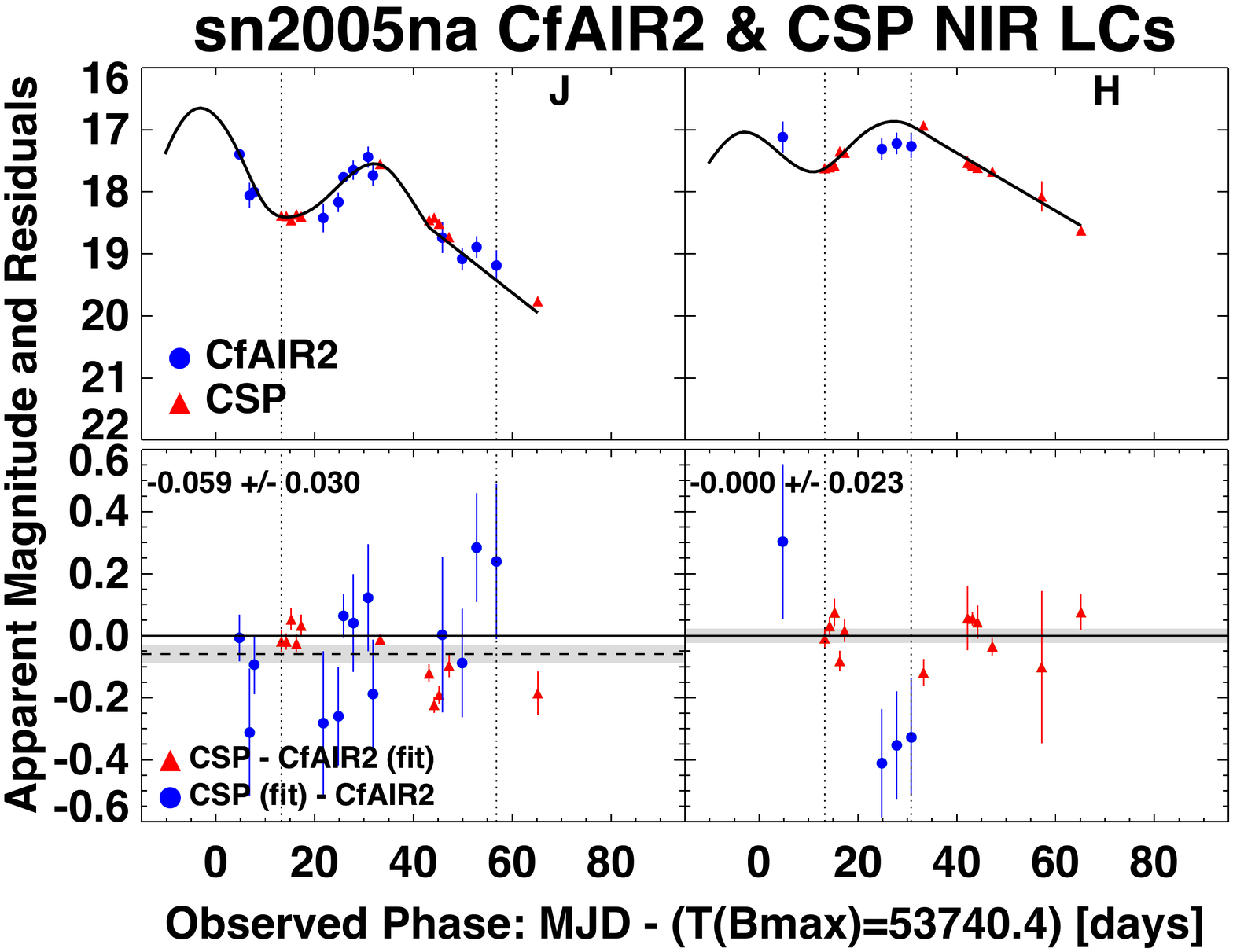} &

\includegraphics[width=\scale\linewidth,angle=0]
{\colordir/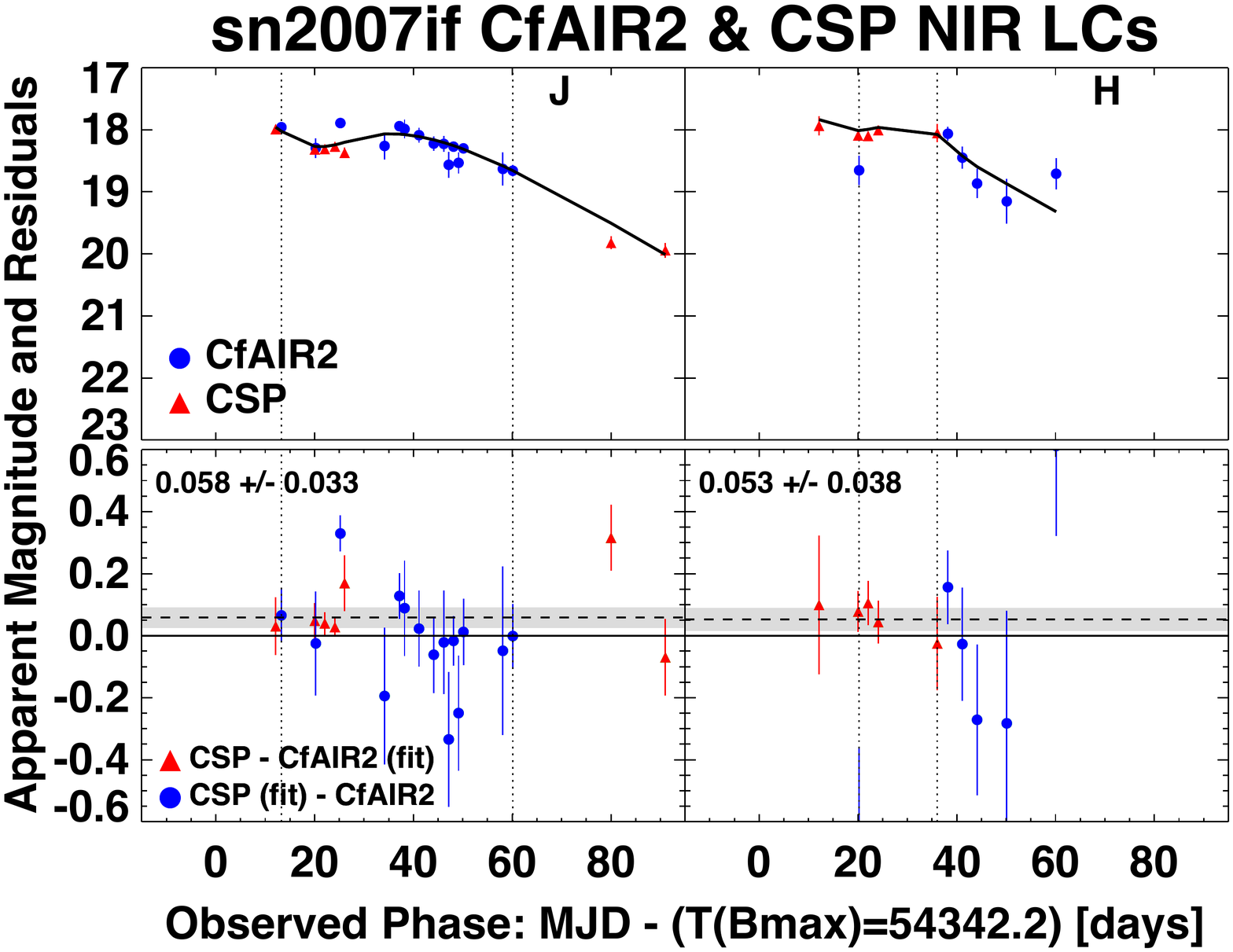} \\

\end{tabular}
\caption[\F{}/CSP Photometry Comparison]{
Comparing \F{} to CSP Photometry 
\\
\\
\tiny
(Color online) 
{\bf (Top panels)} Plot shows 6 example NIR \snIa{} LCs out of the \nptelcsp{} \F{} objects observed by both \PTL{} and CSP.
\jhk{} \snIa{} LCs are shown from PAIRITEL \F{} galaxy subtracted photometry (blue circles) and CSP LCs (red triangles) after applying color terms from \eq{}~\ref{eq:friedman_colorterms} of this paper (see \S\ref{sec:ptel_csp_color_terms}).  Vertical dotted lines show regions of temporal overlap for both LCs. The black line is a cubic spline model fit to the joint PAIRITEL+CSP data with a simple linear fit applied $\gtrsim 30$-$40$ days in specific cases.  For normal \snIa{}, the \WV{} mean template LC is used to help fit for missing data.   
CSP $K_s$-band is missing for all the above SN.  
\\
\\
{\bf (Bottom panels)} CSP - \F{} residuals are computed as either (CSP data minus \F{} joint model fit) or (CSP joint model fit - \F{} data) for each epoch, using the same plot symbols as above for differences computed using CSP or \F{} data. While the CSP (fit) - \F{} residuals (blue circles) are above the zero residual line when the corresponding \F{} data point has a larger magnitude value than the joint model fit in the top row panels, since we are computing CSP - \F{} residuals, the CSP - \F{} (fit) (red triangles) residuals behave in the opposite sense. For example, when the CSP data has a larger magnitude than the joint model fit in the top row panels, the corresponding residual lies {\it below} the zero residual line.
Weighted mean residuals and 1-$\sigma$ uncertainties for CSP - \F{} data in the phase range $[-10,60]$ days, as listed in Table~\ref{tab:ptel_csp_overlap}, are also shown in the upper left corner of each panel and indicated by the dashed line and the gray strip, respectively.
}
\label{fig:f12_csp3}
\end{figure*}

We can also test whether \F{} and CSP are consistent for the entire overlap sample, rather than just individual objects. \fig{}~\ref{fig:f12_csp_res_agg} shows aggregated residuals in the phase range $[-15,100]$ days after applying color terms from \eq{}~\ref{eq:friedman_colorterms} to the CSP data. Using 433, 390, and 218 \F{} LC points, and 275, 257, and 42 CSP LC points, each in \jhk{}, respectively, we find the global weighted mean of the aggregated residuals is consistent with zero in each case (see \fig{}~\ref{fig:f12_csp_res_agg}).  Applying color terms from C10 (or no color terms) did not affect the results.  We conclude that both for individual LCs and for the global aggregated sample, \PTL{} \F{} photometry and CSP photometry show satisfactory overall agreement.

\renewcommand{\scale}1}
\begin{figure}
\centering
\begin{tabular}{@{}c@{}}

\includegraphics[width=\scale\linewidth,angle=0]
{\colordir/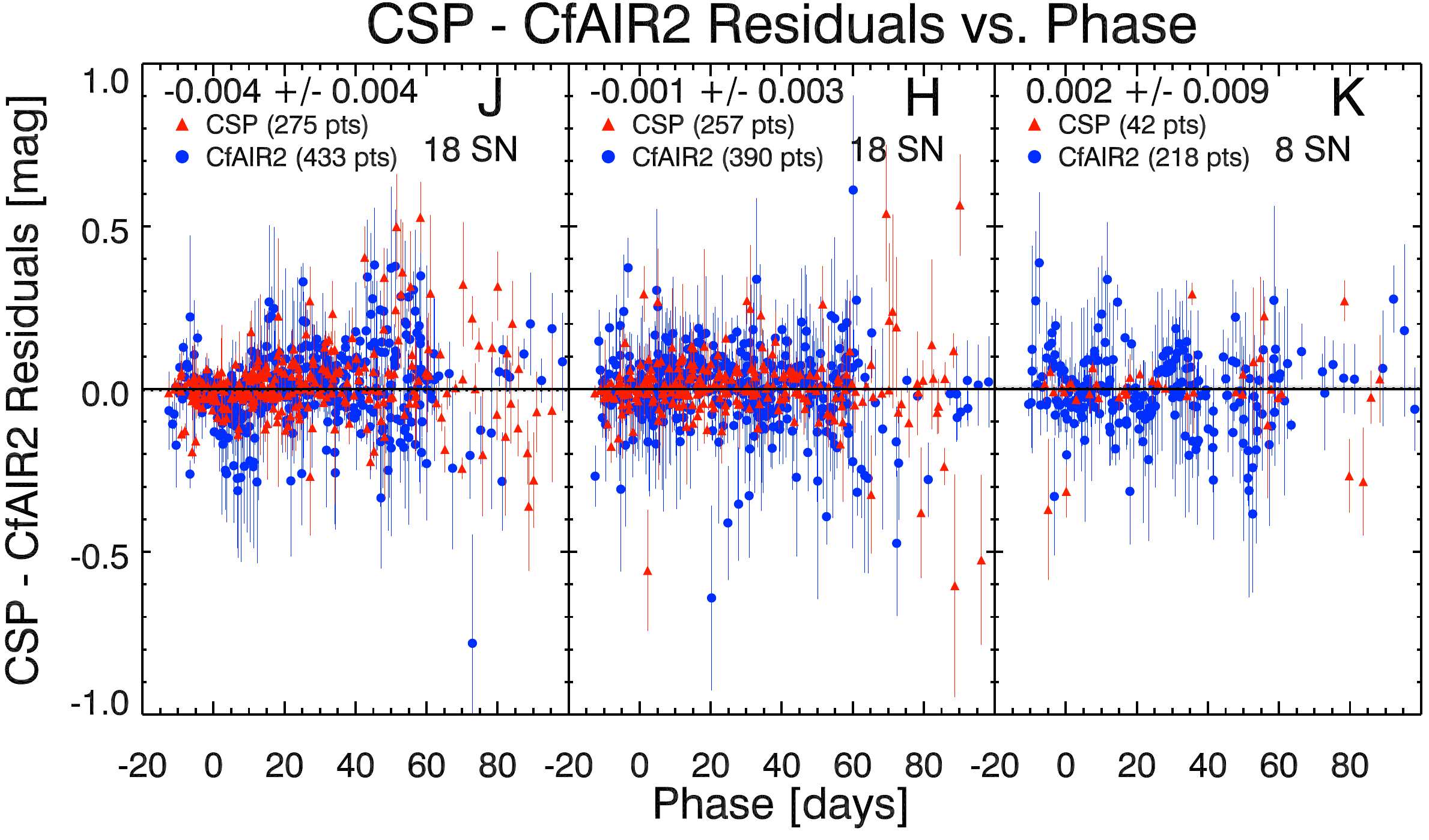}  \\

\end{tabular}
\vspace{-0.2cm}
\caption[\F{}/CSP Aggregated Residuals]{\F{}/CSP Aggregated Residuals
\\
\\
\scriptsize
(Color online) 
Aggregated residuals and errors from LC model fits in \S\ref{sec:csp_comp1}, Figs.~\ref{fig:f12_csp}-\ref{fig:f12_csp3}, for CSP (red filled triangles) and \F{} (blue filled circles) data from $[-15,100]$ days after applying the color terms from \eq{}~\ref{eq:friedman_colorterms} to CSP data. Outlier residuals from bad fits were removed with conservative 10-$\sigma$ clipping.  
There are 18 SN with joint $JH$ data and 8 with $K_s$ data.  Aggregated residuals include the following number of data points for \F{}: 433, 390, and 218, and CSP: 275, 257, and 42, in \jhk{}, respectively. The weighted means of the aggregated CSP - \F{} residuals are $-0.004 \pm 0.004$, $-0.001 \pm 0.003$, and $0.002 \pm 0.009$ for \jhk{}, respectively. Applying the C10 color terms from \eq{}~\ref{eq:contreras_colorterms} or applying no color terms had a negligible effect on the results. In all cases, differences between the \jhk{} CSP and \F{} global weighted mean residuals have absolute values of only $\sim0.001-0.004$ mag, and are consistent with zero to within 1-$\sigma$, where the 1-$\sigma$ error is given by the standard error on the mean. 
\PTL{} \F{} data thus show excellent global agreement with CSP.
}
\label{fig:f12_csp_res_agg}
\end{figure}

\section{Final \F{} Data Set}
\label{sec:results}

\renewcommand{\arraystretch}{0.5}
\renewcommand{\tabcolsep}{1pt}
\begin{table}
\begin{center}
\caption[\PTL{} \F{} \jhk{} Photometry]{\PTL{} \F{} \jhk{} Photometry (Stub)} 
\tiny
\begin{tabular}{@{}ccccccrrll@{}}
\hline
\multicolumn{1}{c}{SN}              & \multicolumn{1}{c}{Type} & \multicolumn{1}{c}{Telescope}     & \multicolumn{1}{c}{Band}   & \multicolumn{1}{c}{Date}                &  \multicolumn{1}{c}{MJD}                                 & \multicolumn{1}{c}{$f_{25}$} & \multicolumn{1}{c}{$\sigma_{f_{25}}$} & \multicolumn{1}{c}{\jhk}                                   & \multicolumn{1}{c}{$\sigma_{\jhk}$} \\
                    &           &                         &              &                        &  \multicolumn{1}{c}{[days]\tablenotemark{a}} &  \multicolumn{1}{c}{\tablenotemark{b}}     & \multicolumn{1}{c}{\tablenotemark{c}}       & \multicolumn{1}{c}{[mag]\tablenotemark{d}} & \multicolumn{1}{c}{[mag]\tablenotemark{d}} \\
\hline
\sn{}2005ao    & Ia      & PAIRITEL     & $J$         & 2005Mar22    & 53451.48    & 227.592     & 17.306       & 19.11    & 0.08    \\
\sn{}2005ao    & Ia      & PAIRITEL     & $J$         & 2005Apr02    & 53462.51    & 255.056     & 21.694       & 18.98    & 0.09    \\
\sn{}2005ao    & Ia      & PAIRITEL     & $J$         & 2005Apr04    & 53464.39    & 263.369     & 29.603       & 18.95    & 0.12    \\
\sn{}2005ao    & Ia      & PAIRITEL     & $J$         & 2005Apr05    & 53465.39    & 266.528     & 72.947       & 18.94    & 0.30    \\
\sn{}2005ao    & Ia      & PAIRITEL     & $J$         & 2005Apr07    & 53467.39    & 311.257     & 40.449       & 18.77    & 0.14    \\
\sn{}2005ao    & Ia      & PAIRITEL     & $J$         & 2005Apr09    & 53469.42    & 341.932     & 12.230       & 18.67    & 0.04    \\
\sn{}2005ao    & Ia      & PAIRITEL     & $J$         & 2005Apr10    & 53470.38    & 343.194     & 25.402       & 18.66    & 0.08    \\
\sn{}2005ao    & Ia      & PAIRITEL     & $J$         & 2005Apr11    & 53471.38    & 395.464     & 65.052       & 18.51    & 0.18    \\
\sn{}2005ao    & Ia      & PAIRITEL     & $J$         & 2005Apr20    & 53480.35    & 259.901     & 17.128       & 18.96    & 0.07    \\
\sn{}2005ao    & Ia      & PAIRITEL     & $H$         & 2005Mar22    & 53451.48    & 535.150     & 44.485       & 18.18    & 0.09    \\
\sn{}2005ao    & Ia      & PAIRITEL     & $H$         & 2005Apr02    & 53462.51    & 416.466     & 50.697       & 18.45    & 0.13    \\
\sn{}2005ao    & Ia      & PAIRITEL     & $H$         & 2005Apr04    & 53464.39    & 393.065     & 120.604      & 18.51    & 0.34    \\
\sn{}2005ao    & Ia      & PAIRITEL     & $H$         & 2005Apr05    & 53465.39    & 475.528     & 75.989       & 18.31    & 0.18    \\
\sn{}2005ao    & Ia      & PAIRITEL     & $H$         & 2005Apr07    & 53467.39    & 526.212     & 113.705      & 18.20    & 0.24    \\
\sn{}2005ao    & Ia      & PAIRITEL     & $H$         & 2005Apr09    & 53469.42    & 596.101     & 72.917       & 18.06    & 0.13    \\
\sn{}2005ao    & Ia      & PAIRITEL     & $H$         & 2005Apr10    & 53470.38    & 695.897     & 83.084       & 17.89    & 0.13    \\
\sn{}2005ao    & Ia      & PAIRITEL     & $H$         & 2005Apr13    & 53473.36    & 713.816     & 114.068      & 17.87    & 0.18    \\
\sn{}2005ao    & Ia      & PAIRITEL     & $K_s$       & 2005Mar22    & 53451.48    & 833.517     & 126.880      & 17.70    & 0.17    \\
\sn{}2005ao    & Ia      & PAIRITEL     & $K_s$       & 2005Mar27    & 53456.43    & 723.626     & 127.287      & 17.85    & 0.19    \\
\sn{}2005ao    & Ia      & PAIRITEL     & $K_s$       & 2005Apr02    & 53462.51    & 622.584     & 126.942      & 18.01    & 0.22    \\
\sn{}2005ao    & Ia      & PAIRITEL     & $K_s$       & 2005Apr04    & 53464.39    & 550.997     & 88.049       & 18.15    & 0.18    \\
\sn{}2005ao    & Ia      & PAIRITEL     & $K_s$       & 2005Apr06    & 53466.39    & 862.798     & 125.926      & 17.66    & 0.16    \\
\sn{}2005ao    & Ia      & PAIRITEL     & $K_s$       & 2005Apr09    & 53469.42    & 871.012     & 138.486      & 17.65    & 0.17    \\
\sn{}2005ao    & Ia      & PAIRITEL     & $K_s$       & 2005Apr10    & 53470.38    & 1004.776    & 132.201      & 17.49    & 0.14    \\
\sn{}2005ao    & Ia      & PAIRITEL     & $K_s$       & 2005Apr11    & 53471.38    & 776.477     & 73.523       & 17.77    & 0.10    \\
\sn{}2005ao    & Ia      & PAIRITEL     & $K_s$       & 2005Apr13    & 53473.36    & 354.654     & 56.674       & 18.63    & 0.18    \\
\sn{}2005ao    & Ia      & PAIRITEL     & $K_s$       & 2005Apr20    & 53480.35    & 446.927     & 102.060      & 18.37    & 0.25    \\
\hline
\end{tabular}
\tablecomments{ \tiny 
\\
({\bf A full machine-readable Table is available online in the electronic version of this paper. A portion is shown here for guidance}).
\\
\scriptsize
{\bf (a)} Modified Julian Date.
\\
{\bf (b)} $f_{25}$: Flux normalized to a magnitude of 25.  $JHK_s$ mag = $-2.5$ log$_{10}$($f_{25}$) + $25$ mag.
\\
{\bf (c)} $\sigma_{f_{25}}$: Symmetric 1-$\sigma$ error on $f_{25}$, computed as the error weighted standard deviation of the flux measurements for each host galaxy subtraction on a given night, weighted by photometric errors corrected for \dophot{} underestimates. See Table~\ref{tab:nnt_err} and Appendix \ref{sec:nnt_math}. 
\\
$\sigma_{JHK_s}$ mag $ = [   -2.5 {\rm log}_{10}$($f_{25} - \sigma_{f_{25}}$) $+ 2.5 {\rm log}_{10}$($f_{25} + \sigma_{f_{25}}$)  ]/2.
\\
{\bf (d)} \jhk{} magnitude and 1-$\sigma$ uncertainty.
}
\label{tab:cfair2_Ia_lcs}
\end{center}
\end{table}
\renewcommand{\tabcolsep}{6pt}
\renewcommand{\arraystretch}{1}

\renewcommand{\scale}{1}
\begin{figure*}
\centering
\begin{tabular}{@{}c@{}}

\includegraphics[width=\scale\linewidth,angle=0]
{\colordir/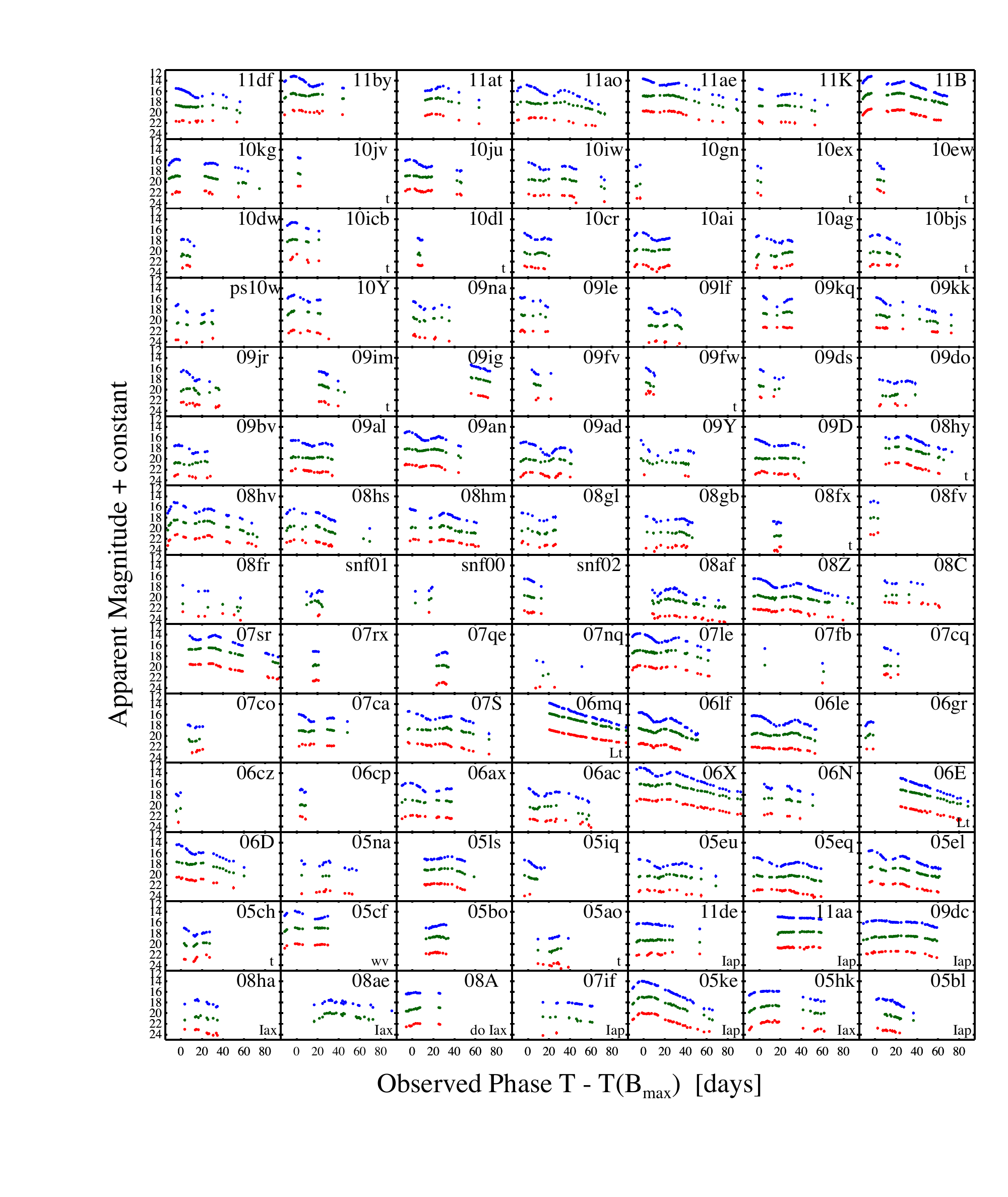} \\
\end{tabular}
\vspace{-0.1cm}
\caption
[\PTL{} \F{} NIR LCs: \nsnIacfair{} \snIa{} and \nsnIaxcfair{} \snIax{}]
{\PTL{} \F{} NIR LCs: \nsnIacfair{} \snIa{} and \nsnIaxcfair{} \snIax{}
\\
\\
\scriptsize
(Color online)
\nsnIacfair{} \F{} NIR \snIa{} and 4 \snIax{} LCs. Data points in magnitudes are shown for $J$ (blue), $H + 3$ (green), and $K_s + 6$ (red).  Uncertainties are comparable to the sizes of the plot symbols.  Plots are for the \nsnIanorm{} spectroscopically normal \snIa{} {\it except} for 6 peculiar \snIa{} and \nsnIaxcfair{} \snIax{} (also see \fig{}~\ref{fig:f12_Ia_pec_oplot_templ}) marked in the lower right of each panel with {\bf Iap} or {\bf Iax}, which are displayed last starting with \sn{}2011de.  
\\
\\
\tiny
See notes below for the lower right corner of some LC plots: 
\\
{\bf t}: \tbmax{} estimated from optical spectra and cross checked with NIR LC features in lieu of early-time optical photometry (see Table~\ref{tab:jhkdata}).
\\
{\bf Lt}: \sn{}2006E and \sn{}2006mq were discovered late, so lack precise \tbmax{} estimates (see Table~\ref{tab:jhkdata}). 
\\
{\bf Iap}: Peculiar objects, which clearly differ from the mean \jhk{} LC templates (see \fig{}~\ref{fig:f12_Ia_pec_oplot_templ}).
\\ 
{\bf Iax}: See \citealt{foley13} for a description of this distinct class of objects.
\\
{\bf wv}: \sn{}2005cf is included in \F{} but uses the same forced \dophot{} LC as in \WV{}, without host galaxy subtraction.
\\
{\bf do}: \sn{}2008A used forced \dophot{} photometry, not the \nnt{} host galaxy subtraction used for all other \F{} LCs except \sn{}2005cf.
 }
\label{fig:f12_Ia_all}
\end{figure*}

\renewcommand{\scale}{1}
\begin{figure}
\centering
\begin{tabular}{@{}c@{}}

\includegraphics[width=\scale\linewidth,angle=0]
{\colordir/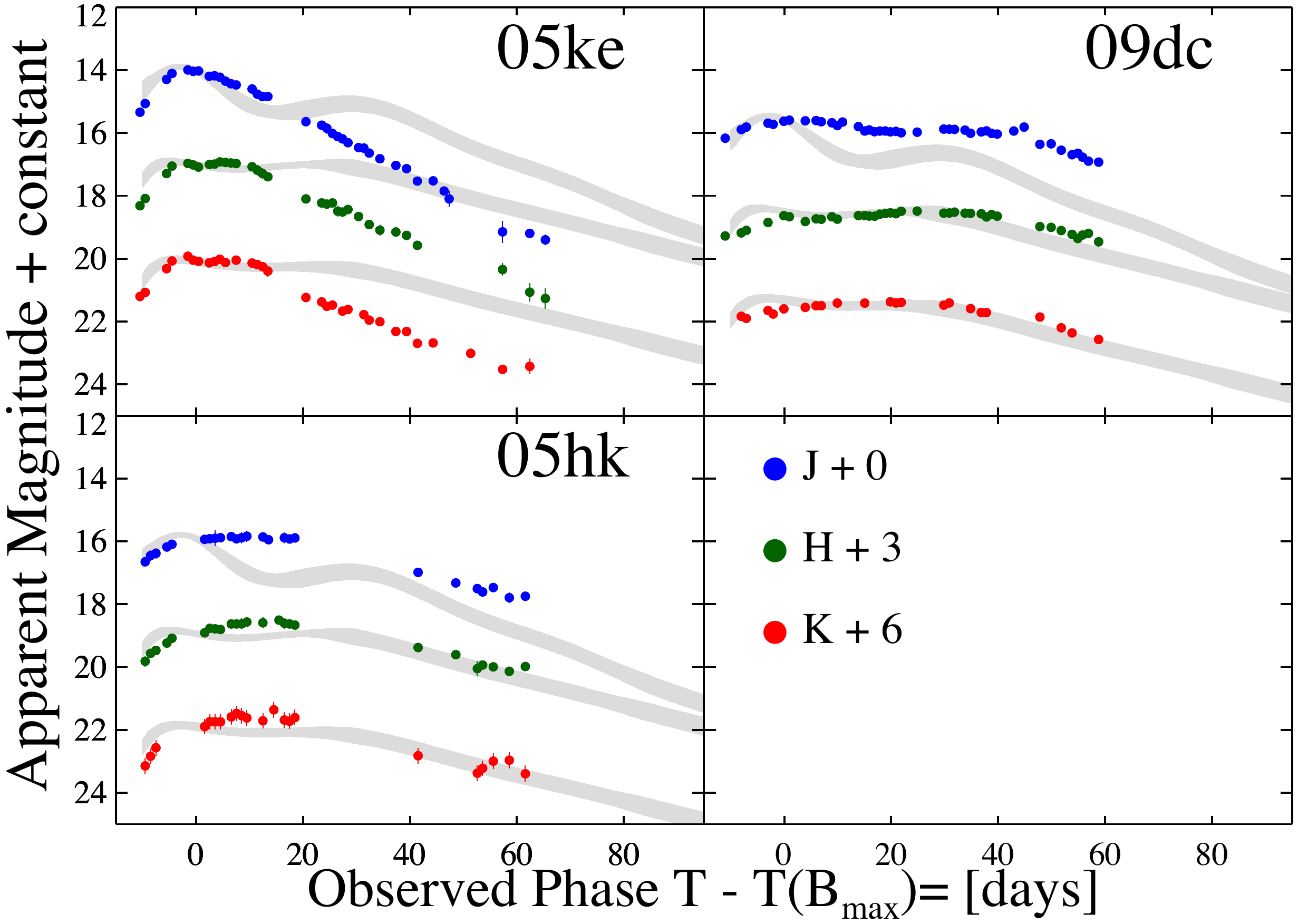} \\

\end{tabular}
\caption
[Peculiar \snIa{} or \snIax{} NIR LC Morphology]
{Peculiar \snIa{} or \snIax{} NIR LC Morphology 
\\
\\
\scriptsize
(Color online)
\F{} NIR LCs of 2 peculiar \snIa{} (\sn{}2005ke, \sn{}2009dc), and 1 \snIax{} (\sn{}2005hk) with the \WV{} mean \jhk{} LC templates for spectroscopically normal \snIa{} overplotted. Such objects can easily be distinguished from normal \snIa{} based on NIR LC morphology alone.
}
\label{fig:f12_Ia_pec_oplot_templ}
\end{figure}

Final, host galaxy subtracted \jhk{} LCs for \nsnIacfair{} spectroscopically normal and peculiar \F{} \snIa{} and \nsnIaxcfair{} \snIax{} are presented in \fig{}~\ref{fig:f12_Ia_all} and Table~\ref{tab:cfair2_Ia_lcs}.\footnote{Only \snIax{} \sn{}2008A and the \sn{}2005cf LC from \WV{} used forced \dophot{} photometry, without galaxy subtraction.} No K-corrections or Milky Way dust extinction corrections have been applied to the final \F{} LCs (see \S\ref{sec:disc}). \PTL{} flux and magnitude measurements and errors are listed in Table~\ref{tab:cfair2_Ia_lcs} (see \S\ref{sec:galsub_syserr}). \fig{}~\ref{fig:f12_Ia_pec_oplot_templ} shows \F{} data for 2 peculiar \snIa{} and 1 \snIax{} with the \WV{} mean LC template shown to emphasize how easily these objects can be distinguished from normal \snIa{} using NIR LC shape alone.
A new mean normal \snIa{} NIR LC template using \F{} and literature data will be presented elsewhere.  Preliminary results show that the mean template using only \F{} data is very similar to the \WV{} template. We thus felt the \WV{} template LC was sufficient for the purposes of this work, where it was used only to help fit  \PTL{} and CSP LCs for comparing normal \snIa{} (\S\ref{sec:csp_comp}) and to provide a visual comparison to peculiar objects (\fig{}~\ref{fig:f12_Ia_pec_oplot_templ}).

Table~\ref{tab:jhkdata} shows fits of the observed \jhk{} properties for \nsnIanorm{} 
\F{} spectroscopically normal \snIa{}. 
We determined \tbmax{} and the LC shape parameter $\Delta$ using \texttt{MLCS2k2.v007}~\citep{jha07} fits to our own CfA optical CCD observations~(\citealt{hicken09,hicken09a,hicken09b,hicken12}, \CFAFIVE{}) combined with other optical data from the literature where available (e.g., \citealt{ganeshalingam10}; C10; S11) and approximate \tbmax{} estimates from optical spectra in discovery and follow up notices as needed (See Table~\ref{tab:jhkdata}). Table~\ref{tab:jhkdata} also lists the CMB frame redshift, $z_{\rm CMB}$, the \jhk{} apparent magnitudes at the brightest LC point, and the number of epochs in each LC. 

Note that the \jhk{} magnitudes listed in Table~\ref{tab:jhkdata} are {\it not} necessarily the apparent magnitudes at \tbmax{} or the relevant NIR first peaks, but simply the apparent magnitude of the brightest observed data point, which is very sensitive to data coverage. Also note that the $z_{\rm CMB}$ values in Table~\ref{tab:jhkdata} have {\it not} been corrected for any local flow models, which would provide more accurate redshift estimates for objects with $z_{\rm CMB} \lesssim 0.01$ ($\lesssim$ 3000 km s$^{-1}$).  The apparent magnitudes and redshifts in Table~\ref{tab:jhkdata} should thus not be naively used to estimate galaxy distances or naively combined with high-redshift data to estimate cosmological parameters. 

\renewcommand{\arraystretch}{0.0001}
\renewcommand{\tabcolsep}{3pt}
\begin{table*}
\begin{center}
\caption[\jhk{} Light Curve Properties for \nsnIanorm{} \PTL{} \F{} \snIa{}]{\jhk{} Light Curve Properties for \nsnIanorm{} Spectroscopically Normal \PTL{} \F{} \snIa{} \\} 
\tiny
\begin{tabular}{@{}llrcrrrrrrrrrrrrrr@{}}
\hline
SN & \tbmax{}\tablenotemark{a} & $\sigma_{t_{B}}$\tablenotemark{a} & Optical Ref.\tablenotemark{b} &
$z_{\rm CMB}$\tablenotemark{c} & $\sigma_{z_{\rm CMB}}$\tablenotemark{c} & 
$\Delta$\tablenotemark{d} & $\sigma_{\Delta}$\tablenotemark{d} & 
$J_p$\tablenotemark{e} & $\sigma_{J_p}$\tablenotemark{e} & $H_p$\tablenotemark{e} & $\sigma_{H_p}$\tablenotemark{e} & ${K}_p$\tablenotemark{e} & $\sigma_{{K}_p}$\tablenotemark{e} & 
$N_J$\tablenotemark{f} & $N_H$\tablenotemark{f} & $N_{K}$\tablenotemark{f}
\\
\hline
SN2005ao           & 53442       & 2       & IAUC 8492                 & 0.03819    & 0.00099    &  \nodata     &  \nodata     & 18.51    & 0.18    & 17.87    & 0.18    & 17.49    & 0.14    & 9     & 8     & 10    \\
SN2005bo           & 53477.99    & 0.53    & C10, G10                  & 0.01502    & 0.00108    & 0.146        & 0.060        & 16.33    & 0.04    & 15.55    & 0.07    & 15.53    & 0.12    & 15    & 16    & 13    \\
SN2005cf           & 53533.56    & 0.11    & CfA3, Pa07a, WX09, G10    & 0.00702    & 0.00109    & -0.108       & 0.032        & 13.83    & 0.04    & 13.96    & 0.02    & 13.97    & 0.02    & 17    & 17    & 15    \\
SN2005ch           & 53536       & 3       & CBET 167                  & 0.02782    & 0.00103    &  \nodata     &  \nodata     & 17.02    & 0.03    & 16.79    & 0.08    & 16.05    & 0.06    & 13    & 11    & 8     \\
SN2005el           & 53646.33    & 0.17    & CfA3, G10                 & 0.01490    & 0.00100    & 0.256        & 0.044        & 15.46    & 0.03    & 15.54    & 0.05    & 15.24    & 0.05    & 35    & 34    & 24    \\
SN2005eq           & 53653.73    & 0.19    & CfA3, G10                 & 0.02837    & 0.00098    &  \nodata     &  \nodata     & 16.83    & 0.03    & 17.08    & 0.06    & 16.77    & 0.18    & 31    & 33    & 29    \\
SN2005eu           & 53659.70    & 0.16    & CfA3, G10                 & 0.03334    & 0.00098    & -0.153       & 0.039        & 17.14    & 0.07    & 16.96    & 0.10    & 16.81    & 0.17    & 23    & 23    & 14    \\
SN2005iq           & 53687.45    & 0.22    & CfA3, C10                 & 0.03191    & 0.00097    & 0.157        & 0.049        & 17.26    & 0.08    & 17.14    & 0.09    & 17.68    & 0.18    & 12    & 9     & 2     \\
SN2005ls           & 53714       & 2       & CfA3                      & 0.02051    & 0.00097    &  \nodata     &  \nodata     & 16.61    & 0.17    & 15.85    & 0.25    & 15.67    & 0.25    & 21    & 19    & 19    \\
SN2005na           & 53740.57    & 0.36    & CfA3, C10, G10            & 0.02683    & 0.00102    &  \nodata     &  \nodata     & 17.40    & 0.07    & 17.12    & 0.25    & 16.80    & 0.26    & 13    & 4     & 10    \\
SN2006D            & 53757.30    & 0.21    & CfA3, C10, G10            & 0.00965    & 0.00113    &  \nodata     &  \nodata     & 14.34    & 0.02    & 14.61    & 0.04    & 14.45    & 0.06    & 23    & 21    & 17    \\
SN2006E            & 53729       & 10      & ATEL 690                  & 0.00364    & 0.00134    &  \nodata     &  \nodata     & 14.91    & 0.01    & 14.08    & 0.01    & 14.22    & 0.06    & 30    & 29    & 25    \\
SN2006N            & 53760.44    & 0.50    & CfA3                      & 0.01427    & 0.00100    & 0.468        & 0.066        & 16.02    & 0.09    & 15.65    & 0.23    & 15.49    & 0.04    & 14    & 12    & 7     \\
SN2006X            & 53785.90    & 0.11    & CfA3, C10, WX08, G10      & 0.00627    & 0.00121    & -0.040       & 0.030        & 12.92    & 0.01    & 12.90    & 0.02    & 12.81    & 0.03    & 45    & 44    & 37    \\
SN2006ac           & 53781.38    & 0.30    & CfA3, G10                 & 0.02412    & 0.00104    & 0.230        & 0.062        & 16.82    & 0.12    & 17.03    & 0.11    & 16.55    & 0.17    & 22    & 15    & 16    \\
SN2006ax           & 53826.98    & 0.14    & CfA3, C10                 & 0.01797    & 0.00107    &  \nodata     &  \nodata     & 15.82    & 0.01    & 15.92    & 0.17    & 15.87    & 0.08    & 19    & 15    & 16    \\
SN2006cp           & 53896.76    & 0.14    & CfA3, G10                 & 0.02332    & 0.00105    & -0.166       & 0.048        & 16.96    & 0.08    & 16.84    & 0.08    & 16.06    & 0.14    & 5     & 5     & 3     \\
SN2006cz           & 53903       & 3       & CfA3, CBET 550            & 0.04253    & 0.00102    &  \nodata     &  \nodata     & 17.63    & 0.06    & 17.61    & 0.28    & 17.17    & 0.30    & 4     & 2     & 1     \\
SN2006gr           & 54012.07    & 0.15    & CfA3, G10                 & 0.03348    & 0.00097    & -0.257       & 0.032        & 17.30    & 0.25    & 16.61    & 0.18    & 16.43    & 0.16    & 7     & 5     & 2     \\
SN2006le           & 54047.36    & 0.14    & CfA3, G10                 & 0.01727    & 0.00099    & -0.219       & 0.031        & 16.14    & 0.02    & 16.36    & 0.08    & 16.04    & 0.08    & 39    & 36    & 31    \\
SN2006lf           & 54044.79    & 0.13    & CfA3, G10                 & 0.01297    & 0.00098    & 0.304        & 0.059        & 15.57    & 0.17    & 15.53    & 0.06    & 15.35    & 0.25    & 40    & 41    & 28    \\
SN2006mq           & 54031       & 10      & CBET 724, CBET 731        & 0.00405    & 0.00125    &  \nodata     &  \nodata     & 13.82    & 0.01    & 12.78    & 0.01    & 12.81    & 0.00    & 45    & 45    & 45    \\
SN2007S            & 54143.25    & 0.17    & CfA3, S11                 & 0.01505    & 0.00108    & -0.303       & 0.028        & 15.36    & 0.02    & 15.32    & 0.25    & 15.18    & 0.04    & 29    & 27    & 25    \\
SN2007ca           & 54226.80    & 0.15    & CfA3, S11, G10            & 0.01511    & 0.00107    &  \nodata     &  \nodata     & 15.92    & 0.25    & 15.77    & 0.07    & 15.47    & 0.18    & 18    & 18    & 10    \\
SN2007co           & 54264.61    & 0.24    & CfA3, G10                 & 0.02657    & 0.00099    & -0.035       & 0.046        & 17.89    & 0.17    & 17.57    & 0.18    & 16.50    & 0.22    & 7     & 6     & 5     \\
SN2007cq           & 54280.50    & 0.25    & CfA3, G10                 & 0.02503    & 0.00095    &  \nodata     &  \nodata     & 16.40    & 0.04    & 16.70    & 0.19    & 15.29    & 0.25    & 6     & 6     & 6     \\
SN2007fb           & 54287       & 3       & CfA4, CBET 993            & 0.01681    & 0.00093    & 0.348        & 0.076        & 16.58    & 0.25    & 16.70    & 0.18    & 17.03    & 0.28    & 2     & 2     & 1     \\
SN2007le           & 54398.83    & 0.14    & CfA4, S11, G10            & 0.00551    & 0.00082    & -0.111       & 0.033        & 13.76    & 0.02    & 13.91    & 0.01    & 13.76    & 0.18    & 35    & 31    & 25    \\
SN2007nq           & 54396.94    & 0.47    & CfA4, S11                 & 0.04390    & 0.00098    & 0.361        & 0.063        & 18.84    & 0.17    & 18.36    & 0.06    & 17.76    & 0.19    & 3     & 2     & 3     \\
SN2007qe           & 54428.87    & 0.15    & CfA3, G10                 & 0.02286    & 0.00095    & -0.215       & 0.035        & 17.22    & 0.06    & 16.71    & 0.05    & 16.91    & 0.17    & 8     & 8     & 7     \\
SN2007rx           & 54441       & 3       & CfA4, CBET 1157           & 0.02890    & 0.00096    & -0.249       & 0.080        & 17.10    & 0.07    & 16.56    & 0.06    & 16.45    & 0.08    & 5     & 5     & 5     \\
SN2007sr           & 54447.92    & 0.51    & CfA3, S08, G10            & 0.00665    & 0.00122    & -0.083       & 0.040        & 14.06    & 0.02    & 13.44    & 0.03    & 13.39    & 0.03    & 30    & 32    & 32    \\
SN2008C            & 54464.79    & 0.59    & CfA4, S11, G10            & 0.01708    & 0.00103    & -0.038       & 0.046        & 16.89    & 0.31    & 16.46    & 0.17    & 14.89    & 0.25    & 8     & 4     & 12    \\
SN2008Z            & 54514.66    & 0.19    & CfA4, G10                 & 0.02183    & 0.00104    & -0.176       & 0.038        & 16.45    & 0.03    & 16.55    & 0.18    & 16.16    & 0.10    & 45    & 44    & 32    \\
SN2008af           & 54500.47    & 1.02    & CfA3                      & 0.03411    & 0.00102    & 0.275        & 0.092        & 18.16    & 0.25    & 17.24    & 0.25    & 17.01    & 0.25    & 23    & 31    & 21    \\
SNf20080514-002    & 54611.55    & 0.42    & G10                       & 0.02306    & 0.00104    & 0.275        & 0.068        & 16.51    & 0.11    & 16.61    & 0.12    & 16.47    & 0.18    & 9     & 9     & 8     \\
SNf20080522-000    & 54621.28    & 0.48    & CfA4                      & 0.04817    & 0.00102    & -0.137       & 0.075        & 18.06    & 0.17    & 17.17    & 0.25    & 16.79    & 0.30    & 4     & 3     & 1     \\
SNf20080522-011    & 54617       & 2       & CfA4                      & 0.04026    & 0.00101    & -0.141       & 0.053        & 18.68    & 0.08    & 17.59    & 0.12    & 17.24    & 0.18    & 8     & 9     & 2     \\
SN2008fr           & 54732       & 2       & CfA4                      & 0.04793    & 0.00098    & -0.126       & 0.046        & 17.72    & 0.05    & 18.18    & 0.32    & 16.68    & 0.17    & 5     & 6     & 6     \\
SN2008fv           & 54749.80    & 0.20    & CfA5, Bi12                & 0.00954    & 0.00102    &  \nodata     &  \nodata     & 14.91    & 0.25    & 14.98    & 0.25    & 14.84    & 0.25    & 3     & 3     & 3     \\
SN2008fx           & 54729       & 3       & CBET 1525                 & 0.05814    & 0.00099    &  \nodata     &  \nodata     & 18.72    & 0.12    & 18.37    & 0.10    & 17.50    & 0.18    & 6     & 5     & 5     \\
SN2008gb           & 54745.42    & 1.09    & CfA4                      & 0.03643    & 0.00098    & -0.093       & 0.073        & 17.78    & 0.21    & 17.67    & 0.25    & 17.19    & 0.28    & 19    & 14    & 12    \\
SN2008gl           & 54768.13    & 0.27    & CfA4                      & 0.03297    & 0.00097    & 0.311        & 0.081        & 17.14    & 0.17    & 17.08    & 0.18    & 16.45    & 0.17    & 9     & 12    & 10    \\
SN2008hm           & 54804.33    & 0.41    & CfA4                      & 0.01918    & 0.00098    & -0.122       & 0.052        & 16.36    & 0.03    & 16.48    & 0.21    & 16.06    & 0.18    & 26    & 22    & 23    \\
SN2008hs           & 54812.64    & 0.15    & CfA4                      & 0.01664    & 0.00096    & 0.927        & 0.070        & 16.37    & 0.07    & 16.49    & 0.05    & 16.17    & 0.12    & 20    & 21    & 17    \\
SN2008hv           & 54816.91    & 0.11    & CfA4, S11                 & 0.01359    & 0.00108    & 0.376        & 0.051        & 15.14    & 0.25    & 15.44    & 0.04    & 15.15    & 0.08    & 26    & 29    & 24    \\
SN2008hy           & 54803       & 5       & AAVSO 392, CBET 1610      & 0.00821    & 0.00097    &  \nodata     &  \nodata     & 15.67    & 0.03    & 14.72    & 0.02    & 14.68    & 0.06    & 27    & 23    & 20    \\
SN2009D            & 54842       & 2       & CfA4, CBET 1647           & 0.02467    & 0.00099    & -0.106       & 0.058        & 16.31    & 0.01    & 16.78    & 0.25    & 16.29    & 0.25    & 27    & 24    & 19    \\
SN2009Y            & 54875.89    & 0.48    & CfA4                      & 0.01007    & 0.00108    & -0.116       & 0.051        & 16.52    & 0.22    & 16.93    & 0.25    & 16.98    & 0.25    & 11    & 15    & 3     \\
SN2009ad           & 54886.05    & 0.24   & CfA4                      & 0.02834    & 0.00100    &  \nodata     &  \nodata     & 16.82    & 0.08    & 16.92    & 0.10    & 16.51    & 0.14    & 27    & 20    & 19    \\
SN2009al           & 54896.41    & 0.31    & CfA4                      & 0.02329    & 0.00105    &  \nodata     &  \nodata     & 16.52    & 0.03    & 16.55    & 0.04    & 15.84    & 0.14    & 22    & 22    & 19    \\
SN2009an           & 54898.21    & 0.24    & CfA4                      & 0.00954    & 0.00104    & 0.350        & 0.079        & 14.85    & 0.06    & 15.08    & 0.04    & 14.97    & 0.03    & 31    & 29    & 22    \\
SN2009bv           & 54926.33    & 0.38    & CfA4                      & 0.03749    & 0.00102    & -0.180       & 0.056        & 17.34    & 0.07    & 17.43    & 0.09    & 16.91    & 0.20    & 13    & 13    & 8     \\
SN2009do           & 54945       & 2       & CfA4, CBET 1778           & 0.04034    & 0.00102    & 0.079        & 0.072        & 18.12    & 0.13    & 17.84    & 0.18    & 16.64    & 0.25    & 14    & 9     & 5     \\
SN2009ds           & 54960.50    & 0.38    & CfA4                      & 0.02045    & 0.00106    & -0.120       & 0.056        & 16.22    & 0.23    & 16.20    & 0.17    & 15.29    & 0.25    & 6     & 6     & 3     \\
SN2009fw           & 54993       & 3       & CBET 1849                 & 0.02739    & 0.00097    &  \nodata     &  \nodata     & 15.94    & 0.09    & 15.65    & 0.25    & 14.27    & 0.18    & 6     & 5     & 5     \\
SN2009fv           & 54998       & 3       & CfA4, CBET 1846           & 0.02937    & 0.00100    & 0.238        & 0.188        & 16.30    & 0.16    & 15.90    & 0.25    & 15.57    & 0.26    & 6     & 5     & 3     \\
SN2009ig           & 55079.43    & 0.25    & CfA4                      & 0.00801    & 0.00091    & -0.238       & 0.038        & 15.34    & 0.18    & 14.70    & 0.17    & 14.72    & 0.25    & 11    & 9     & 7     \\
SN2009im           & 55074       & 3       & CBET 1934                 & 0.01256    & 0.00096    &  \nodata     &  \nodata     & 16.60    & 0.07    & 16.11    & 0.03    & 16.21    & 0.02    & 9     & 11    & 6     \\
SN2009jr           & 55119.83    & 0.49    & CfA4                      & 0.01562    & 0.00094    & -0.167       & 0.058        & 16.37    & 0.17    & 16.66    & 0.18    & 16.34    & 0.25    & 11    & 14    & 11    \\
SN2009kk           & 55125.83    & 0.73    & CfA4                      & 0.01244    & 0.00097    & 0.237        & 0.069        & 15.72    & 0.05    & 15.96    & 0.06    & 15.33    & 0.18    & 17    & 17    & 16    \\
SN2009kq           & 55154.61    & 0.35    & CfA4                      & 0.01236    & 0.00107    & -0.030       & 0.052        & 15.47    & 0.18    & 15.38    & 0.04    & 15.27    & 0.17    & 10    & 11    & 11    \\
SN2009le           & 55165.41    & 0.23    & CfA4                      & 0.01704    & 0.00096    & -0.096       & 0.106        & 15.68    & 0.06    & 15.85    & 0.17    & 15.81    & 0.18    & 9     & 7     & 8     \\
SN2009lf           & 55148       & 2       & CfA4, CBET 2025           & 0.04409    & 0.00098    & 0.338        & 0.085        & 17.70    & 0.08    & 17.81    & 0.18    & 17.86    & 0.36    & 18    & 16    & 7     \\
SN2009na           & 55201.31    & 0.28    & CfA4                      & 0.02202    & 0.00105    & 0.052        & 0.060        & 16.47    & 0.25    & 16.44    & 0.18    & 16.61    & 0.17    & 11    & 10    & 8     \\
SN2010Y            & 55247.76    & 0.14    & CfA4                      & 0.01126    & 0.00103    & 0.826        & 0.063        & 15.23    & 0.02    & 15.20    & 0.18    & 15.82    & 0.23    & 15    & 10    & 12    \\
PS1-10w            & 55248.01    & 0.11    & R14                       & 0.03176    & 0.00102    &  \nodata     &  \nodata     & 17.00    & 0.06    & 17.34    & 0.17    & 17.35    & 0.34    & 10    & 10    & 5     \\
PTF10bjs           & 55256       & 3       & CfA4, ATEL 2453           & 0.03055    & 0.00102    &  \nodata     &  \nodata     & 16.95    & 0.06    & 17.09    & 0.07    & 16.48    & 0.17    & 11    & 12    & 10    \\
SN2010ag           & 55270.23    & 0.63    & CfA4                      & 0.03376    & 0.00100    & -0.249       & 0.051        & 17.13    & 0.01    & 17.14    & 0.26    & 16.50    & 0.25    & 15    & 15    & 9     \\
SN2010ai           & 55276.84    & 0.13    & CfA4                      & 0.01927    & 0.00105    & 0.358        & 0.074        & 16.56    & 0.04    & 16.67    & 0.11    & 16.49    & 0.10    & 22    & 17    & 17    \\
SN2010cr           & 55315       & 3       & CfA4, ATEL 2580           & 0.02253    & 0.00104    &  \nodata     &  \nodata     & 16.65    & 0.01    & 17.24    & 0.17    & 16.80    & 0.17    & 15    & 12    & 8     \\
SN2010dl           & 55341       & 3       & CBET 2298                 & 0.02892    & 0.00096    &  \nodata     &  \nodata     & 17.58    & 0.11    & 17.35    & 0.18    & 16.59    & 0.28    & 5     & 3     & 5     \\
PTF10icb           & 55360       & 3       & Pa11                      & 0.00905    & 0.00105    &  \nodata     &  \nodata     & 14.63    & 0.02    & 14.80    & 0.17    & 14.58    & 0.25    & 12    & 12    & 7     \\
SN2010dw           & 55357.75    & 0.65    & CfA4                      & 0.03870    & 0.00102    & -0.146       & 0.088        & 17.78    & 0.05    & 17.55    & 0.25    & 16.66    & 0.25    & 6     & 6     & 4     \\
SN2010ew           & 55379       & 3       & CBET 2356                 & 0.02504    & 0.00098    &  \nodata     &  \nodata     & 16.53    & 0.25    & 16.59    & 0.25    & 15.39    & 0.25    & 5     & 4     & 4     \\
SN2010ex           & 55386       & 3       & CBET 2353                 & 0.02164    & 0.00095    &  \nodata     &  \nodata     & 17.06    & 0.11    & 16.79    & 0.23    & 16.10    & 0.17    & 2     & 2     & 2     \\
SN2010gn           & 55399       & 3       & CfA5, CBET 2386           & 0.03638    & 0.00100    & 0.023        & 0.099        & 16.85    & 0.18    & 17.42    & 0.23    & 17.09    & 0.37    & 3     & 3     & 2     \\
SN2010iw           & 55492       & 6       & CfA5, CBET 2511           & 0.02230    & 0.00104    & -0.169       & 0.056        & 16.38    & 0.05    & 16.41    & 0.10    & 16.31    & 0.17    & 18    & 18    & 13    \\
SN2010ju           & 55523.80    & 0.44    & CfA5                      & 0.01535    & 0.00101    & 0.054        & 0.110        & 15.83    & 0.02    & 15.84    & 0.06    & 15.32    & 0.18    & 21    & 20    & 19    \\
SN2010jv           & 55516       & 3       & CBET 2550                 & 0.01395    & 0.00104    &  \nodata     &  \nodata     & 15.44    & 0.05    & 15.42    & 0.10    & 14.82    & 0.17    & 3     & 3     & 2     \\
SN2010kg           & 55543.48    & 0.13    & CfA5                      & 0.01644    & 0.00099    & 0.281        & 0.069        & 15.76    & 0.07    & 15.86    & 0.11    & 15.71    & 0.17    & 25    & 27    & 12    \\
SN2011B            & 55582.92    & 0.13    & CfA5                      & 0.00474    & 0.00101    & 0.142        & 0.054        & 13.21    & 0.17    & 13.33    & 0.18    & 13.34    & 0.18    & 46    & 43    & 37    \\
SN2011K            & 55578       & 3       & CfA5, CBET 2636           & 0.01438    & 0.00099    & -0.138       & 0.076        & 15.54    & 0.01    & 15.63    & 0.09    & 15.59    & 0.28    & 16    & 16    & 8     \\
SN2011ae           & 55619    & 3    & CfA5                      & 0.00724    & 0.00120    & -0.235       & 0.063        & 13.69    & 0.02    & 13.70    & 0.03    & 13.65    & 0.25    & 32    & 32    & 26    \\
SN2011ao           & 55638.26    & 0.15    & CfA5                      & 0.01162    & 0.00109    & -0.157       & 0.037        & 14.83    & 0.03    & 14.99    & 0.03    & 14.95    & 0.08    & 28    & 29    & 16    \\
SN2011at           & 55635       & 5       & CfA5, CBET 2676           & 0.00787    & 0.00116    & 0.321        & 0.398        & 15.04    & 0.02    & 14.21    & 0.17    & 14.25    & 0.04    & 13    & 14    & 10    \\
SN2011by           & 55690.60    & 0.15    & CfA5                      & 0.00341    & 0.00120    & -0.007       & 0.046        & 13.17    & 0.14    & 13.37    & 0.03    & 13.55    & 0.18    & 28    & 27    & 13    \\
SN2011df           & 55716.08    & 0.41    & CfA5                      & 0.01403    & 0.00096    & -0.157       & 0.070        & 15.49    & 0.03    & 15.62    & 0.06    & 15.50    & 0.17    & 24    & 25    & 11    \\
\\
\hline
\end{tabular}
\tablecomments{ \tiny
\\
{\bf (a)} MJD of \tbmax{} and error from \texttt{MLCS2k2.v007}~\citep{jha07} fits to $B$-band LCs from the CfA or the literature, where available. \tbmax{} fits from the literature are used for \sn{}2008fv (\citealt{biscardi12}), PS1-10w (\citealt{rest14}).  For objects with no optical data or bad MLCS fits with reduced $\chi^2 > 3$, \tbmax{} is estimated from any or all of: the MJD of the brightest point (rounded to the nearest day), optical spectra in listed CBET/IAUC/ATEL notices, and cross checked with fitted phases of NIR LC features, where possible (see F12).  This applies to all SN in Table~\ref{tab:jhkdata} with \tbmax{} and error rounded to the nearest day, with most assuming a $\pm 2-3$ day uncertainty.
Of these we observed \sn{}2009fw, \sn{}2010ew, \sn{}2010ex, and \sn{}2010jv at the CfA but do not have successfully reduced optical LCs for these objects, which are marked CfA? and may or may not be included in CfA5. 
2 objects, \sn{}2006E, \sn{}2006mq were discovered several weeks after maximum and have only late time optical data, and only rough \tbmax{} estimates from optical spectra (these assume a $\pm 10$ day uncertainty).
Other objects with \tbmax{} from early optical data but with only late time NIR data where the first \PTL{} observation is at a phase $\gtrsim 20$ days after \tbmax{} include \sn{}2007qe, \sn{}2009ig, \sn{}2009im. 
\\
{\bf (b)} Optical LC References:
CfA5: in preparation; CfA4: \citet{hicken12}; CfA3: \citet{hicken09b}; CfA2: \citet{jha06}; CfA1: \citet{riess99}; F09: \citet{foley09b}; R14: \citealt{rest14}; Br12: \citet{bryngelson12}; Bi12: \citet{biscardi12}; S11: \citet{stritzinger11}; Pa11: \citet{parrent11}; C10: \citet{contreras10}; 
G10: \citet{ganeshalingam10}; WX09: \citet{wangx09}; WX08: \citet{wangx08}; S08: \citet{schweizer08}.
\\
{\bf (c)} Redshift $z_{\rm CMB}$ and error converted to CMB frame with apex vectors from \citealt{fixsen96} (see NED: \url{http://ned.ipac.caltech.edu/help/velc\_help.html}).  Redshifts have not been corrected with any galactic flow models. Heliocentric redshifts (and references) and galactic coordinates are in Tables~\ref{tab:general}-\ref{tab:general3}.
\\
{\bf (d)} $\Delta$ parameter from \texttt{MLCS2k2.v007}~\citep{jha07} fits to optical data from the CfA and/or the literature, where available.  Only fits with reduced $\chi^2 < 3$ are included.  The following objects were not run through \texttt{MLCS2k2}: PS1-10w (PanSTARRS1: \tbmax{} from \texttt{SALT} fit in \citealt{rest14}), \sn{}2008fv (\tbmax{} in \citealt{biscardi12}). PTF10icb (\citealt{parrent11}, PTF) have unpublished optical data;
PTF10bjs (PTF) has unpublished data and is in CfA4, but only in the r' i' natural system and not standard system magnitudes (\citealt{hicken09b});  \sn{}2006E (\citealt{bryngelson12}) and \sn{}2006mq (CfA3) have only late time optical data.
\\
{\bf (e)} Magnitudes and 1-$\sigma$ uncertainties in \jhk{} LCs at the brightest LC point (This is not necessarily the \jhk{} magnitude at the first NIR maximum or at \tbmax{}).
\\
{\bf (f)} Number of epochs w/ S/N$>3$ in the \jhk{} lightcurves, respectively. 
}
\label{tab:jhkdata}
\end{center}
\end{table*}
\renewcommand{\tabcolsep}{6pt}
\renewcommand{\arraystretch}{1}

\section{Discussion}
\label{sec:disc}

The \nsnIacfair{} \F{} NIR \snIa{} and \nsnIaxcfair{} \snIax{} LCs obtained in the northern hemisphere with \PTL{}
are matched only by the comparable, excellent quality, southern hemisphere CSP data set, which includes 72 \snIa{} LCs (and 1 \snIax{}) with at least 1 band of published $YJHK_s$ data (see Table~\ref{tab:ptel_csp_census}). 
The \F{} and CSP data sets are quite complementary, observing mostly different objects with varying observation frequencies in individual NIR bandpasses (see \S\ref{sec:csp_comp}). \F{} includes more than twice as many $JH$ observations and more than ten times as many $K_s$ observations as CSP.  By contrast, the CSP $Y$-band observations form a unique data set, since no CfA telescopes at FLWO currently have $Y$-band filters (see Table~\ref{tab:ptel_csp_census}). 

While \F{} presents more total NIR \snIa{} and \snIax{} LCs than the CSP (98 vs. \ncspir{}), more unique LCs (78 vs. \ncspir{}), and includes $\sim 3$--$4$ times the number of individual NIR observations, CSP photometric uncertainties are typically $\sim 2-3$ times smaller than for \F{} (see Table~\ref{tab:ptel_csp_census}), as a result of key differences between the NIR capabilities at CfA and CSP observing sites (see Table 2.1 of F12). These include better seeing at LCO vs. FLWO, a newer, higher resolution camera on the Swope 1.0-m telescope compared to the 2MASS south camera on the \PTL-1.3m telescope, and CSP host galaxy template images sometimes taken with the 2.5-m du Pont telescope compared to \F{} template images taken with the 1.3-m \PTL{} using an undersampled camera.
Overall, the CSP \jhk{} photometric precision for observations of the {\it same objects} at the brightest LC point is generally a factor of $\sim 2-3$ better than \PTL{}, with median \jhk{} uncertainties of $\sim 0.01$--$0.02$ mag for CSP and $\sim 0.02-0.05$ mag for \PTL{} (see Table~\ref{tab:ptel_csp_census}). 
More specifically, while CSP has fewer $K_s$-band measurements, the peak photometric precision is $\sim 3$ times better than \PTL{} mainly because the CSP $K_s$ filter is on the duPont 2.5-m telescope, as compared to the \PTL{} 1.3-m.  What the CSP lacks in quantity compared to \F{}, it makes up for in quality.   

However, unlike the CSP NIR data, since \PTL{} photometry is {\it already on} the standard 2MASS \jhk{} system, no zero point offsets or color term corrections (e.g., \citealt{carpenter01,leggett06}) or S-corrections based on highly uncertain NIR \snIa{} SEDs (e.g., \citealt{stritzinger02}) are needed to transform \F{} data to the 2MASS passbands. Avoiding additional systematic uncertainty from S-corrections is a significant advantage for \PTL{} \F{} data, since the published spectral sample of only 75 NIR spectra of 33 \snIa{} is still quite limited (\citealt{hsiao07,marion09,boldt14}).  This advantage also applies to future cosmological uses of \PTL{} data that would employ state-of-the-art NIR K-corrections to transform LCs to the rest-frame 2MASS filter system as the current world NIR spectral sample is increased. Even for relatively nearby $z \sim 0.08$ objects, NIR K-corrections in $YJHK_s$ currently contribute uncertainties of $\sim0.04$-$0.10$ mag to distance estimates \citep{boldt14}. Since NIR K-corrections at $z \sim 0.08$ can themselves have values ranging from $\sim -0.8$ to $\sim 0.4$ mag, depending on the filter and phase, they can yield significant systematic distance errors if ignored \citep{boldt14}.

\begin{table}
\centering
\caption[\PTL{} and CSP NIR Data Census]{\PTL{} and CSP NIR Data Census \\}
\scriptsize 
\begin{tabular}{@{}c | c | c | c | c | c  | c | c@{} }
\hline
Project & SN$^a$ & NIR$^b$ & $Y$$^b$ & $J$$^b$ & $H$$^b$ & $K_s$$^b$& $\sigma$ [mag]$^c$ \\
\hline
\F{}       &   98    &    \nptelobs  &  0                 &    \nptelobsj  &     \nptelobsh   &    \nptelobsk  &    0.02-0.05   \\
CSP     &    \ncspir               &    \ncspobsir & \ncspobsy   &    \ncspobsj  &     \ncspobsh   &     \ncspobsk  &   0.01-0.03 \\
\hline
\end{tabular}
\tablecomments{ \scriptsize
\\
({\it a}) Number of \snIa{} and \snIax{} with NIR $YJH$ or $K_s$ observations in \F{} (this paper) or CSP (C10; S11; \citealt{phillips07,schweizer08,taubenberger11,stritzinger10}).
\\
({\it b}) Number of epochs of photometry.
\\
({\it c}) Median magnitude uncertainties for \F{} and CSP for same objects at the brightest LC pt.
}
\label{tab:ptel_csp_census}
\end{table}

\section{Conclusions}
\label{sec:conc}

This work presents the \F{} data set, including \nsnIacfair{} NIR \jhk{}-band \snIa{} and \nsnIaxcfair{} \snIax{} LCs observed from 2005--2011 with \PTL{}. The \nptelobs{} individual \F{} data points represent the largest homogeneously observed and reduced set of NIR \snIa{} and \snIax{} observations to date, nearly \obsmulting{} the number of individual \jhk{} photometric observations 
from the CSP, surpassing the number of unique CSP objects, and increasing the total number of spectroscopically normal \snIa{} with published NIR LCs in the literature by $\sim65$\%.\footnote{Including revised photometry for 12 \PTL{} objects with no CSP or other NIR data.  } \F{} presents revised photometry for 20 out of \nsnIaredo{} \WV{} objects (and \sn{}2008ha from \citealt{foley09b}) with more accurate flux measurements and increased agreement for the subset of \F{} objects also observed by the CSP, as a result of greatly improved data reduction and photometry pipelines, applied nearly homogeneously to all \F{} SN.\footnote{With the exception of \sn{}2005cf and \sn{}2008A (see \S\ref{sec:reduction}-\ref{sec:phot}). SN of other types were also reduced using the same mosaicking and photometry pipelines as the \F{} data set and are presented elsewhere (e.g., \citealt{bianco14}).} 

Previous studies have presented evidence that \snIa{} are more standard in NIR luminosity than at optical wavelengths, less sensitive to dimming by host galaxy dust, and crucial to reducing systematic galaxy distance errors as a result of the degeneracy between intrinsic supernova color variation and reddening of light by dust, the most dominant source of systematic error in \snIa{} cosmology (K04a; \WV; M09; F10; \citealt{burns11}; M11; K12; \citealt{burns14}).  Combining \PTL{} \WV{} \snIa{} data with optical and NIR data from the literature has already demonstrated that including NIR data helps to break the degeneracy between reddening and intrinsic color, making distance estimates less sensitive to model assumptions of individual LC fitters (M11; \citealt{mandel14a}).  \F{} photometry will allow the community to further test these conclusions.  

The addition of \F{} to the literature presents clear new opportunities. 
A next step for the community is combine \F{}, CSP, and other NIR and optical low redshift \snIa{} LC databases together using S-corrections, or color terms like those derived in this paper, to transform all the LCs to a common filter system. This optical and NIR data can be used to compute optical-NIR colors, derive dust and distance estimates, and construct optical and NIR Hubble diagrams for the nearby universe that are more accurate and precise than studies with optical data alone (e.g., M11).
Empirical LC fitting and \snIa{} inference methods that handle both optical and NIR data (e.g., \bayesn{}: M09; M11 and \snoopy{}: \citealt{burns11}) can be extended to utilize low and high-$z$ \snIa{} samples to obtain cosmological inferences and dark energy constraints that take full advantage of \F{}, CSP and other benchmark NIR data sets.

Increasingly large, homogeneous, data sets like \F{}, raise hopes that \snIa{}, especially in the rest-frame $YH$ bands, can be developed into the most precise and accurate of cosmological distance probes. 
This hope is further bolstered by complementary progress modeling \snIa{} NIR LCs theoretically (e.g., \citealt{kasen06,jack12}) and empirically (M09; M11; \citealt{burns11}).
Combining future $IJHYK_s$ data with $\gtrsim \nirIa$ NIR \snIa{} LCs from \F{}, the CSP, and the literature, will provide a growing low-$z$ training set to study the intrinsic NIR properties of nearby \snIa{}. 
This NIR data can better constrain the parent populations of host galaxy dust and extinction, elucidating the properties of dust in external galaxies, and allowing researchers to disentangle \snIa{} reddening from dust and intrinsic color variation (M11).

\F{} data should be further useful for a number of cosmological and other applications. 
Improved NIR distance measurements could also allow mapping of the local velocity flow independent of cosmic expansion to understand how peculiar velocities in the nearby universe affect cosmological inferences from \snIa{} data (\citealt{turnbull11,davis11}). NIR data should also provide the best \snIa{} set with which to augment existing optical measurements of the Hubble Constant (e.g., \citealt{riess11}). See \citealt{cartier14} for a specific use of NIR \snIa{} data to measure $H_0$.  Future work can compare NIR LC features and host-galaxy properties, which have been shown to correlate with Hubble diagram residuals for optical \snIa{} \citep{kelly10}. Adding NIR spectroscopy to optical and infrared photometry can also help test physical models of exploding white dwarf stars (e.g.,~\citealt{kasen06,jack12}), and investigate NIR spectral features that correlate with \snIa{} luminosity, helping to achieve improved \snIa{} distance estimates, similar to what has already been demonstrated with optical spectra (\citealt{bailey09,blondin11,mandel14a}). 

Our work emphasizing the intrinsically standard and relatively dust insensitive nature of NIR \snIa{} has highlighted the rest-frame NIR as a promising wavelength range for future space based cosmological studies of \snIa{} and dark energy, where reducing systematic uncertainties from dust extinction and intrinsic color variation become more important than simply increasing the statistical sample size (e.g., \citealt{gehrels10,beaulieu10,astier11}).  Although ground-based NIR data can be obtained for low redshift objects, limited atmospheric transmission windows require that rest-frame NIR observations of high-z \snIa{} be done from space. 
Currently, rest-frame \snIa{} Hubble diagrams of high-$z$ \snIa{} have yet to be constructed beyond the $I$ band \citep{freedman05,nobili05,freedman09}, with limited studies of \snIa{} and their host galaxies conducted in the mid-infrared with Spitzer \citep{chary05,gerardy07}. Our nearby NIR observations at the CfA with \PTL{} have been recently augmented by {\bf RAISIN}: Tracers of cosmic expansion with {\bf SN IA} in the {\bf IR}, an ongoing Hubble Space Telescope (HST) program (begun in Cycle 20) to observe $\sim25$ \snIa{} at $z\sim0.35$ in the rest-frame NIR with WFC3/IR.
 
Along with current and future NIR data, \F{} will provide a crucial low-$z$ anchor for future space missions capable of high-$z$ \snIa{} cosmology in the NIR, including WFIRST (the Wide-Field Infrared Survey Telescope; a candidate for JDEM, the NASA/DOE Joint Dark Energy Mission; \citealt{gehrels10}), The European Space Agency's EUCLID mission \citep{beaulieu10}, and the NASA James Webb Space Telescope (JWST; \citealt{clampin11}).
To fully utilize the standard nature of rest-frame \snIa{} in the NIR and ensure the most precise and accurate extragalactic distances, the astronomical community should strongly consider space-based detectors with rest-frame NIR capabilities toward as long a wavelength as possible. 

Until the launch of next generation NIR space instruments, continuing to observe \snIa{} in the NIR from the ground with observatories like \PTL{}
 and from space with HST programs like RAISIN is the best way to reduce the most troubling fundamental uncertainties in \snIa{} cosmology as a result of dust extinction and intrinsic color variation. Ultimately, the \F{} sample of nearby, low-redshift, NIR \snIa{} will help lay the groundwork for next generation ground-based cosmology projects and space missions that observe very distant \snIa{} at optical and NIR wavelengths to provide increasingly precise and accurate constraints on dark energy and its potential time variation over cosmic history. NIR \snIa{} observations thus promise to play a critical role in elucidating the nature of one of the most mysterious discoveries in modern astrophysics and cosmology. 

\acknowledgments
%
{\tiny \singlespace
The Peters Automated Infrared Imaging Telescope (\PTL{}) is operated by the Smithsonian Astrophysical Observatory (SAO) and was enabled by a grant from the Harvard University Milton Fund, the camera loan from the University of Virginia, and continued support of the SAO and UC Berkeley. 
Partial support for \PTL{} operations and this work comes from National Aeronautics and Space Administration (NASA) {\it Swift} Guest Investigator grant NNG06GH50G (``\PTL{}: Infrared Follow-up for Swift Transients''). \PTL{} support and processing is conducted under the auspices of a DOE SciDAC grant (DE-FC02-06ER41453), which provides support to J.S.B.'s group. J.S.B. thanks the Sloan Research Fellowship for partial support as well as NASA grant NNX13AC58G.
We gratefully made use of the NASA/IPAC Extragalactic Database (NED). The NASA/IPAC Extragalactic Database (NED) Is operated by the Jet Propulsion Laboratory, California Institute of Technology, under contract with NASA. 
This publication makes use of data products from the 2MASS Survey, funded by NASA and the US National Science Foundation (NSF). IAUC/CBET were very useful. 
A.S.F. acknowledges support from an NSF STS Postdoctoral Fellowship (SES-1056580), a NSF Graduate Research Fellowship, and a NASA Graduate Research Program Fellowship.
M.W.V. is funded by a grant from the US National Science Foundation (AST-057475).  
R.PK. acknowledges NSF Grants AST 12-11196, AST 09-097303, and AST 06-06772. 
M.M. acknowledges support in part from the Miller Institute at UC Berkeley, from Hubble Fellowship grant HST-HF-51277.01-A, awarded by STScI, which is operated by AURA under NASA contract NAS5-26555, and from the NSF CAREER award AST-1352405.
A.A.M. acknowledges support for this work by NASA from a Hubble Fellowship grant HST-HF-51325.01, awarded by STScI, operated by AURA, Inc., for NASA, under contract NAS 5-26555. Part of the research was carried out at the Jet Propulsion Laboratory, California Institute of Technology, under a contract with the National Aeronautics and Space Administration.
A.S.F, R.P.K, and M.M. thank the Kavli Institute for Theoretical Physics at UC Santa Barbara, which is supported by the NSF through grant PHY05-51164. 
C.B. acknowledges support from the Harvard Origins of Life Initiative. 
Computations in this work were run on machines supported by the Harvard Astronomy Computation Facility including the CfA Hydra cluster and machines supported by the Optical and Infrared Astronomy Division of the CfA. Other crucial computations were performed on the Harvard Odyssey cluster, supported by the Harvard FAS Science Division Research Computing Group.
We thank the anonymous referee for a thorough and fair report that significantly helped to improve the paper.
}




{\scriptsize {\it Facilities:} \facility{FLWO}, \facility{PAIRITEL}, \facility{2MASS}.}





\begin{appendix}
\label{sec:appendix}

\section{NNT Uncertainties}
\label{sec:nnt_math}

We compute the estimated mean flux $\tilde{f}$ and uncertainty $\sgnnt$ for a given night using the \nnt{} host galaxy subtraction method in the following manner. For a night with $\nt$ successful host galaxy template subtractions, we have $\nt$ LC points with flux $f_i$ each with corrected \dophot{} flux uncertainties $\sigma_{f_{{\rm do},i}}$, where $i = \{1,2,\ldots,\nt\}$ indexes the $\nt$ subtractions that are implicitly summed over for every summation symbol $\Sigma$ below.  The estimated flux on this night is simply given by the weighted mean:
\begin{equation}
\tilde{f} = \frac{ \Sigma f_{i}  w_{f_i} }{ \Sigma w_{f_i} }
\label{eq:nntavg}
\end{equation}
\noindent with weights given by $w_{f_i} = 1/\sigma^2_{f_{{\rm do},i}}$. We choose to conservatively estimate the uncertainty on $\tilde{f}$ using the error weighted sample standard deviation of the $\nt$ flux measurements, which has the advantage of being a function of both the input fluxes $f_i$ and corrected \dophot{} flux errors $\sigma_{f_{{\rm do},i}}$ via the weights $w_{f_i} = 1/\sigma^2_{f_{{\rm do},i}}$, given by:
\begin{equation}
\sigma_{\tilde{f}} = \sqrt{ \frac{ \Sigma w_{f_i} ( f_i - \tilde{f} )^2 }{ \Sigma w_{f_i}  } } 
\label{eq:w_stddev_flux_err}
\end{equation}
\noindent However, to correct bias as a result of small sample sizes, which is appropriate here, since $\nt \sim 3-12$, we refine Eq.~\ref{eq:w_stddev_flux_err} and instead use an appropriate unbiased estimator of the weighted sample standard deviation, given by:
\begin{equation}
\sgnnt = \sqrt{  \left[ \frac{\Sigma w_{f_i}}{(\Sigma w_{f_i})^2 - \Sigma w^2_{f_i}}\right] \Sigma w_{f_i} ( f_i - \tilde{f} )^2  } 
\label{eq:w_stddev_unbiased_flux_err}
\end{equation}
\noindent We use Eq.~\ref{eq:w_stddev_unbiased_flux_err} to compute our final $\nnt$ error estimate $\sgnnt$ on the flux averaged over several subtractions on an individual night.  To account for nights with only $\nt=1$ or $2$ successful subtractions, we further implement a systematic error floor with a conservative magnitude cutoff as described in \S\ref{sec:galsub_err} (see Table~\ref{tab:nnt_err}).

\end{appendix}

\vspace{0.3cm}





\bibliography{../../../../../bib/sn,../../../../../bib/sngroup,../../../../../bib/selection,../../../../../bib/mypapers}



\end{document}